\RequirePackage{snapshot}
\documentclass[a4paper, 11pt]{article}
\pdfoutput=1
\usepackage{jcappub}
\usepackage{bbold}
\usepackage{mathtools}
\usepackage{ulem}
\usepackage{float}
\usepackage{caption}
\usepackage{subcaption}
\usepackage{enumitem}
\usepackage{diagbox}
\def\laq{~\raise 0.4ex\hbox{$<$}\kern -0.8em\lower 0.62ex\hbox{$\sim$}~}
\def\gaq{~\raise 0.4ex\hbox{$>$}\kern -0.7em\lower 0.62ex\hbox{$\sim$}~}

\newcommand{\avg}[1]{{\langle{#1}\rangle}}

\newcommand{\deltastd}{{\Delta_\mathrm{std}}}

\def \la {\lambda}

\def \La {\Lambda}
\def \De {\Delta}
\def \de {\delta}

\def \si {\sigma}

\def \Om {\Omega}

\def\laq{~\raise 0.4ex\hbox{$<$}\kern -0.8em\lower 0.62ex\hbox{$\sim$}~}
\def\gaq{~\raise 0.4ex\hbox{$>$}\kern -0.7em\lower 0.62ex\hbox{$\sim$}~}

\def \La {\Lambda}

\def \Om {\Omega}

\newcommand{\bk}{{\bf k}}
\newcommand{\bn}{{\bf n}}

\newcommand{\dd}{\partial}

\newcommand{\HH}{\mathcal H}

\newcommand{\bx}{\boldsymbol{x}}

\newcommand{\rS}{\rm{S}}
\newcommand{\rP}{\rm{P}}

\def\laq{~\raise 0.4ex\hbox{$<$}\kern -0.8em\lower 0.62ex\hbox{$\sim$}~}
\def\gaq{~\raise 0.4ex\hbox{$>$}\kern -0.7em\lower 0.62ex\hbox{$\sim$}~}

\def\be{\begin{equation}}
\def\ee{\end{equation}}
\def\bea{\begin{eqnarray}}
\def\eea{\end{eqnarray}}
\def\bean{\begin{eqnarray*}}
\def\eean{\end{eqnarray*}}

\def \cd {\cdot}

\newcommand{\ft}{\tilde f}
\newcommand{\bt}{\tilde b}
\newcommand\integrand{\ensuremath{\mathrm{d}}}

\title{
On the importance of lensing for galaxy clustering in photometric and spectroscopic surveys
}

\author{{Goran Jelic-Cizmek},}
\author{{Francesca Lepori},}
\author{{Camille Bonvin} and}
\author{{Ruth~Durrer}}

\affiliation{
Universit\'e de Gen\`eve, D\'epartement de Physique Th\'eorique and CAP,
24 quai Ernest-Ansermet, CH-1211 Gen\`eve 4, Switzerland
}

\emailAdd{goran.jelic-cizmek@unige.ch}
\emailAdd{francesca.lepori@unige.ch}
\emailAdd{camille.bonvin@unige.ch}
\emailAdd{ruth.durrer@unige.ch}

\abstract{
We study the importance of gravitational lensing in modelling the number counts of galaxies {for the first time in spectroscopic surveys}. We confirm previous results for photometric surveys, showing that lensing cannot be neglected in a survey like LSST since it would infer a significant shift of standard cosmological parameters. For a spectroscopic survey like SKA2, we find that neglecting lensing in the monopole, quadrupole and hexadecapole of the correlation function can also induce an important shift of cosmological parameters. For $\Lambda$CDM parameters, the shift is moderate, of the order of $0.6\sigma$ or less. However, for a model-independent analysis, that measures the growth rate of structure in each redshift bins, neglecting lensing introduces a shift of up to $2.3\sigma$ at high redshift. Since the growth rate is directly used to test the theory of gravity, such a strong shift would wrongly be interpreted as the breakdown of General Relativity. This shows the importance of including lensing in the analysis of future surveys. For a survey like DESI, we find on the other hand that lensing is not important, mainly due to the value of the magnification bias parameter of DESI, $s(z)$, which strongly reduces the lensing contribution at high redshift. {This result relies on our theoretical modelling of $s(z)$ in DESI and should therefore be confirmed with measurements of $s(z)$ in simulations.}
We also propose a way of improving the analysis of spectroscopic surveys, by including the cross-correlations between different redshift bins (which is neglected in spectroscopic surveys) from the spectroscopic survey or from a different photometric sample. We show that including the cross-correlations in the SKA2 analysis does not improve the constraints. On the other hand replacing the cross-correlations from SKA2 by cross-correlations measured with LSST improves the constraints by {10$\%$ to $20\%$.}
{The marginal improvement is mainly due to the density correlations between nearby bins and, therefore, does not
strongly depend on our knowledge of the magnification bias. }
Interestingly, for standard cosmological parameter estimation, the photometric survey LSST in its 12 redshift bin configuration
{and the spectroscopic SKA2 survey are highly complementary, since they are affected by different degeneracies between parameters:}
LSST yields the tightest constraints on {$\Omega_{\rm cdm}, h$ and $n_s$, while SKA2 better constrains $\Omega_{\rm baryon}, A_s$ and the bias.}
}

\begin{document}

\maketitle

\section{Introduction}\label{s:intro}
At present, the most important data set to constrain cosmological models comes from observations of the cosmic microwave background (CMB) anisotropies and polarization~\cite{Aghanim:2018eyx}.
However, to optimally probe the growth of structure at late time, the CMB needs to be complemented by observations at low redshift.
The 3-dimensional distribution of galaxies provides a powerful way of measuring the growth rate of structure in the late Universe, which is highly sensitive to the theory of gravity.
This growth rate has been successfully measured by several surveys like SDSS~\cite{10.1093/mnras/stu197,10.1093/mnras/sty506}, WiggleZ~\cite{Blake_2011} and BOSS~\cite{10.1093/mnras/sty453,Li_2016}, allowing us to test the consistency of General Relativity over a wide range of scales and redshifts.
This success has triggered the construction of various large-scale structure (LSS) surveys, that are planed for the coming decade, like Euclid~\cite{Laureijs:2011gra,Amendola:2016saw,Blanchard:2019oqi}, LSST~\cite{Abate:2012za,Abell:2009aa}, DESI~\cite{Aghamousa:2016zmz}, SKA~\cite{Maartens:2015mra,Santos:2015gra}, SphereX~\cite{Korngut2018SPHERExAA} and WFIRST~\cite{akeson2019wide}.
These surveys will observe much higher redshifts and larger volumes, improving the precision on tests of gravity.

Even though LSS data are more difficult to interpret than CMB data (non-linearities are relevant on small scales; we observe luminous galaxies made out of baryons, while we compute matter over-densities), on large scales or at higher redshifts we expect that linear perturbation theory or weakly non-linear schemes are sufficient to describe cosmic structure.
The fact that LSS is a three dimensional dataset, and that it allows us to measure the cosmic density field, velocity field and the gravitational potential independently, makes it highly complementary to the CMB.
For these reasons it is very important that we make best use of the LSS data soon to come.

To extract information from the distribution of galaxies, data have to be compressed using adequate estimators. Past and current surveys have mainly focused on the galaxy two-point correlation function in redshift-space and on its Fourier transform, the power spectrum, see e.g.~\cite{Reid:2015gra,Satpathy:2016tct,Alam:2016hwk}. Since redshift-space distortions (RSD) break the isotropy of the correlation function, the optimal way of extracting information is to fit for a monopole, quadrupole and hexadecapole in the two-point function of galaxies (or equivalently in the power spectrum). In the linear regime, these three multipoles contain all the information about density and RSD. These multipoles have been measured successfully in various surveys and have been used to constrain the growth of structure and to determine the baryon acoustic oscillation (BAO) scale, see e.g.~\cite{Aghanim:2018eyx} for an overview of present constraints.

However, it is well known that LSS observations are not only affected by RSD (due to the motion of galaxies)~\cite{Kaiser1987}, but that they are also affected by gravitational lensing due to foreground structures~\cite{Matsubara:2004fr,Scranton:2005ci,Duncan:2013haa}. Gravitational lensing modifies the observed size of the solid angle in which we count how many galaxies we detect, consequently diluting the number of galaxies per unit of solid angle. In addition, it increases the apparent luminosity of galaxies, enhancing the number of galaxies above the magnitude threshold of a given survey. These two effects combine to distort the number counts of galaxies. As a consequence, gravitational lensing contributes to the observed two-point correlation function of galaxies.

In this paper, we investigate {for the first time} the relevance of gravitational lensing for the coming generation of {spectroscopic} LSS surveys.
Various previous analyses have studied the impact of lensing for photometric surveys using the angular power spectrum, $C_\ell(z,z')$~\cite{Montanari:2015rga, Raccanelli:2015vla, Cardona:2016qxn, DiDio:2016ykq, Lorenz:2017iez, Villa_2018, Tanidis:2019fdh}.
These studies have shown that neglecting lensing in future photometric surveys like LSST and Euclid will significantly shift the determination of a host of cosmological parameters.
In this paper, we focus instead on spectroscopic surveys. {This is the first study of} the impact of lensing on the measurement of the growth rate of structure, using the monopole, quadrupole and hexadecapole of the {full sky relativistic correlation function computed with the public code COFFE~\cite{Tansella:2018sld}}.
We show that neglecting lensing in the modelling of these multipoles significantly shifts the determination of cosmological parameters for a survey like SKA2~\cite{Bull:2014rha}.
In particular, if we assume $\Lambda$CDM and constrain the standard parameters $(\Omega_{\rm baryon}, \Omega_{\rm cdm}, h, n_s, A_s)$ and the galaxy bias, we find that neglecting lensing generates a shift in the determination of these parameters, which ranges from 0.1$\sigma$ to 0.6$\sigma$.
More importantly, if we treat the growth rate $f$ and the galaxy bias, as free parameters in each redshift bins (which is what is routinely done in RSD surveys), we find a shift as large as 2.3$\sigma$ for $f$, and 3.1$\sigma$ for the galaxy bias $b(z)$ in the highest redshift bin of SKA2.
Since the growth rate is used to test the theory of gravity, such a large shift could lead us to wrongly conclude that gravity is modified.
Including lensing in the analysis, we moreover find that our imperfect knowledge of the magnification bias parameter, $s(z)$, degrades the constraints on the bias and the growth rate by up to 25\%.
This analysis therefore shows that lensing cannot be neglected in the modelling of multipoles for a survey like SKA2 and a good knowledge of the magnification bias, $s(z)$, is important.

We also perform a similar analysis for a survey like DESI~\cite{Aghamousa:2016zmz}. In this case we find that neglecting lensing has almost no impact on the determination of cosmological parameters. This is mainly due to the fact that at high redshift, where lensing could be relevant, the dilution of the number of galaxies (due to volume effects) almost exactly cancels with the increase in the number of detected galaxies due to the amplification of the luminosity. Since this effect is directly related to the population of galaxies under consideration (more precisely to the slope of the luminosity function {which determines the value of $s(z)$}), it would be important to confirm this result using precise specifications for the various galaxy populations in DESI.

In addition to these analyses, we study how gravitational lensing changes the constraining power of spectroscopic and photometric surveys. {The analysis of the constraining power from the combination of spectroscopic and photometric surveys is new as well.}
We first compare the constraints on cosmological parameters, by including or omitting lensing in the modelling.
We find that, for spectroscopic surveys, including lensing does not improve the constraints on cosmological parameters, even in the optimistic case where we assume that the amplitude of lensing (and therefore the magnification bias) is perfectly known.
For a photometric survey like LSST~\cite{Abell:2009aa}, including lensing helps break the degeneracy between the bias and the primordial amplitude of perturbations $A_s$, and consequently improves the constraints on these parameters significantly (by a factor 3 to 9, depending on the size of the redshift bins), if the magnification bias parameter and the amplitude of the lensing potential are perfectly known.
This improvement increases if we decrease the number of redshift bins used in the analysis, because this strongly reduces the RSD contribution, which also helps breaking the degeneracy between $A_s$ and the bias.
On the other hand, if the magnification bias parameter $s(z)$, or the amplitude of the lensing potential $A_\mathrm{L}$, are considered as free parameters, we find only a mild (few \%) improvement with respect to the case with no lensing.

Comparing constraints from SKA2 and LSST, including lensing in both cases, we find that, within $\Lambda$CDM, LSST provides better constraints than SKA2 {on $\Omega_{\rm cdm}, h$ and $n_s$.} {This is mainly due to the higher number of galaxies in LSST}. On the other hand, SKA2 yields better constraints on
$\Om_{\text{baryon}}$ (which relies on a good resolution of the baryon acoustic oscillations), $A_s$ and the bias.
This is due to the fact that $A_s$ and the bias are degenerate in the density contribution.
RSD break this degeneracy in spectroscopic surveys, but only marginally in photometric surveys, where they are subdominant.
Lensing helps breaking this degeneracy in photometric surveys, but to a lesser extent than RSD.
Interestingly, we also find that the amplitude of the lensing potential, $A_{\rm L}$, is better constrained in SKA2 than in LSST, even though the signal-to-noise ratio of lensing is significantly larger in LSST.
This is related to the fact that, in LSST, $A_{\rm L}$ is degenerate with $A_s$ and the bias, whereas in SKA2 this degeneracy is broken by RSD.
Spectroscopic and photometric surveys are therefore highly complementary.
In general, the main advantage of spectroscopic surveys over photometric surveys is their ability to measure the growth of structure{, which is directly proportional to the RSD signal,} in a model-independent way, which is something that is not at all straightforward with photometric surveys using the angular power spectrum with relatively poor redshift resolution.

Finally, we propose a combined analysis using the multipoles of the correlation function in each redshift bin and the angular power spectrum between different bins, including lensing in both cases.
We perform this analysis first for SKA2 alone, and then combining the correlation function from SKA2 with the angular power spectrum from LSST.
Such an analysis provides an efficient way of combining RSD constraints (from the two-point correlation functions) with {the information coming from the cross-correlations between different bins.}
Using only the specifications of the SKA2 survey, we find that including cross-correlations between different bins does not improve the constraints on the standard $\Lambda$CDM parameters.
This indicates that RSD have already enough constraining power and that therefore adding lensing does not add new information.
Note however that this result is specific to $\Lambda$CDM and will probably change in modified gravity models, where the relations between the gravitational potentials, and the density and velocity are modified.

Combining SKA2 and LSST, we find on the other hand that the cross-correlation between bins improves parameter constraints {by 10-20\%}.
The size of this improvement depends weakly on the assumptions made about the magnification bias parameters $[s_\mathrm{LSST}(z), s_\mathrm{SKA2}(z)]$.
{This is due to the fact that the improvement in $\La$CDM parameters mainly comes from the density cross correlations which are still relevant in neighboring bins. Lensing is not very important for the five $\La$CDM parameters, but becomes of essence once we aim at determining the growth factor in each redshift bin.}

Throughout this paper we make the following assumptions.
Firstly, we use linear perturbation theory, which we assume to be valid above separations of $30$\,{\rm Mpc}/$h$ at $z=0$.
We therefore limit our analysis to scales $d>d_{\rm NL}(z)$ with $d_{\rm NL}(0)=30\,{\rm Mpc}/h$.
Adding non-linear effects may change the specific form of the correlation function and of the lensing contribution.
However, we do not expect non-linearities to {significantly change the main results of this paper, i.e.\ to drastically reduce or increase the shift induced by lensing on the analysis.}

Secondly, we use a Fisher matrix analysis, and our estimates therefore always assume Gaussian statistics.
We are aware that this approximation typically underestimates the error bars.
{
The impact of the Gaussian approximation depends on how far from Gaussianity the likelihood function is. In Refs. \cite{Wolz:2012sr, Takada:2008fn} it is shown that, for observables that trace structure formation, such as weak-lensing and tomographic galaxy clustering analysis, the underestimation of the $1\sigma$ errors ranges between $10\%$ and $20\%$. Including parameters whose likelihood is significantly asymmetrical in the cosmological model would make the results of a Fisher analysis less reliable. This is the case, for example, when the neutrino mass is included as there is an asymmetry in the likelihood of this parameter due to a physically forbidden region (negative neutrino masses). We do not consider this model in our work.
The second assumption that our approach implicitly adopts is that the shift induced by neglecting lensing is well below $1\sigma$. As we will show in the remainder of the paper, this approximation is often not valid.
In Ref.~\cite{Cardona:2016qxn} it has been found that, whenever the Fisher matrix analysis leads to large parameter shifts, the more appropriate MCMC analysis also leads to large shifts at least in some of the parameters.
}
When the shift becomes larger than $\sim 1\sigma$, the Fisher analysis is not {quantitatively} reliable anymore and an MCMC analysis should be performed instead to determine the value of the shift precisely.    
However, the fact that Fisher analyses find a large shift indisputably means that lensing cannot be neglected in a survey like SKA2. {Even if the exact value of the shift which we obtain cannot be trusted, we can trust that there is a large shift in some of the parameters.}

Finally, we also neglect large scale relativistic effects in the number counts of galaxies, that have been derived in~\cite{2009PhRvD..80h3514Y,Yoo:2010ni,Bonvin:2011bg,Challinor:2011bk}.
It has been shown in several papers that, while the large scale relativistic effects are very nearly degenerate with effects of primordial non-Gaussianity in the real space power spectrum~\cite{Bruni:2011ta,Camera:2014bwa,Alonso:2015uua,Baker:2015bva}, they are rather difficult to detect in galaxy catalogs and usually require more sophisticated statistical methods like the use of different populations of galaxies~\cite{DiDio:2013sea,Bonvin:2013ogt,Bonvin:2014owa}.
We have checked that large scale relativistic effects do not influence the results reported in this work in a noticeable way.

The remainder of this paper is organized as follows.
In the next section, we briefly review the expression for the number counts of galaxies. {This very short section does not contain any new material and can by skipped by the experts.}
In Section~\ref{s:cor-fctn} we focus on the multipoles of the correlation function for spectroscopic surveys like DESI and SKA2. {This is the first Fisher matrix analysis using the relativistic full sky correlation function code {COFFE}~\cite{Tansella:2018sld}.}
In Section~\ref{s:Cls} we study the angular power spectrum, for a photometric survey like LSST. {The new ingredient in this section is the comparison of the analysis of a photometric survey with the power spectra $C_\ell(z,z')$, and the analysis of a spectroscopic survey with the relativistic correlation function, $\xi(\bar z, r,\mu)$.}
In Section~\ref{s:main_result} we combine the correlation function and the angular power spectrum, and we conclude in Section~\ref{s:con}. A theoretical modelling of the magnification bias parameter $s(z)$, which is essential to assess the impact of lensing, is presented in the appendices, where we also provide more detail about the specifications of the surveys, and on the photometric analysis. {The modelling of $s(z)$ for DESI provided there is new.}

\vspace{0.2cm}

{\bf Notation:} We set the speed of light $c=1$. We work in a flat Friedmann Universe with scalar perturbations in longitudinal gauge such that the metric is given by
\be
ds^2 = a^2(t)\left[-(1+2\Psi)dt^2 +(1-2\Phi)\de_{ij}dx^idx^j\right]\,.
\ee
The functions $\Phi$ and $\Psi$ are the Bardeen potentials.
The variable $t$ denotes conformal time and $\HH=\dot a/a= aH$ is the conformal Hubble parameter while $H$ is the physical Hubble parameter. The derivative with respect to $t$ is denoted by an overdot.

\section{The galaxy number counts}\label{s:numbercounts}

When we count galaxies, we observe them in a given direction and at a given redshift. The expression from linear perturbation theory for the over-density of galaxies at redshift $z$ and in direction $\bn$ is given by~\cite{2009PhRvD..80h3514Y,Bonvin:2011bg,Challinor:2011bk}
\bea
\label{e:DezNF}
\Delta(z, \mathbf{n})&=&b\cdot \delta+\frac{1}{\mathcal{H}}\partial_r^2V
+\left(5s-2\right)\int_0^{r}\integrand r' \frac{r-r'}{2rr'}\Delta_\Omega(\Phi+\Psi)\\
&&-\left(1-5s-\frac{\dot{\mathcal{H}}}{\mathcal{H}^2}+\frac{5s-2}{r\mathcal{H}} +f_{\rm evo} \right)\partial_rV-\frac{1}{\mathcal{H}}\partial_r\dot{V}+\frac{1}{\mathcal{H}}\partial_r\Psi\nonumber\\
&&+\frac{2-5s}{r}\int_0^{r}\integrand r'(\Phi+\Psi)+(f_{\rm evo}-3)\mathcal{H}V+\Psi+(5s-2)\Phi\nonumber\\
&&+\frac{1}{\mathcal{H}}\dot{\Phi}+\left(\frac{\dot{\mathcal{H}}}{\mathcal{H}^2}+\frac{2-5s}{r\mathcal{H}}+5s -f_{\rm evo} \right)\left[\Psi+\int_0^{r}\integrand r'(\dot{\Phi}+\dot{\Psi})\right]\, ,\nonumber
\eea
where $r=r(z)$ is the comoving distance to redshift $z$.
The functions $b(z)$, $s(z)$ and $f_{\rm evo}(z)$ are the galaxy bias, the magnification bias and the evolution bias respectively. They depend on the specifications of the catalog (which types of galaxies have been included) and on the instrument (what is the flux limit of the instrument in which frequency band). In Appendix~\ref{ap:surveyspec} we specify these functions for the surveys studied in this work.

The first term in Eq.~\eqref{e:DezNF}, $\de$, is the matter density fluctuation in comoving gauge: on small scales it reduces to the Newtonian density contrast. The second term in the first line encodes the effect of RSD. The velocity is given by $v_i=-\dd_iV$, where $V$ is the peculiar velocity potential in longitudinal gauge. These first two terms are the ones that are currently used in the modelling of the two-point correlation function. The last term in the first line is the lensing contribution, that we are investigating in this paper. It contains the angular Laplacian, $\Delta_\Om$, transverse to the direction of observation $\bn$. The last three lines are the so-called large-scale relativistic effects. The second line is suppressed by one power $\HH/k$ with respect to the first line, and the last two lines are suppressed by $(\HH/k)^2$ with respect to the first line. More details on these terms are given in~\cite{DiDio:2013sea, DiDio:2013bqa, Bonvin:2014owa}.
On scales which are much smaller than the horizon scales, $k\gg \HH$, only the first line contributes significantly and therefore we include only these terms in the present analysis.
We have checked that including the other, large-scale terms, does not alter our results.

\section{The correlations function -- spectroscopic surveys}\label{s:cor-fctn}

In spectroscopic surveys, the standard estimators used to extract information from maps of galaxies are the multipoles of the correlation function.
{
Here we use, for the first time, the fully relativistic full sky correlation function, derived in~\cite{Tansella:2018sld}, and implemented in the public code {COFFE}, for a Fisher matrix analysis.
Even though in the end we neglect the large scale relativistic effects, since they are too small to affect our results, we use the full sky correlation function and, more importantly, the lensing term which is not available in the standard correlation function.
}
The first three multipoles encode all the information about density and RSD in the linear regime, and they allow us to measure the growth rate of structure and the galaxy bias in a model-independent way (up to an arbitrary distance $r(z)$ which has to be fixed in each redshift bin and which in a given model is the comoving distance out to redshift $z$).

{For the standard terms,} the multipoles of the power spectrum are also routinely used. However, the power spectrum suffers from two limitations, that make it ill-adapted to future spectroscopic surveys. First, the power spectrum is constructed using the flat-sky approximation, which breaks down at large scales. Alternative power spectrum estimators have been constructed to include wide-angle effects, but they are not straightforward since they require to vary the line-of-sight for each pixel~\cite{Yamamoto:2005dz}.
Second, the lensing contribution in $\Delta$ cannot be consistently accounted for in the power spectrum, since to calculate the power spectrum one needs to know the Fourier transform of the galaxy number counts, $\Delta$, over 3-dimensional hypersurfaces, whereas lensing can only be computed along our past light-cone. In contrast, the correlation function can consistently account for wide-angle effects and for gravitational lensing. A general expression for the correlation function without assuming the flat-sky approximation has been derived in~\cite{Szalay:1997cc,Szapudi:2004gh,Papai:2008bd,Campagne:2017wec} for density and RSD, and extended to lensing and large scale relativistic effects in~\cite{Tansella:2018sld, Bertacca:2012tp, Raccanelli:2013gja}.
The correlation function is therefore the relevant observable to use in future spectroscopic surveys that go to high redshift (where lensing is important) and cover large parts of the sky (where the flat-sky approximation is not valid).

In Section~\ref{s:Cls}, we will discuss the use of the angular power spectrum, $C_\ell$, to extract information from galaxy surveys.
The angular power spectrum has the advantage over the correlation function that it does not require a fiducial cosmology to translate angles and redshifts into distances.
However, this problem can be circumvented in the correlation function by including rescaling parameters, that account for a difference between the fiducial cosmology and the true cosmology.

The angular power spectrum is not ideal for spectroscopic surveys, since it requires too many redshift bins in order to optimally profit from the redshift resolution. Indeed, in the $C_\ell$'s, the sensitivity to RSD is related to the size of the redshift bins, whereas in the correlation function it is related to the size of the pixels, that are typically of $\sim 2-8$\,Mpc/$h$. The split of the data into many redshift bins does very significantly enhance the shot noise per bin when one uses the $C_\ell$'s. Furthermore this makes the computation of the covariance matrix for the full set of angular power spectra challenging to invert. Finally, in the $C_\ell$'s, density and RSD are completely mixed up and distributed over the whole range of multipoles, whereas in the correlation function these terms can be separated by measuring the monopole, quadrupole and hexadecapole.
As a consequence, the correlation function is mainly used for spectroscopic surveys, and the angular power spectrum for photometric surveys, where RSD are subdominant and the number of bins is small.

We now study how the lensing contribution affects the monopole, quadrupole and hexadecapole of the correlation function.
For this, we use the code COFFE~\cite{Tansella:2018sld}, which computes the multipoles of the correlation function and their covariance matrices, and which is publicly available at
\href{https://github.com/JCGoran/coffe}{https://github.com/JCGoran/coffe}.
We consider two spectroscopic surveys, one with DESI-like specifications, and a more futuristic one with SKA2-like specifications.
The galaxy bias, magnification bias and the redshift distribution for these surveys are given in Appendix~\ref{ap:surveyspec}. As explained there, for DESI we use a weighted mean of the three different types of galaxies used in this survey to determine $b(z)$ and $s(z)$ in each redshift bin.

In order to mimic a real survey, we apply a window function to our galaxy density field, in the form of a spherical top-hat filter in real space:
\bea
\Delta (R, \mathbf x)
=
\int \integrand \mathbf x' W(R; \mathbf x - \mathbf x') \Delta (\mathbf x')\, ,
\eea
with a pixel size $R \equiv L_p = 5$ Mpc/$h$. As our analysis is based on linear perturbation theory, in order to curtail the effects of non-linearities, we implement a \textit{non-linear cutoff scale}, $d_\mathrm{NL}(z)$, below which we assume we cannot trust our results, and we do not consider the correlation function on scales smaller than $d_\mathrm{NL}(z)$ in our Fisher matrix analysis.
We parametrize the comoving non-linearity scale as follows:
\bea \label{e:rNL}
d_\mathrm{NL}(z)
=
\frac{d_\mathrm{NL}(z = 0)}{(1 + z)^\frac{2}{3}}\, ,
\eea
where for the present cutoff scale we assume the value $d_\mathrm{NL}(z = 0) = 30$ Mpc/$h$.%
\footnote{
This can be obtained by considering the evolution of $\delta$: in linear theory, $\delta(k, z) \sim D_1(z) \delta(k, 0) \sim \delta(k, 0) / (1 + z)$, hence $P(k, z) \sim P(k, 0) / (1 + z)^2$.
On the other hand, in linear theory we also know that $P(k) \sim k^{-3}$ for large values of $k$, and thus combining these two expressions we obtain $k_\mathrm{NL} \sim (1 + z)^{2/3}$, and from $k_\mathrm{NL} \sim 1 / d_\mathrm{NL}$ we obtain the scaling behaviour as described in the text.
}

\begin{table}[h!]
\caption{The cosmological parameters used for the Fisher matrix analysis as well as their fiducial values. \label{table:planck_params}}
\begin{center}
\begin{tabular}{|ccccc|}
\hline
$\Omega_\mathrm{baryon}$ & $\Omega_\mathrm{cdm}$ & $h$ & $n_s$ & $\ln(10^{10}A_s)$ \\
\hline
0.04841  & 0.25793 & 0.6781 & 0.9677 & 3.062 \\
\hline
\end{tabular}
\end{center}
\end{table}

We use the best fit values from Planck~\cite{Aghanim:2018eyx} for our fiducial values of the parameters, see Table~\ref{table:planck_params}.
The galaxy bias for the different surveys has been modelled as described in Appendix~\ref{ap:surveyspec}.
To account for our limited knowledge of the bias, we multiply this redshift dependent bias by a parameter $b_0$, with fiducial value $b_0=1$, that we also vary in our Fisher forecasts.
Finally, when estimating the impact of large scale relativistic effects, for the evolution bias $f_\mathrm{evo}$ we take a fiducial value $f_\mathrm{evo}=0$, but we have seen that varying $f_\mathrm{evo}$ in the interval $f_\mathrm{evo}\in [-5,5]$ does not affect the results.

\subsection{Signal-to-noise ratio}

As a first estimate of the relevance of lensing as well as relativistic and wide-angle effects for future spectroscopic surveys, we compute the signal-to-noise ratio (SNR) of these contributions, for the monopole $(\ell=0)$, quadrupole $(\ell=2)$ and hexadecapole $(\ell=4)$ of the correlation function. {For this we fix all cosmological parameters. This SNR ratio therefore does not tell us how well we can measure the lensing or the relativistic terms, but it is simply an indication whether they affect the correlation function at all. If the SNR of a contribution is significantly smaller than 1, it is safe to neglect the corresponding terms.}  The SNR in a redshift bin with mean $\bar z$ for the contribution $X=\{{\rm lens, rel}\}$ is given by
\be
{\rm SNR}^X=\sum_{ij\ell m}\xi^X_\ell(d_i, \bar z){\rm cov}[\xi^{\rm std}_\ell, \xi^{\rm std}_m]^{-1}(d_i,d_j,\bar z)\xi^X_m(d_j, \bar z)\, ,
\label{eq:snr}
\ee
where the sum runs over the multipoles $\ell, m=0,2,4$ and over separations $d_i, d_j$ between $d_{\rm NL}$ and $d_{\rm max}$, where $d_{\rm max}$ is the largest separation available inside the redshift bin, ranging from 185\,Mpc/$h$ {at $\bar z=0.2$} to 440\,Mpc/$h$ {at $\bar z=1.85$}.
Lensing does generate higher multipoles in the correlation function~\cite{Tansella:2018sld}, however since those are not usually measured in spectroscopic surveys, we do not include them in our calculation of the SNR.

The lensing correlation function is defined as
\bea
\xi^{\rm lens}(d,\bar z)
\equiv
\avg{\Delta_{\rm lens}(z,\bn)\Delta_{\rm lens}(z',\bn')} + \avg{\deltastd(z,\bn)\Delta_{\rm lens}(z',\bn')} + \avg{\Delta_{\rm lens}(z,\bn)\deltastd(z',\bn')}\, ,
\label{eq:lensing_contrib}
\eea
where $\deltastd = \Delta_\mathrm{den} + \Delta_\mathrm{rsd}$ and $\Delta_{\rm lens}$ is given by the last term in the first line of Eq.~\eqref{e:DezNF}. The redshifts $z$ and $z'$ inside the given redshift bin around $\bar z$ and $\cos\theta=\bn\cd\bn'$ are such that $d=|r(z)\bn-r(z')\bn'| = \sqrt{r(z)^2 + r(z')^2-2r(z)r(z')\cos\theta}$.

The large scale relativistic and wide-angle correlation function is defined as
\bea
\xi^{\rm rel}(d,\bar z)
&\equiv&
\avg{\Delta_{\rm rel}(z,\bn)\Delta_{\rm rel}(z',\bn')} + \avg{\deltastd(z,\bn)\Delta_{\rm rel}(z',\bn')} + \avg{\Delta_{\rm rel}(z,\bn)\deltastd(z',\bn')}
\label{eq:rel_2pcf}
\\
&&+ \avg{\Delta_{\rm std}(z,\bn)\Delta_{\rm std}(z',\bn')}^{\rm full-sky}-
\avg{\Delta_{\rm std}(z,\bn)\Delta_{\rm std}(z',\bn')}^{\rm flat-sky}\,, \nonumber
\eea
where the first line contains the large scale relativistic effects without lensing, $\Delta_{\rm rel}$ is given by the last three lines of Eq.~\eqref{e:DezNF}, while the second line contains the wide-angle effects, i.e.\ the difference between the standard terms calculated in the full-sky and in the flat-sky. The 'full-sky' standard terms are calculated using the first two terms of Eq.~\eqref{e:DezNF}, while the 'flat-sky' approximation is obtained by Fourier transforming the Kaiser-approximation~\cite{Kaiser} to the power spectrum, given by:

\begin{equation}
P(\mu,k) = P_D(k,z)(b(z) + f(z)\mu^2)^2 \,,
\end{equation}

where $P_D$ is the linear matter power spectrum, $b$ denotes the galaxy bias, and $f=\text{d}\log D_1/\text{d}\log a$ is the growth factor.
In Eq.~\eqref{eq:rel_2pcf}, we neglect the lensing contribution; we expect that the contributions from $\langle \Delta_\text{rel} \Delta_\text{std} \rangle$ are much larger than $\langle \Delta_\text{rel} \Delta_\text{len} \rangle$ and, therefore, including these terms would not affect our signal-to-noise analysis.

Note that eq.~\eqref{eq:snr} is just a special case of a $1 \times 1$ Fisher matrix, where, as our signal, we take $\xi_\ell(d, \bar z) = \xi_\ell^\text{std} + A\, \xi_\ell^X$, with $X \in \{\text{lens}, \text{rel}\}$, and take the derivative with respect to the artificial parameter $A$, so that $\partial \xi_\ell / \partial A = \xi_\ell^X$, and, as expected, the Fisher matrix for the SNR doesn't depend on the value of $A$.

\begin{figure}
\centering
\begin{minipage}[b]{0.45\linewidth}
\centering
{\large\bf lensing}\\
\centering
\includegraphics[width=\linewidth]{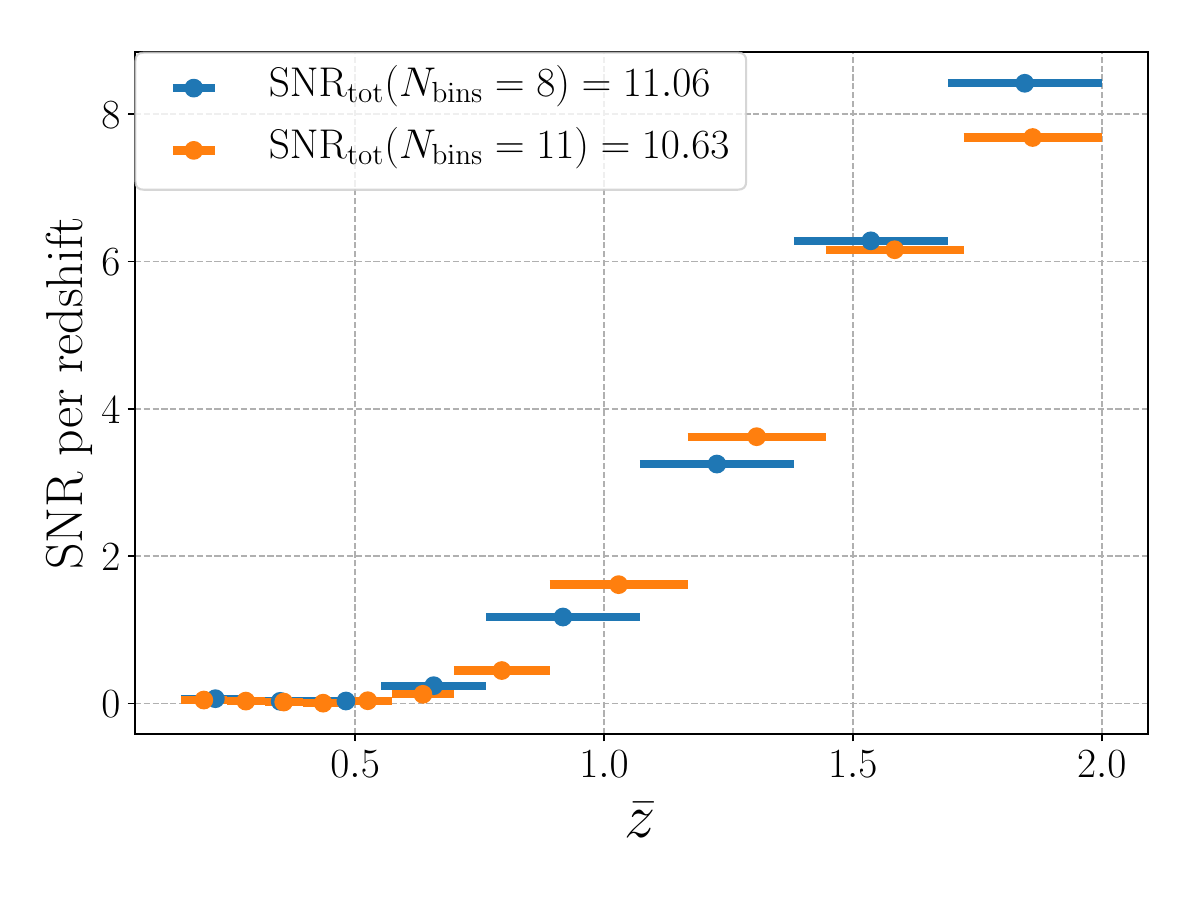}
\end{minipage}
\begin{minipage}[b]{0.45\linewidth}
\centering
{\large\bf relativistic effects}\\
\centering
\includegraphics[width=\linewidth]{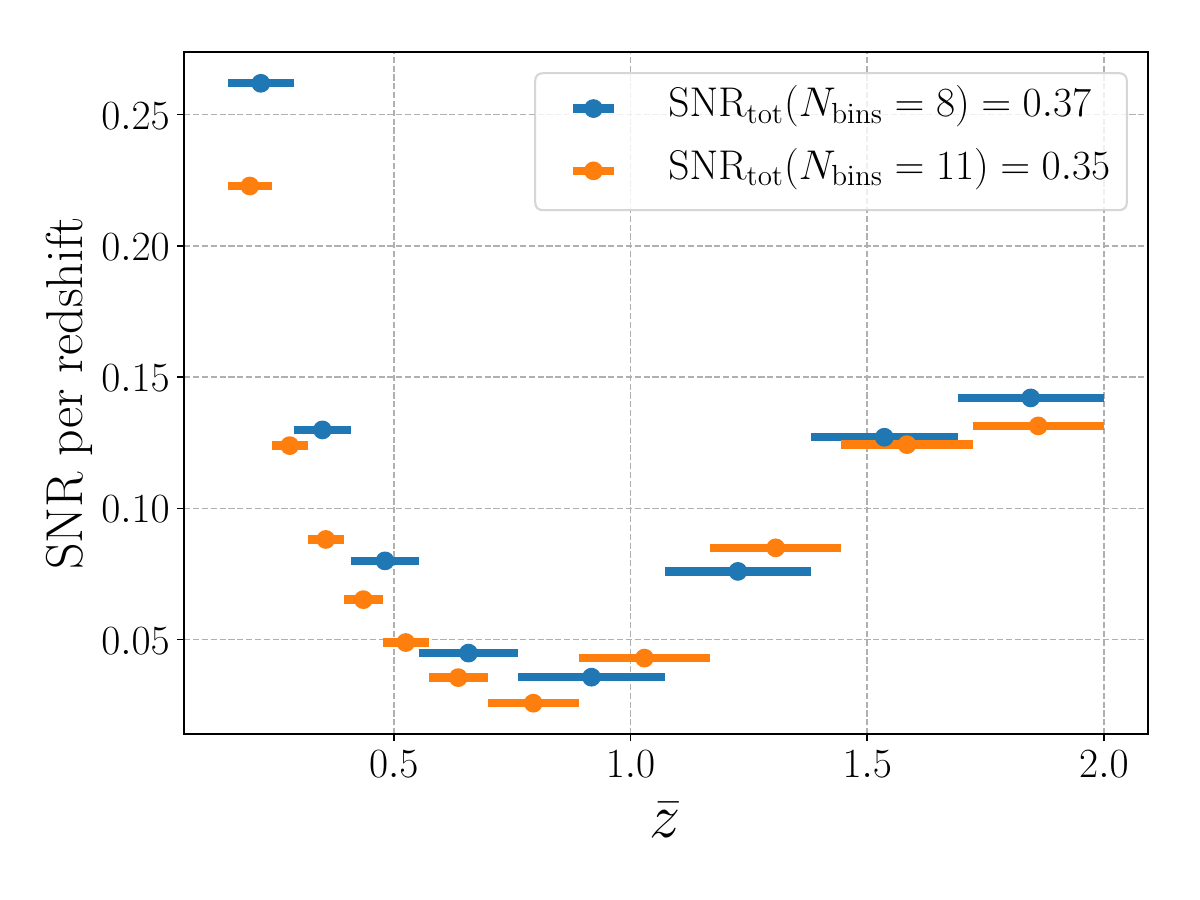}
\end{minipage}
\caption{
Signal-to-noise ratio (SNR) in each redshift bin for SKA2, using 8 redshift bins (blue) and 11 redshift bins (orange). The left panel shows the SNR for the lensing contribution and the right panel the SNR for the large scale relativistic effects, {including wide-angle effects}.
The horizontal lines depict the width of the redshift bins. We also indicate the cumulative SNR over all redshift bins, SNR$_{\rm tot}$.
}
\label{fig:ska2_snr}
\end{figure}

\begin{figure}
\centering
\begin{minipage}[b]{0.45\linewidth}
\centering
{\large\bf lensing}\\
\centering
\includegraphics[width=\linewidth]{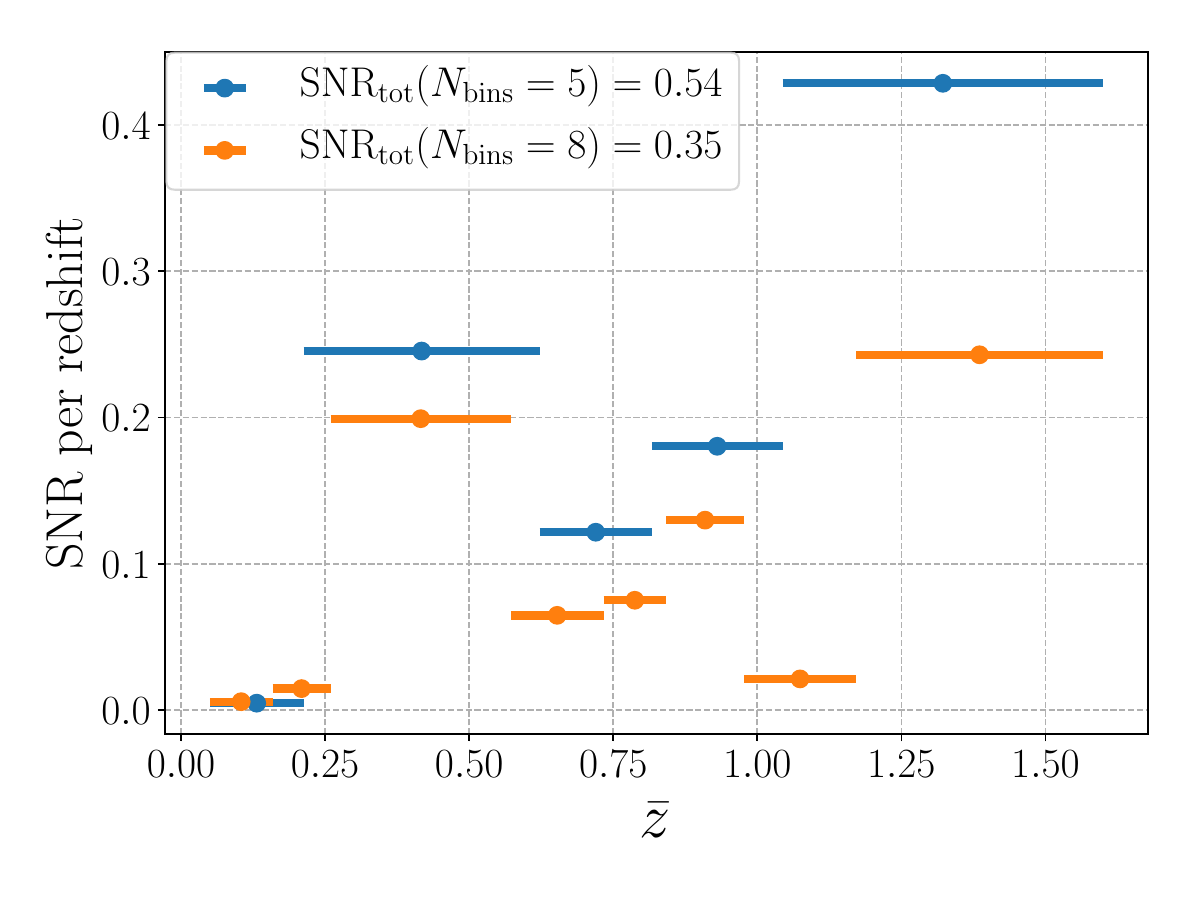}
\end{minipage}
\begin{minipage}[b]{0.45\linewidth}
\centering
{\large\bf relativistic effects}\\
\centering
\includegraphics[width=\linewidth]{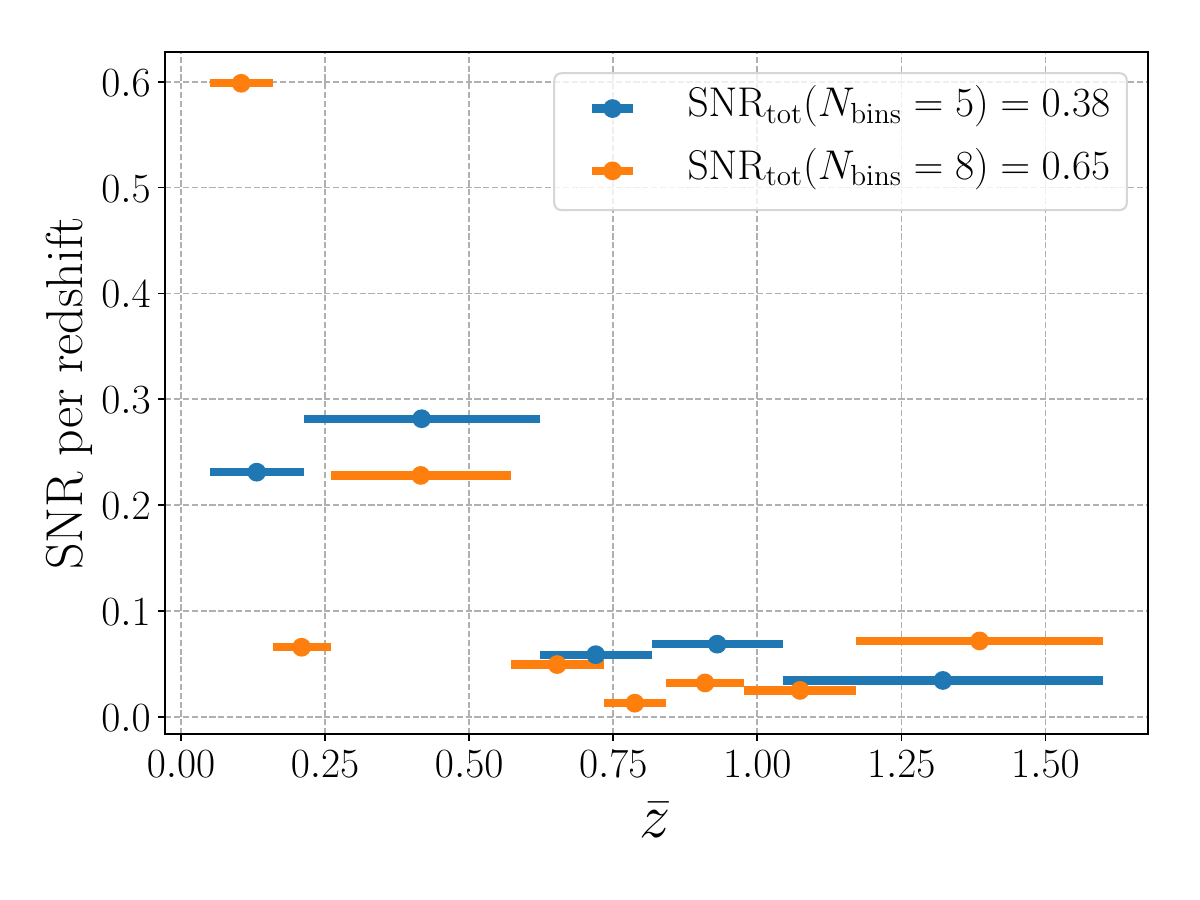}
\end{minipage}
\caption{
Signal-to-noise ratio (SNR) in each redshift bin for DESI, using 5 redshift bins (blue) and 8 redshift bins (orange). The left panel shows the SNR for the lensing contribution and the right panel the SNR for the large scale relativistic effects, {including wide-angle effects}.
The horizontal lines depict the width of the redshift bins. We also indicate the cumulative SNR over all redshift bins, SNR$_{\rm tot}$.
}
\label{fig:desi_snr}
\end{figure}

The covariance matrix contains both shot noise and cosmic variance. In the calculation of the cosmic variance we include only the standard terms, since those are the dominant ones. We have tested that including the lensing term in the covariance changes the $1\sigma$ error bars on the parameters by less than 1\% for an angular power spectrum analysis. Therefore, neglecting lensing in the covariance is a good approximation to the full result.
Moreover, we use the flat-sky approximation to calculate the covariance matrix, since even for large separations most of the covariance comes from pixels that are close to each other. The covariance matrix accounts for both correlations between different separations, $d_i\neq d_j$, and for correlations between different multipoles, $\ell\neq m$. We consider different redshift bin configurations in order to test the sensitivity of our analyses to the binning.
For SKA2, we use an 8 bin and an 11 bin configuration, and for DESI a 5 bin and an 8 bin configuration (see Appendix~\ref{a:SKAII} and~\ref{a:DESI} for more detail on these configurations).
In the calculation of the cumulative SNR over all redshift bins, we do not include the correlations between different bins, since they are very small due to the relatively large size of the bins ($\Delta z \geq 0.08$).

The results of the SNR analysis for SKA2 are shown in Fig.~\ref{fig:ska2_snr}.
The cumulative SNR for lensing (left panel) is larger than 10 so that the lensing is clearly detectable in SKA2.
Note that around $z=0.4$ the magnification bias parameter $s(z=0.4)\simeq 0.4$, and therefore the lensing contribution almost exactly vanishes.
At higher redshifts, $s$ becomes much larger (see Fig.~\ref{fig:ska2_biases} in Appendix~\ref{a:SKAII}), and at the same time the integral along the photons' trajectory significantly increases, such that the lensing contribution becomes more and more important. At $z=1$, the SNR becomes larger than one, however the bulk of the contribution to the SNR comes from higher redshifts.

The SNR of the large scale relativistic effects (right panel of Fig.~\ref{fig:ska2_snr}) always remains significantly below one.
This indicates that these terms do not affect parameter estimation in any appreciable way and can be safely neglected in the data analysis.
At low redshifts, the impact of large scale relativistic effects is larger due to one of the Doppler terms which is enhanced by a factor $1/(r(z)\HH(z))$ and whose contribution to the correlation function scales therefore as
\be
\langle\Delta^{\rm Dopp}\Delta^{\rm Dopp} \rangle\sim
1/(r\HH)^2(\HH/k)^2\langle\Delta^{\rm dens}\Delta^{\rm dens}\rangle
\sim(d/r)^2\langle\Delta^{\rm dens}\Delta^{\rm dens}\rangle\, ,
\ee
where we have used that $k\sim 1/d$.
Therefore, at very low redshifts and large separations this term cannot be neglected, as has been already discussed in~\cite{Papai:2008bd,Raccanelli_2010, Samushia_2012,Tansella:2017rpi}.

The same analysis using DESI specifications is shown in Fig.~\ref{fig:desi_snr}.
On the left panel, we see that the SNR for the lensing term remains always well below one, and that even the cumulative SNR over all redshift bins does not exceed one.
This means that lensing will not be detectable with a survey like DESI.
This is due to the fact that at high redshift ($z\geq 1$), when the integral along the photons' trajectory becomes important, the magnification bias parameter becomes close to 0.4, such that $2-5s(z)$ is very small, see Fig.~\ref{fig:sz-galaxysamples} in Appendix~\ref{a:DESI}.
Consequently, the lensing contribution to the number counts is strongly suppressed at high redshifts.
This result is very sensitive to the value of the magnification bias parameter $s(z)$, which we have computed in Appendix~\ref{a:DESI}, using a Schechter function to model the luminosity function, with parameters fitted to similar galaxy samples.
This gives us a crude approximation of $s(z)$ for the different galaxy populations that will be detected by DESI.
A more precise determination of $s(z)$ would be needed before one can definitely conclude that lensing is irrelevant for DESI.

On the right panel of Fig.~\ref{fig:desi_snr}, we show the SNR for large scale relativistic effects in DESI. Like for SKA2, the SNR remains well below one at all redshifts. Note however that since DESI starts observing at lower redshift than SKA2, the SNR in the lowest bin is larger.

For both surveys, large scale relativistic effects cannot be detected. These results confirm that in order to detect relativistic effects, we need {alternative estimators constructed from different populations of galaxies.} In the case of multiple populations, relativistic effects generate indeed odd multipoles in the correlation function, that have an SNR of the order of 7 for DESI~\cite{Bonvin:2015kuc} and 46 for SKA2~\cite{Bonvin:2018ckp}, making them clearly detectable with these surveys.

\subsection{Shift of $\Lambda$CDM parameters}
\label{s:shiftLCDM}

\begin{figure}
\centering
\includegraphics[width=0.9\linewidth]{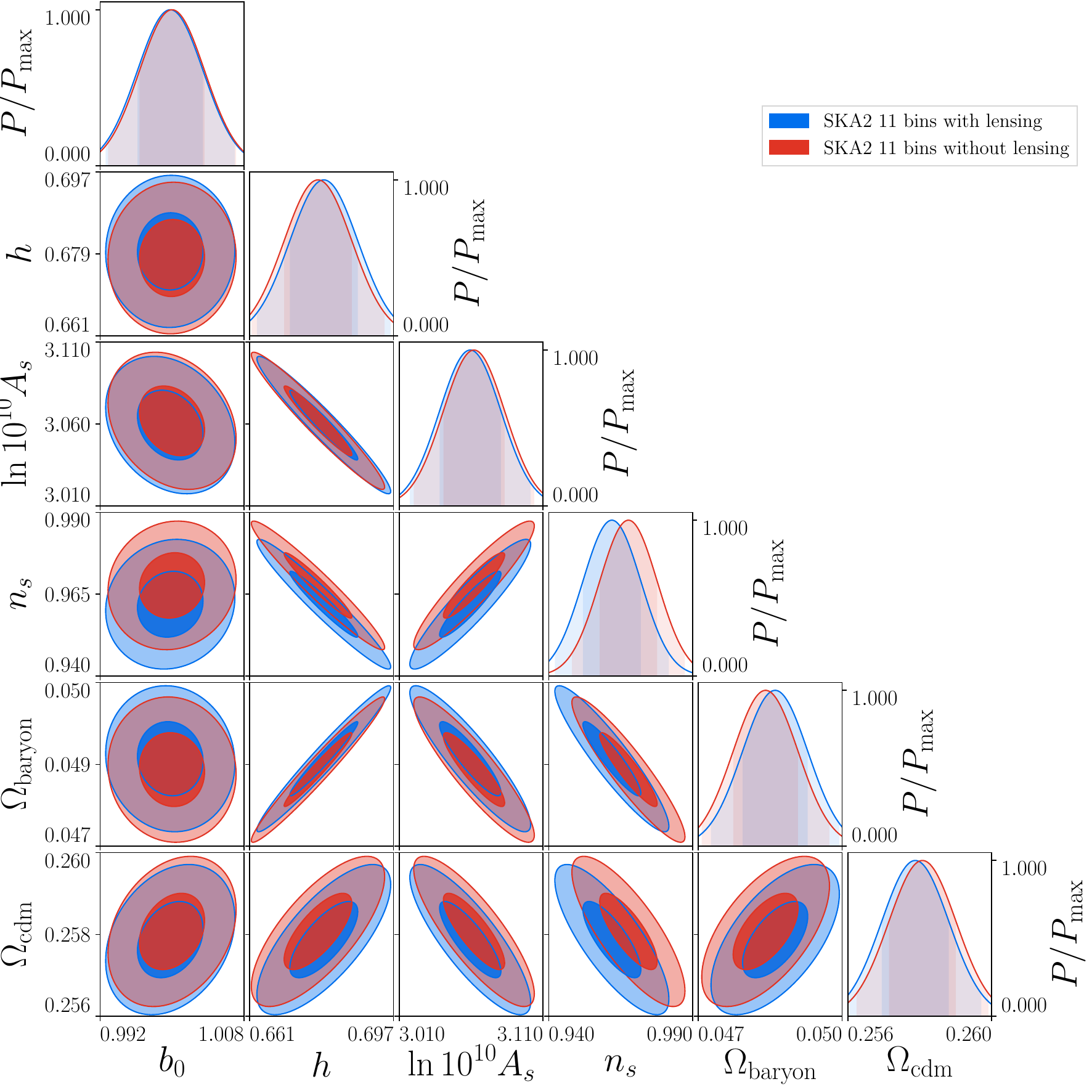}
\caption{
Contour plot for $\Lambda$CDM parameters from the SKA2 Fisher analysis, using 11 redshift bins. Blue contours show the constraints on parameters when we consistently include lensing in our theoretical model. Red contours show the constraints on the parameters when we neglect lensing magnification. In the latter case, the best-fit parameters are shifted with respect to the fiducial values.
In both cases we indicate 1$\sigma$ and $2\sigma$ contours.
\label{fig:ska2_param_errors_std_bins11}}
\end{figure}

Since lensing is detectable by a survey like SKA2, we now study its impact on parameter estimation. We use a Fisher matrix analysis to determine the error bars on each of the parameters $\theta_a\in (\Omega_{\rm baryon}, \Omega_{\rm cdm}, h, n_s, A_s, b_0)$. The Fisher matrix element for the parameters $\theta_a, \theta_b$ is given by
\be
F_{ab}=\sum_{\bar{z}}\sum_{ij\ell m}\frac{\partial\xi_\ell(d_i, \bar z)}{\partial\theta_a}{\rm cov}[\xi^{\rm std}_\ell, \xi^{\rm std}_m]^{-1}(d_i,d_j,\bar z)\frac{\partial\xi_m(d_j, \bar z)}{\partial\theta_b}\, . \label{fisher-xi}
\ee
We perform two analyses: one where the multipoles of the correlation function $\xi_\ell$ contains only the standard density and RSD contributions, and one where they also contain the lensing contribution. Comparing the error bars in these two cases allows us to understand if lensing brings additional constraining power. We also compute the shift of the central value of the parameters due to the fact that lensing is neglected in the modelling, whereas it is present in the signal.
This is given by (see e.g.~\cite{Cardona:2016qxn}, appendix B):
\be
\Delta(\theta_a)
=
\sum\limits_{b} (\tilde{F}^{-1})_{ab} B_b\, .
\ee
Here we define
\be
B_b
\equiv
\sum\limits_{\bar z}\sum\limits_{i j \ell m}
\Delta \xi_\ell(d_i, \bar z) \mathrm{cov}[\xi^{\rm std}_\ell, \xi^{\rm std}_m]^{-1}(d_i,d_j,\bar z) \frac{\partial \tilde{\xi}_{m}(d_j, \bar z)}{\partial \theta_b}\, ,
\label{eq:shift_vector_1}
\ee
where $\Delta \xi_\ell$ denotes the difference between the model and the signal (i.e.\ in our case, the lensing term~\eqref{eq:lensing_contrib}), and the quantities with a tilde ($\sim$) are computed without lensing, i.e.\ according to the wrong model.

\begin{table}[!t]
\caption{
The constraints and the shifts on $\Lambda$CDM parameters for SKA2, with 11 redshift bins
}
\begin{center}
\begin{tabular}{|c|c|cccccc|}
\hline
& parameter & $b_0$ & $\Omega_\mathrm{baryon}$ & $\Omega_\mathrm{cdm}$ & $h$ & $n_s$ & $\ln 10^{10} A_s$
\\
\hline
without lensing & $\sigma(\theta_i) / \theta_i$ ($\%$) & 0.37 & 1.32 & 0.37 & 1.26 & 1.05 & 0.70
\\
\hline
with lensing & $\sigma(\theta_i) / \theta_i$ ($\%$) & 0.36 & 1.31 & 0.37 & 1.25 & 1.04 & 0.70
\\
\hline
shift & $\Delta(\theta_i) / \sigma(\theta_i)$ & -0.05 & 0.29 & -0.21 & 0.17 & -0.57 & -0.12
\\
\hline
\end{tabular}
\end{center}
\label{table:ska2_fisher_11bins}
\end{table}

The results for SKA2 are shown in Table~\ref{table:ska2_fisher_11bins} and in Fig.~\ref{fig:ska2_param_errors_std_bins11}. Comparing the error bars with and without lensing, we see that the improvement brought by lensing is extremely small. This is due to the fact that density and RSD are significantly stronger than the lensing contribution and that they are very efficient at constraining the standard $\Lambda$CDM parameters. This is very specific to $\Lambda$CDM, in which the degeneracy between the bias $b_0$ and the primordial amplitude $A_s$ is broken by RSD, as can be seen in Fig.~\ref{fig:ska2_param_errors_std_bins11}.
This result would change if we have a modified gravity model, where the relation between density, velocity and gravitational potentials is altered and for which lensing brings complementary information.
We have also tested that our limited knowledge of the magnification bias parameter, $s(z)$, does not degrade the constraints on $\Lambda$CDM parameters.
For this we parametrize $s(z)$ with four parameters (see Eq.~\eqref{e:magSKA2}) that we include in the Fisher forecasts.
We find that the constraints on $\Lambda$CDM parameters are the same up to 1\% as in the situation where $s(z)$ is fixed.

From the third line of Table~\ref{table:ska2_fisher_11bins}, we see that the parameter which is the most significantly shifted when lensing is neglected in SKA2 is $n_s$, which experiences a shift of $-0.57\si(n_s)$.
For $\Om_\mathrm{baryon}$ and $\Omega_\mathrm{cdm}$ the shifts are smaller but also not negligible. A shift of less than one $\sigma$ is not catastrophic, but it still indicates that the analysis can be significantly improved by including lensing in the modelling. For example, such a shift could hide deviations from General Relativity if those deviations are in the opposite direction as the shift. Note also that the Fisher analysis that we have performed is not precise for a shift which approaches $1\sigma$. A more robust analysis using MCMC may give an even larger shift. Finally, we have checked that the results are very similar if we reduce the number of redshift bins from 11 to 8.

\begin{table}[h!]
\caption{
The constraints and the shifts on $\Lambda$CDM parameters for DESI, with 8 redshift bins.
}
\begin{center}
\begin{tabular}{|c|c|cccccc|}
\hline
&parameter & $b_0$ & $\Omega_\mathrm{baryon}$ & $\Omega_\mathrm{cdm}$ & $h$ & $n_s$ & $\ln 10^{10} A_s$
\\
\hline
without lensing &$\sigma(\theta_i) / \theta_i$ ($\%$) & 0.75 & 3.50 & 0.85 & 3.30 & 2.65 & 1.75
\\
\hline
with lensing &$\sigma(\theta_i) / \theta_i$ ($\%$) & 0.75 & 3.50 & 0.85 & 3.30 & 2.65 & 1.75
\\
\hline
shift &$\Delta(\theta_i) / \sigma(\theta_i)$ & 0.014 & -0.005 & -0.017 & -0.008 & -0.0004 & 0.005
\\
\hline
\end{tabular}
\end{center}
\label{table:desi_fisher_8bins}
\end{table}

A corresponding analysis for the DESI spectroscopic survey confirms the results already indicated by the small SNR of the lensing term: neglecting lensing shifts all cosmological parameter by less than 0.02$\sigma$, meaning that lensing can be neglected in the analysis.
The results are summarized in Table ~\ref{table:desi_fisher_8bins}.

\subsection{Shift of the growth rate}\label{s:growth}

One of the main motivations to measure the multipoles of the correlation function is that they provide a model-independent way of measuring the growth of structure $f$ given by
\be
f = \frac{d\ln D_1}{d\ln a} = \frac{\dot D_1}{\HH D_1}\, ,
\ee
where $D_1$ is the linear growth function that encodes the evolution of density fluctuations: $\delta(\bk,z)=D_1(z)/D_1(z=0)\delta(\bk,z=0)$. This growth rate is very sensitive to the theory of gravity. It has been measured in various surveys and is used to test the consistency of General Relativity and to constrain deviations from it.

In $\Lambda$CDM, $f$ is fully determined by the matter density $\Omega_\mathrm{m} = \Omega_{\rm baryon}+\Omega_{\rm cdm}$. In modified gravity theories, the growth rate depends directly on the parameters of the theory. Here we take an agnostic point of view and simply consider $f$ in each redshift bin as a free parameter.

The monopole of the correlation function can be written as
\be
\xi_0(d,z)=\left(b^2(z)+\frac{2b(z)f(z)}{3}+\frac{f^2(z)}{5} \right)\frac{1}{2\pi^2}\int \integrand k k^2 P_\delta(k,z)j_0(kd)\, ,
\ee
where $P_\delta$ is the density power spectrum and $j_0$ is the spherical Bessel function of order 0. Assuming that the growth of structure is scale-independent, we can relate the power spectrum at redshift $z$ to its value at a high redshift, $z_\star$, which we choose to be well in the matter era, before the acceleration of the Universe has started
\be
P_\delta(k,z)=\left(\frac{D_1(z)}{D_1(z_\star)}\right)^2P_\delta(k,z_\star)\, .
\ee
Similarly the parameter $\sigma_8(z)$, which denotes the amplitude of the mean matter fluctuation in a sphere of radius $8h^{-1}$Mpc evolves as
\be
\sigma_8(z)=\frac{D_1(z)}{D_1(z_\star)}\sigma_8(z_\star)\, .
\ee
With this, the monopole of the correlation function can be written as
\be
\xi_0(d,z)=\left(\tilde b(z)^2+\frac{2\tilde b(z)\tilde f(z)}{3}+\frac{\tilde f^2(z)}{5} \right)\mu_0(d,z_\star)\, , \label{e:mono}
\ee
where
\be
\tilde f(z)\equiv f(z)\sigma_8(z)\quad\mbox{and}\quad\tilde b(z)\equiv b(z)\sigma_8(z)\, ,
\ee
and
\be
\mu_\ell(d,z_\star)=\frac{1}{2\pi^2}\int \integrand k k^2 \frac{P_\delta(k,z_\star)}{\sigma_8(z_\star)^2}j_\ell(kd)\, ,\quad \ell=0,2,4\, .
\label{e:muell}
\ee
Similarly the quadrupole and hexadecapole of the correlation function take the form
\bea
\xi_2(d,z)&=&-\left(\frac{4\tilde b(z)\tilde f(z)}{3}+\frac{4\tilde f^2(z)}{7} \right)\mu_2(d,z_\star)\, ,\label{e:quad}\\
\xi_4(d,z)&=&\frac{8\tilde f^2(z)}{35}\mu_4(d,z_\star)\, .\label{e:hexa}
\eea
{
The relevant lensing contribution to the correlation function, given by eq.~\eqref{eq:lensing_contrib}, and its multipoles, is quite involved, and its explicit form can be found in Appendix~\ref{s:lensing_2pcf}.
We emphasize that, to obtain the terms in eq.~\eqref{eq:shift_vector_1}, we do not need to compute the derivatives of eq.~\eqref{eq:lensing_contrib}  w.r.t. the parameters $\tilde b$ and $\tilde f$, since the lensing only enters through the term $\Delta \xi_\ell(d, \bar z)$.
Lastly, using the flat-sky approximation for the standard terms and ignoring relativistic effects is justified, because, as discussed above, we have verified that the SNR of the relativistic and wide-angle effects are negligible for the surveys considered here.
}

The functions $\mu_\ell(d, z_\star)$ encode the shape of the multipoles and they depend only on the physics of the early Universe, before acceleration has started. Here we assume that this physics has been determined with high precision by CMB measurements and we take these functions as fixed to their fiducial values. The parameters $\tilde f$ and $\tilde b$ govern the amplitude of the multipoles, and they depend on the growth of structure {at late time, when the expansion of the Universe is accelerating. By combining measurements of the monopole, quadrupole and hexadecapole, these two parameters can be measured directly. The can then be used to test the theory of gravity. This procedure is valid for any model of gravity or dark energy that does not affect the evolution of structures at early time before acceleration has started, i.e.\ that leaves the functions $\mu_\ell(d,z_\star)$ unchanged.} In this case we can treat $\tilde f$ and $\tilde b$ as free in each of the bins of the survey.

\begin{figure}
\includegraphics[width=0.45\linewidth]{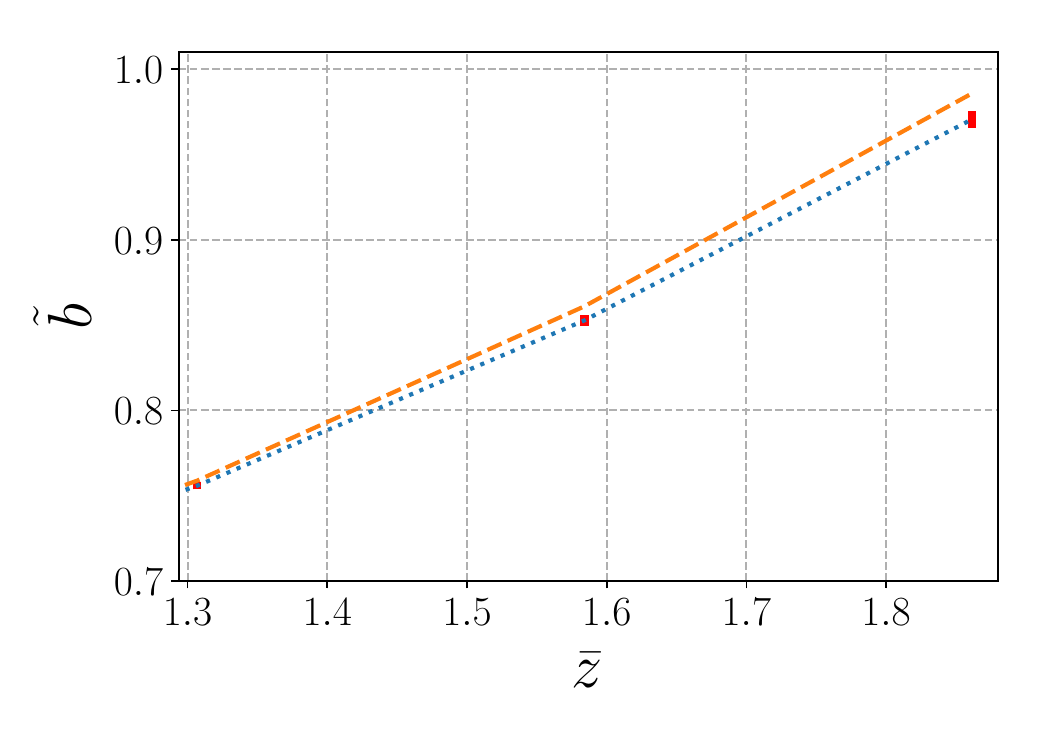}
\includegraphics[width=0.45\linewidth]{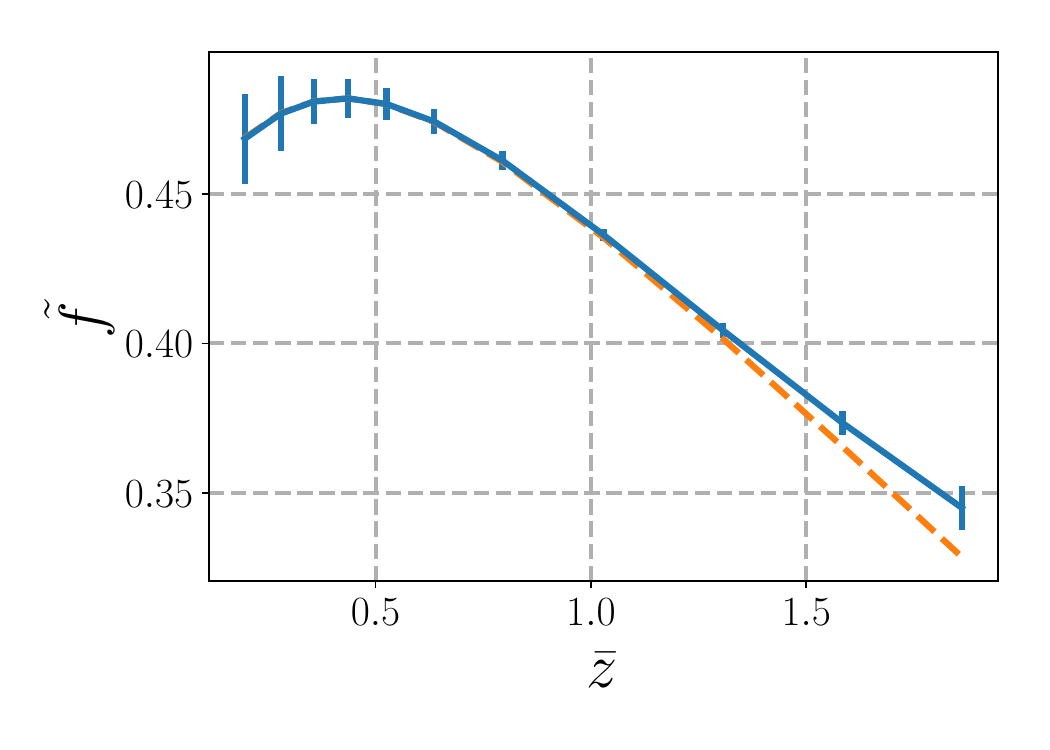}
\caption{
The parameters $\tilde{b}$ (left panel) and $\tilde f$ (right panel) plotted as a function of the mean redshift of the bin without lensing (blue), and with lensing (dashed orange), for SKA2 with 11 redshift bins. We also show the error bars, that are highlighted in red for visibility
in the left plot.
}
\label{fig:ska2_tildes_plot1}
\end{figure}

As discussed before, lensing generates a new contribution to Eqs.~\eqref{e:mono}, \eqref{e:quad} and~\eqref{e:hexa}. We now determine how this new contribution shifts the measurement of $\ft$ and $\bt$ in each redshift bin.
In Fig.~\ref{fig:ska2_tildes_plot1}, we show $\bt$ and $\ft$ inferred with and without the lensing contribution and we compare the difference with the size of the error bars. At high redshift, $z>1.2$ the shift from neglecting lensing is clearly larger than 1$\sigma$. The precise values of the shifts are given in Table~\ref{t:SKA11bin-bf} for each redshift bin. We see that in the highest bin, the shift reaches 3.1$\sigma$ for $\bt$, and -2.3$\sigma$ for the growth rate $\ft$. {The fact that $\bt$ is shifted towards a larger value, whereas $\ft$ is shifted towards a smaller value can be understood in the following way: the lensing contribution to the monopole, quadrupole and hexadecapole is positive. In all cases the signal is therefore larger than the model, which does not include lensing. To account for this, the combinations of $\bt$ and $\ft$ in Eqs.~\eqref{e:mono},~\eqref{e:quad} and~\eqref{e:hexa} need to be shifted towards larger values for all multipoles. This shift is however not the same for all multipoles: in particular the impact of lensing on the quadrupole is larger than on the monopole. One way to account for this is to increase $\bt$ while decreasing $\ft$. Note that this leads to a shift in the wrong direction for the hexadecapole, but since the SNR of the hexadecapole is significantly smaller than that of the monopole and quadrupole, it has a much smaller impact on parameter estimation.}

Our analysis shows that neglecting lensing above redshift 1 is not possible for a survey like SKA2. Since the growth rate of structure is directly used to test the theory of gravity, such large shifts would wrongly be interpreted as a deviation from General Relativity.

\begin{table}[H]
\caption{
Values of the parameters $\tilde b$ (left table) and $\tilde f$ (right table) in each redshift bin, for SKA2. We also show the relative error bars on these parameters, $\sigma/\theta$,
and the ratio of the shift over the error bars, $\Delta/\sigma$, when lensing is neglected in the modelling of the multipoles.\label{t:SKA11bin-bf}
}
\begin{center}
\begin{tabular}{|c|c|c|c|}
\hline
\begin{tabular}{c} $\bar{z}_i$ \end{tabular}
&
\begin{tabular}{c} $\tilde{b}_i$ \end{tabular}
&
\begin{tabular}{c} $\displaystyle\frac{\sigma}{\theta}(\tilde{b}_i) (\%)$ \end{tabular}
&
\begin{tabular}{c} $\displaystyle \frac{\Delta}{\sigma}(\tilde{b}_i)$ \end{tabular}
\\
\hline
0.20 & 0.52 & 2.55 & -0.02
\\
0.28 & 0.53 & 2.06 & -0.01
\\
0.36 & 0.54 & 1.19 & -0.01
\\
0.44 & 0.55 & 0.99 & -0.00
\\
0.53 & 0.57 & 0.80 & 0.01
\\
0.64 & 0.59 & 0.63 & 0.04
\\
0.79 & 0.62 & 0.47 & 0.15
\\
1.03 & 0.68 & 0.25 & 0.55
\\
1.31 & 0.76 & 0.28 & 1.39
\\
1.58 & 0.85 & 0.39 & 2.50
\\
1.86 & 0.97 & 0.50 & 3.10
\\
\hline
\end{tabular}
\quad\quad
\begin{tabular}{|c|c|c|c|}
\hline
\begin{tabular}{c} $\bar{z}_i$ \end{tabular}
&
\begin{tabular}{c} $\tilde{f}_i$ \end{tabular}
&
\begin{tabular}{c} $\displaystyle\frac{\sigma}{\theta}(\tilde{f}_i) (\%)$ \end{tabular}
&
\begin{tabular}{c} $\displaystyle \frac{\Delta}{\sigma}(\tilde{f}_i)$ \end{tabular}
\\
\hline
0.20 & 0.47 & 3.21 & 0.02
\\
0.28 & 0.48 & 2.61 & 0.01
\\
0.36 & 0.48 & 1.58 & 0.01
\\
0.44 & 0.48 & 1.33 & 0.00
\\
0.53 & 0.48 & 1.11 & -0.01
\\
0.64 & 0.47 & 0.91 & -0.04
\\
0.79 & 0.46 & 0.71 & -0.14
\\
1.03 & 0.44 & 0.46 & -0.46
\\
1.31 & 0.40 & 0.63 & -1.12
\\
1.58 & 0.37 & 1.08 & -1.97
\\
1.86 & 0.34 & 2.11 & -2.26
\\
\hline
\end{tabular}
\end{center}
\end{table}

The difference between the analysis presented here and the $\Lambda$CDM analysis of Section~\ref{s:shiftLCDM}, is that in the $\Lambda$CDM analysis most of the constraining power on the parameters comes from small redshift, where shot noise is significantly smaller due to the large number of galaxies. At these small redshifts, lensing is still negligible and has therefore a relatively small impact on the determination of these parameters.
On the other hand, in the model-independent analysis presented here, $\bt$ and $\ft$ are free in each redshift bin, and therefore the constraints on their value comes from the bin in question.
As a consequence, at high redshift, where lensing is important, those constraints are very strongly affected by lensing.

As is clear from our derivation of the shift $\Delta$, the particular value of the shift can only be trusted for values which are (significantly) less that $1\si$.
Hence our results which yield shifts of $1\si$ and more simply indicate that there is a large shift, that cannot be neglected.
The precise value of this shift would have to be determined by an MCMC method which goes beyond the scope of the present work, see e.g.~\cite{Cardona:2016qxn}.

We now study how the constraints and shifts change if instead of fixing $s(z)$ we parametrize it with four parameters (see Eq.~\eqref{e:magSKA2}) and include these parameters in the Fisher analysis.
As can be seen from Table~\ref{t:SKA11bin-bf_magfree}, this degrades the constraints on $\tilde b$ and $\tilde f$ by up to 25\%. We see that the low redshift bins are also affected, even though the lensing contribution is negligible there. This is simply due to the fact that adding extra parameters in the model has an impact on all the other parameters, including the values of $\tilde b$ and $\tilde f$ at small redshift. This degradation of the constraints could be mitigated by an independent measurement of $s(z)$. This parameter is indeed given by the slope of the luminosity function which can be measured from the population of galaxies at each redshift. Again we conclude that a precise measurement of $s(z)$ is crucial for an optimal analysis.

\begin{table}[H]
\caption{
Same as Table~\ref{t:SKA11bin-bf}, except with the magnification bias parameters $\{s_0, s_1, s_2, s_3\}$ marginalized over.\label{t:SKA11bin-bf_magfree}
}
\begin{center}
\begin{tabular}{|c|c|c|c|}
\hline
\begin{tabular}{c} $\bar{z}_i$ \end{tabular}
&
\begin{tabular}{c} $\tilde{b}_i$ \end{tabular}
&
\begin{tabular}{c} $\displaystyle\frac{\sigma}{\theta}(\tilde{b}_i) (\%)$ \end{tabular}
&
\begin{tabular}{c} $\displaystyle \frac{\Delta}{\sigma}(\tilde{b}_i)$ \end{tabular}
\\
\hline
0.20 & 0.52 & 3.17 & -0.01
\\
0.28 & 0.53 & 2.51 & -0.01
\\
0.36 & 0.54 & 1.44 & -0.01
\\
0.44 & 0.55 & 1.26 & -0.00
\\
0.53 & 0.57 & 1.04 & 0.01
\\
0.64 & 0.59 & 0.81 & 0.03
\\
0.79 & 0.62 & 0.57 & 0.13
\\
1.03 & 0.68 & 0.30 & 0.47
\\
1.31 & 0.76 & 0.34 & 1.16
\\
1.58 & 0.85 & 0.47 & 2.07
\\
1.86 & 0.97 & 0.60 & 2.60
\\
\hline
\end{tabular}
\quad\quad
\begin{tabular}{|c|c|c|c|}
\hline
\begin{tabular}{c} $\bar{z}_i$ \end{tabular}
&
\begin{tabular}{c} $\tilde{f}_i$ \end{tabular}
&
\begin{tabular}{c} $\displaystyle\frac{\sigma}{\theta}(\tilde{f}_i) (\%)$ \end{tabular}
&
\begin{tabular}{c} $\displaystyle \frac{\Delta}{\sigma}(\tilde{f}_i)$ \end{tabular}
\\
\hline
0.20 & 0.47 & 3.98 & 0.01
\\
0.28 & 0.48 & 3.20 & 0.01
\\
0.36 & 0.48 & 1.91 & 0.01
\\
0.44 & 0.48 & 1.67 & 0.00
\\
0.53 & 0.48 & 1.37 & -0.01
\\
0.64 & 0.47 & 1.09 & -0.03
\\
0.79 & 0.46 & 0.83 & -0.12
\\
1.03 & 0.44 & 0.51 & -0.41
\\
1.31 & 0.40 & 0.70 & -1.00
\\
1.58 & 0.37 & 1.21 & -1.76
\\
1.86 & 0.34 & 2.29 & -2.08
\\
\hline
\end{tabular}
\end{center}
\end{table}

\begin{figure}
\centering
\begin{minipage}[b]{0.45\linewidth}
\centering
\includegraphics[width=\linewidth]{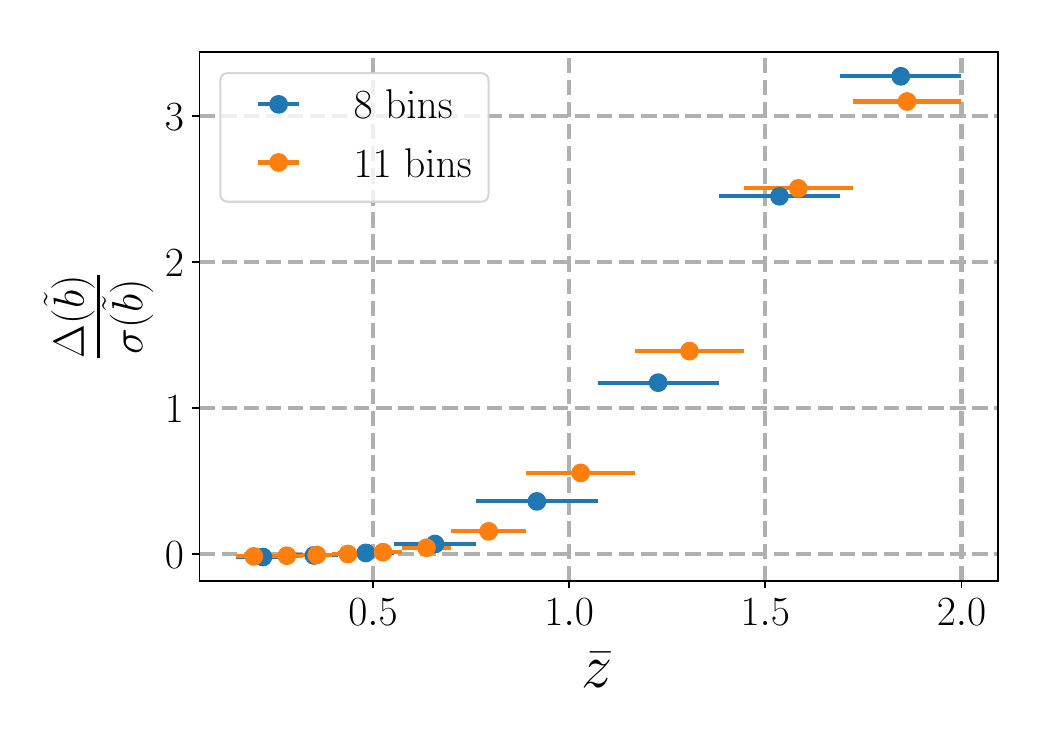}
\end{minipage}
\begin{minipage}[b]{0.45\linewidth}
\centering
\includegraphics[width=\linewidth]{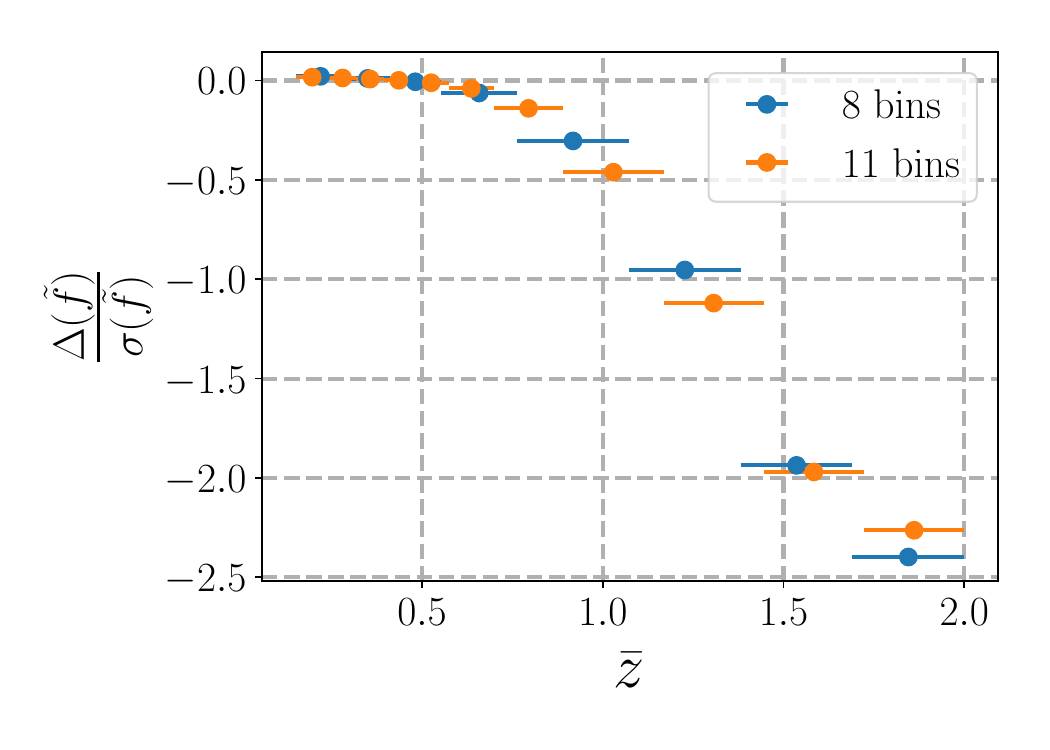}
\end{minipage}
\caption{
The shift divided by the 1$\sigma$ error for the parameters $\tilde{b}$ (left panel) and $\tilde f$ (right panel), plotted as a function of the mean redshift of the bin, for SKA2. The 8 redshift bins configuration is shown in blue and the 11 redshift bins configuration in orange.
The horizontal lines denote the widths of the redshift bins.
}
\label{fig:ska2_tildes_plot3}
\end{figure}

Finally, in Fig.~\ref{fig:ska2_tildes_plot3}, we compare the shift of $\bt$ and $\ft$ for two configurations in SKA2: one with 11 redshift bins, and one with 8 redshift bins. We see that the results are very similar: in both cases the shift becomes larger than 1$\sigma$ above redshift $\sim$ 1.2.

We have performed a similar analysis for DESI, and we found that the shifts on the parameters are completely negligible at all redshifts, as can be seen from Table~\ref{t:desi8bin-bf}.
This has several reasons, all related to the fact that the lensing SNR for DESI is always below 1. First, DESI does not go out to redshifts as high as SKA2 and hence the lensing contribution is smaller. Furthermore, the error bars for DESI are simply larger than the ones for SKA2, due to the smaller sky coverage. But the most relevant point is that the prefactor $5s(z)-2$ which determines the amplitude of the lensing term is always significantly smaller for DESI than for SKA2. It is only for very low redshifts for which the integral along the photons' trajectory is still small, that DESI has a relatively large $s(z)$. As discussed in the previous section, this result is based on our derivation of $s(z)$ for the different populations of galaxies in DESI, and should be confirmed by a more detailed modelling of $s(z)$.

\begin{table}[H]
\caption{
Values of the parameters $\tilde b$ (left table) and $\tilde f$ (right table) in each redshift bin, for DESI with 8 redshift bins. We show also the relative error bars on these parameters, $\sigma/\theta$, and the ratio of the shifts over the error bars, $\Delta/\sigma$, when neglecting lensing in the modelling of the multipoles.\label{t:desi8bin-bf}
}
\begin{center}
\begin{tabular}{|c|c|c|c|}
\hline
\begin{tabular}{c} $\bar{z}_i$ \end{tabular}
&
\begin{tabular}{c} $\tilde{b}_i$ \end{tabular}
&
\begin{tabular}{c} $\displaystyle\frac{\sigma}{\theta}(\tilde{b}_i) (\%)$ \end{tabular}
&
\begin{tabular}{c} $\displaystyle \frac{\Delta}{\sigma}(\tilde{b}_i)$ \end{tabular}
\\
\hline
0.10 & 1.07 & 3.08 & -0.00
\\
0.21 & 1.07 & 1.80 & 0.00
\\
0.42 & 1.21 & 0.46 & 0.05
\\
0.65 & 1.40 & 0.37 & 0.02
\\
0.79 & 0.87 & 0.72 & 0.03
\\
0.91 & 0.77 & 0.55 & 0.05
\\
1.07 & 0.67 & 0.65 & 0.01
\\
1.39 & 0.69 & 0.66 & 0.02
\\
\hline
\end{tabular}
\quad\quad
\begin{tabular}{|c|c|c|c|}
\hline
\begin{tabular}{c} $\bar{z}_i$ \end{tabular}
&
\begin{tabular}{c} $\tilde{f}_i$ \end{tabular}
&
\begin{tabular}{c} $\displaystyle\frac{\sigma}{\theta}(\tilde{f}_i) (\%)$ \end{tabular}
&
\begin{tabular}{c} $\displaystyle \frac{\Delta}{\sigma}(\tilde{f}_i)$ \end{tabular}
\\
\hline
0.10 & 0.45 & 10.17 & 0.00
\\
0.21 & 0.47 & 6.03 & -0.00
\\
0.42 & 0.48 & 2.26 & -0.05
\\
0.65 & 0.47 & 2.81 & -0.02
\\
0.79 & 0.46 & 2.33 & -0.03
\\
0.91 & 0.45 & 1.64 & -0.05
\\
1.07 & 0.43 & 1.68 & -0.01
\\
1.39 & 0.40 & 2.33 & -0.01
\\
\hline
\end{tabular}
\end{center}
\end{table}

\subsection{Determination of the lensing amplitude}

\label{s:SKA2_AL}

\begin{table}[h!]
\caption{
$1\sigma$ relative error (in percent) for the standard $\Lambda$CDM parameters, the bias, and the amplitude of the lensing potential, $A_{\rm L}$, for SKA2 with 11 redshift bins.
}
\begin{center}
\begin{tabular}{| c c c c c c c|}
\hline
$b_0$ & $\Omega_{\rm baryon}$ & $\Omega_{\rm cdm}$ & $h$ & $n_s$ & $\ln{10^{10}\, A_s}$ & $A_\mathrm{L}$ \\
\hline
0.37\% & 1.32\% & 0.37\% & 1.26\% & 1.05\% & 0.70\% & 5.46\%\\
\hline
\end{tabular}
\end{center}
\label{table:ska2_errors_AL}
\end{table}

Finally, we study how well the amplitude of the lensing potential can be measured with SKA2.
We include a parameter $A_{\rm L}$ in front of the lensing potential, with fiducial value $A_{\rm L}=1$, and we let this parameter vary in the Fisher forecasts.
Since the quantity which is measured with lensing is the combination $A_{\rm L}(5s(z)-2)$, we can only measure $A_{\rm L}$ if the magnification bias parameter is known.
The results are shown in Table~\ref{table:ska2_errors_AL}.
We see that SKA2 can measure $A_{\rm L}$ with a precision of 5.46\%, which reflects the relatively large signal-to-noise ratio of lensing in SKA2.
Comparing these results with those in Table~\ref{t:SKA11bin-bf}, with $A_{\rm L}$ fixed, we see that adding this extra free parameter has almost no impact on the other constraints, that are degraded by less than 1\%.
This can be understood by the fact that the constraints on $\Lambda$CDM parameters come exclusively from density and RSD.
Adding lensing does not improve the constraints on these parameters.
Therefore whether $A_{\rm L}$ is fixed or not has no impact on the constraints of the $\La$CDM parameters.

\section{The $C_\ell(z,z')$ angular power spectra -- photometric surveys}\label{s:Cls}

\begin{figure}
\centering
\includegraphics[width=\linewidth]{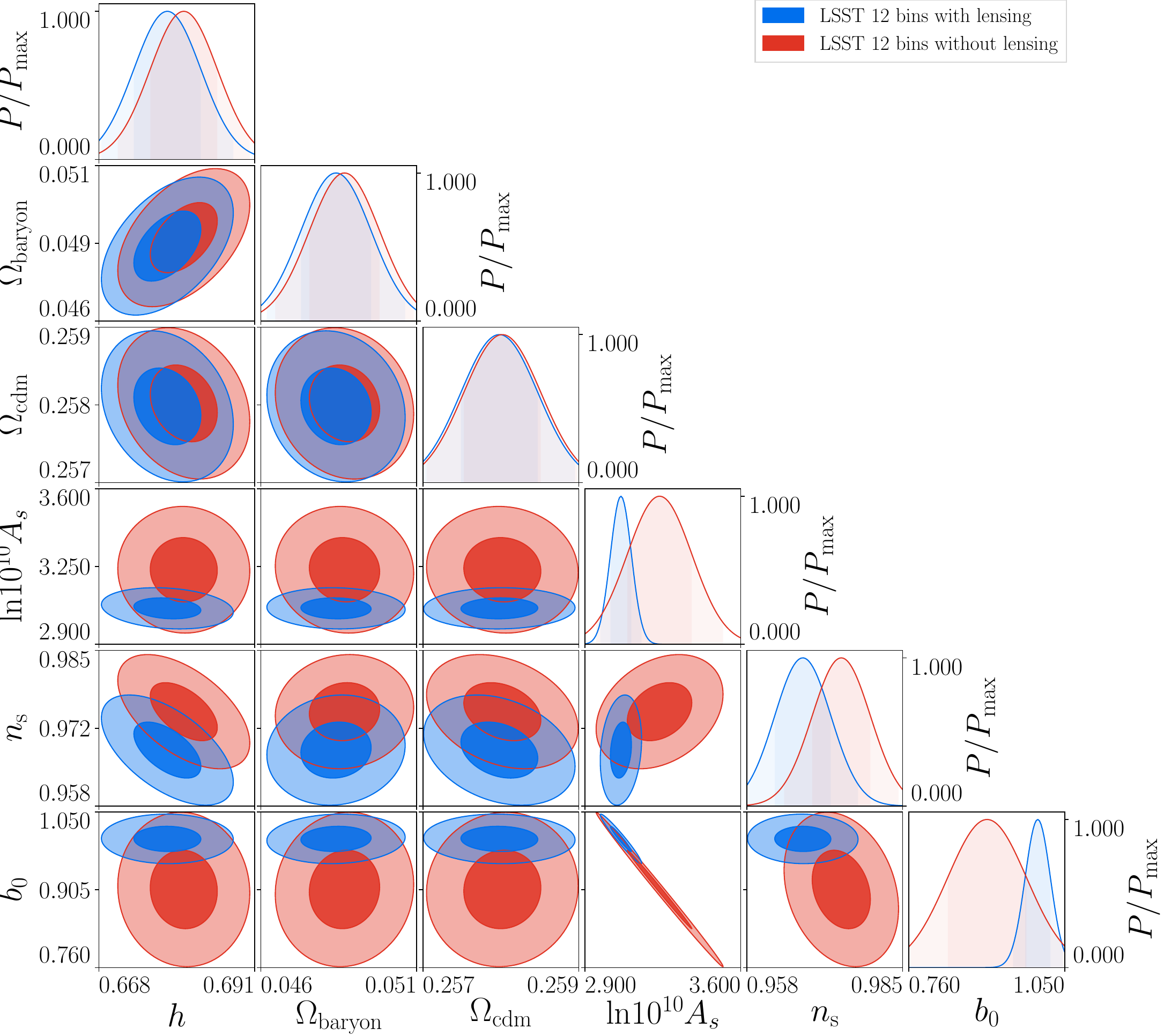}
\caption{Contour plot for the LSST Fisher analysis, in the 12 bins configuration.
Blue contours show the constraints on the parameters when we consistently include lensing in our theoretical model. Red contours show the constraints on the parameters obtained neglecting lensing magnification. In the latter case, the best-fit parameters are shifted with respect to the fiducial values.
}
\label{fig:lsst-fisher-12bins}
\end{figure}

For photometric surveys, where the redshift is not very well known, and the number of redshift bins is not exceedingly large, we base our parameter estimation on the angular power spectra, $C_\ell(z,z')$.
The angular power spectrum is related to the two-point correlation of $\Delta$ through
\be\label{e:Cldef}
\left\langle\De(\bn,z)\De(\bn',z') \right\rangle = \frac{1}{4\pi}\sum_\ell (2\ell+1)C_\ell(z,z')P_\ell(\bn\cd\bn') \,,
\ee
where $P_\ell$ denotes the Legendre polynomial of degree $\ell$.
The $C_\ell(z,z')$'s are well adapted to future surveys since they automatically encode wide-angle effects. Another advantage of the $C_\ell(z,z')$'s is that they refer only to directly measured quantities $z,z'$ and $\theta$ and can therefore be determined from the data in a completely model independent way.
Moreover, lensing is easily included in their modelling.
The $C_\ell(z,z')$'s provide therefore a new route, alternative to the shape measurements of background galaxies, to determine the lensing potential.

We investigate the precision with which a photometric galaxy catalog like LSST will be able to constrain cosmological parameters from the galaxy number counts. We consider three different configurations with 5, 8 and 12 redshift bins and perform a Fisher matrix analysis for the $\Lambda$CDM parameters given in Table~\ref{table:planck_params}, plus the bias with fiducial value $b_0=1$. The evolution of the bias with redshift is given in Appendix~\ref{a:LSST}. We first fix $s(z)$ as given in Appendix~\ref{a:LSST}.

As for spectroscopic surveys, we perform two Fisher analyses: one where we neglect lensing, and one where we include it. This allows us to compare the error bars in these two cases and to determine whether lensing brings additional constraining power.
We also compute the shift of the parameters when lensing is excluded.
In all cases we include both the auto-correlation, $C_\ell(z,z)$, and the cross-correlation between different bins, $C_\ell(z,z')$ with $z\neq z'$. More details are given in Appendix~\ref{ap:photo-alone}.

The results are shown in Fig.~\ref{fig:lsst-fisher-12bins}.
We find that increasing the number of bins significantly reduces the error bars and in the following we show therefore only the optimal case, with 12 redshift bins.
Results for the 5 and 8 bins configurations can be found in Appendix~\ref{ap:photo-alone}.
Comparing the error bars with and without lensing, we see that adding lensing has a small impact on the constraints for most of the $\Lambda$CDM parameters, nearly as for spectroscopic surveys.
The only constraints that are significantly improved when lensing is included are those on the bias $b_0$ and on the primordial amplitude $A_s$.
This can be understood by the fact that the density contribution is sensitive to the combination $b_0^2 A_s$, whereas the lensing contribution is sensitive to $b_0 A_s$ (through the density-lensing correlation) and to $A_s$ (through the lensing-lensing correlation). As a consequence, the lensing contribution helps breaking the degeneracy between $A_s$ and $b_0$.

We have studied how this improvement changes if instead of fixing $s(z)$ we model it by three parameters (see Appendix~\ref{a:LSST}) that we let vary in our Fisher analysis. We find that in this case including lensing provides almost no improvement on the parameters constraints (see Fig.~\ref{fig:lsst-fisher-sbias-margin} in Appendix~\ref{s:LSST_constraints}). A good knowledge of $s(z)$ is therefore crucial.

From Fig.~\ref{fig:lsst-fisher-12bins}, we see that the shifts induced by neglecting lensing are somewhat larger than in spectroscopic surveys.
This is due first to the fact that in photometric surveys the density contribution is smaller than in spectroscopic surveys, due to the large size of the redshift bins, which averages out the small-scale modes.
In spectroscopic surveys, this effect is almost absent since the density contribution is averaged over the size of the pixels, which are usually very small, between $2-8$\,Mpc/$h$.
The lensing on the other hand is almost unaffected by the size of the bins or pixels (see e.g.\ discussion in~\cite{Bonvin:2011bg}), and therefore its relative importance with respect to the density contribution is larger in photometric surveys than in spectroscopic surveys.
The second effect which reduces the importance of lensing in spectroscopic surveys is the fact that only the first three multipoles are measured there.
Since lensing has a complicated dependence on the orientation of the pair of galaxies, these first three multipoles only encode part of the lensing signal, which contributes to much larger multipoles as has been shown in~\cite{Tansella:2017rpi}.
As a consequence, part of the lensing signal is removed in spectroscopic analyses.
This is not the case in photometric surveys, where the angular power spectrum is used, which contains the full lensing contribution.

The fact that the lensing contribution shifts cosmological parameters in photometric surveys is consistent with previous studies~\cite{Raccanelli:2015vla,Alonso:2015uua,Cardona:2016qxn,Lorenz:2017iez}, and it shows that lensing cannot be neglected in a survey like LSST. {The parameter which is the most shifted is} {the spectral index $n_s$, which is shifted by 1.3$\si$ when lensing is neglected (see Table 16 in Appendix~\ref{s:LSST_constraints}). Lensing adds small scale power and if this is interpreted as coming from density fluctuations, a somewhat larger spectral index is inferred.} {The bias $b_0$ and the amplitude $A_s$ are also shifted by more than 1$\sigma$. We see that $A_s$ is shifted toward a larger value, whereas $b_0$ is shifted toward a smaller value. This can be understood in the following way: the lensing term contributes positively to the angular power spectrum at high redshift, when lensing is important. $A_s$ and $b_0$ must therefore be shifted to increase the amplitude of the $C_\ell$'s. Since lensing increases with redshift, this shift cannot be the same at all redshift, and it is therefore not possible to increase both $b_0$ and $A_s$. The solution is then to increase $A_s$ but to decrease $b_0$. By doing that, the density contribution, which is proportional to $A_s b_0^2$, increases less than the RSD contribution which is sensitive to $A_s b_0$ and $A_s$. Since the relative importance of RSD with respect to the density increases with redshift, this is the best way of mimicking the lensing signal. Note that the opposite happens, if instead of considering 12 redshift bins, we consider 5 redshift bins. In this case, $A_s$ is shifted toward a smaller value, while $b_0$ is shifted toward a larger value (see Table 16 in Appendix~\ref{s:LSST_constraints}). For 5 redshift bins, RSD are completely subdominant, and therefore the redshift dependence of the lensing cannot be reproduced by a shift of $A_s$ and $b_0$. What governs the shift of $A_s$ and $b_0$ in this case is probably the redshift bin where lensing affects the constraints most.}
Finally, $h$ is also shifted, by a larger amount than in the spectroscopic case. Only $\Om_{\rm cdm}$ and $\Om_{\rm baryon}$ are virtually unaffected.
On the other hand, we have tested that including or not the large scale relativistic effects does not influence the parameter estimation appreciably.
It is therefore justified to only consider density, RSD and lensing for the analysis.

\begin{figure}
\centering
\includegraphics[width=\linewidth]{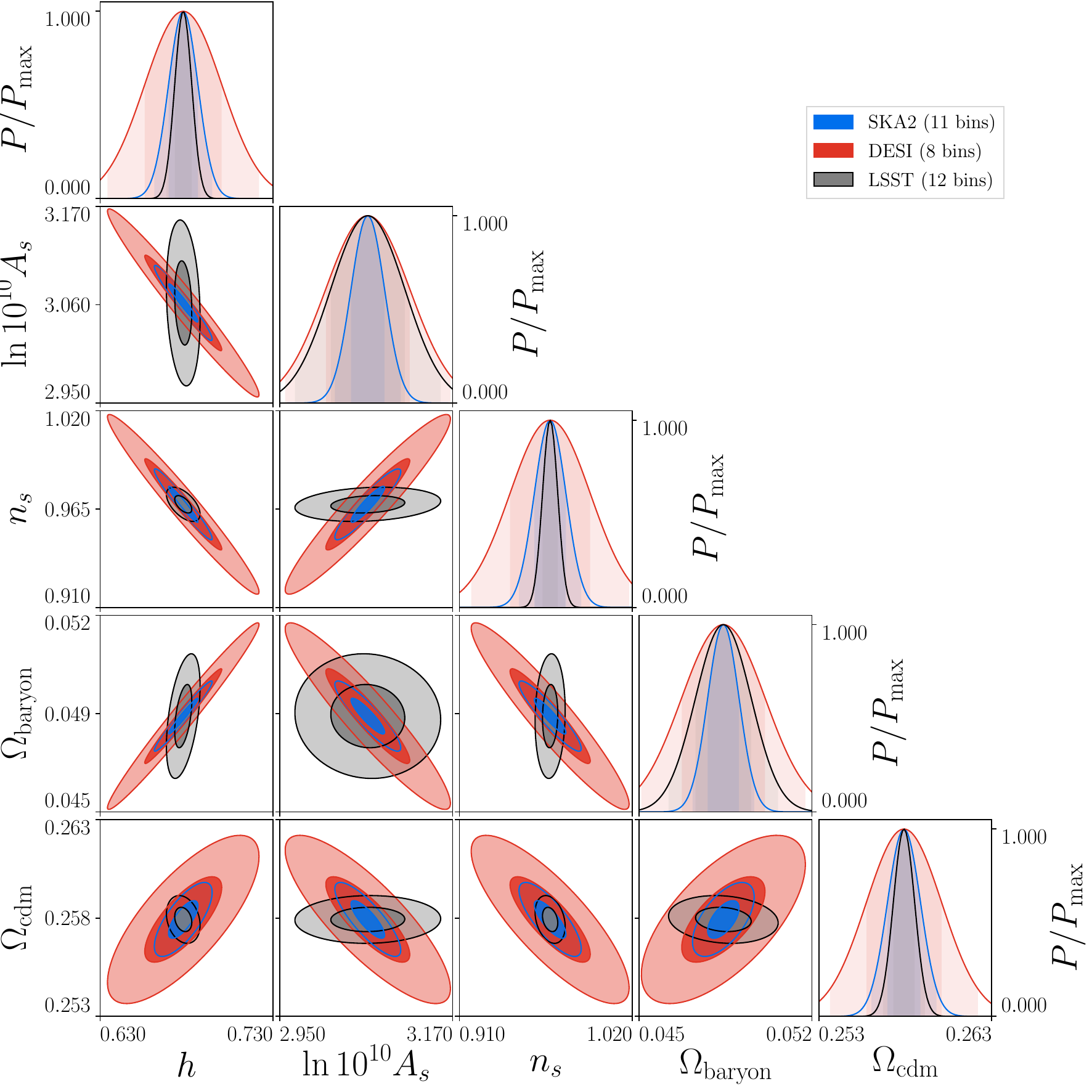}
\caption{
The comparison of constraints on cosmological parameters in $\Lambda$CDM, including lensing, from: SKA2 spectroscopic using 11 bins (blue), DESI spectroscopic using 8 bins (red), and LSST photometric using 12 bins and the redshift range [0, 2.5] and using both auto- and cross-correlations (black), with the galaxy bias parameter $b_0$ marginalized.
}
\label{fig:comparison}
\end{figure}

\begin{table}[h!]
\caption{
Constraints (in percent) on $\Lambda$CDM parameters for DESI (8 bins), SKA2 (11 bins) and LSST (12 bins).
}
\begin{center}
\begin{tabular}{|c|cccccc|}
\hline
\backslashbox{survey}{parameter}
& $b_0$ & $\Omega_\mathrm{baryon}$ & $\Omega_\mathrm{cdm}$ & $h$ & $n_s$ & $\ln 10^{10} A_s$
\\
\hline
DESI & 0.75 & 3.50 & 0.85 & 3.30 & 2.65 & 1.75
\\
\hline
SKA2 & 0.36 & 1.31 & 0.37 & 1.26 & 1.04 & 0.70
\\
\hline
LSST & 2.4 & 2.41 & 0.21 & 0.67 & 0.50 & 1.6
\\
\hline
\end{tabular}
\end{center}
\label{table:comparison_surveys_constraints_3}
\end{table}

In Fig.~\ref{fig:comparison} and Table~\ref{table:comparison_surveys_constraints_3}, we compare the size of the error bars for LSST, DESI and SKA2.
{The error bars on $h$ and $n_s$ in LSST} are a factor of 5 times smaller than in DESI and a factor of 2 times smaller than in SKA2. {The constraints on $\Omega_{\rm cdm}$ are also better in LSST than in SKA2 and DESI.} This is simply due to the larger number of galaxies in photometric samples, which strongly reduces the shot noise compared to spectroscopic samples. {On the other hand, the constraints on $\Omega_{\rm baryons}, b_0$ and $A_s$ are significantly better in SKA2 than in LSST. For $\Omega_{\rm baryon}$ this is due to the fact that the constraints depend on a good resolution of the acoustic oscillations. For $A_s$ and $b_0$, the difference between SKA2 and LSST comes from the fact that these two parameters are degenerate in the density contribution. RSD break this degeneracy very strongly in SKA2, but only marginally in LSST, where they are subdominant.
Lensing helps breaking this degeneracy in LSST, but less efficiently because of the size of this contribution.
Generally, from Fig.~\ref{fig:comparison}, we see that LSST and SKA2 are affected by different degeneracies between parameters, and that they are therefore highly complementary to constrain $\Lambda$CDM parameters.} As discussed in Section~\ref{s:growth}, SKA2 has furthermore the advantage of measuring the growth rate of structure $f$ in a model-independent way, {something which is not achievable with a photometric survey like LSST.}

Finally, we have studied how well we can measure the lensing amplitude $A_{\rm L}$ with LSST.
We find that $A_{\rm L}$ can be determined with a precision of 7.7\% (see Table~\ref{table:lsst_errors} in Appendix~\ref{s:LSST_AL}).
This is worse than the precision obtained on $A_{\rm L}$ with SKA2 (5.5\%), which is somewhat surprising since lensing is stronger in the cross-correlation of the $C_\ell$'s than in the multipoles of the correlation function.
It can however be understood from the fact that, in the $C_\ell$'s, RSD are strongly subdominant, whereas in the correlation function, RSD are extremely well detected.
As a consequence, in the $C_\ell$'s, $A_s$ and $b_0$ are degenerate and this degeneracy is only broken if lensing is added with a known amplitude.
If we do not know the lensing amplitude, adding lensing helps only marginally since lensing is now sensitive to the combinations $A_{\rm L}b_0A_s$ (through the density-lensing correlations) and to $A_{\rm L}^2 A_s$ (through the lensing-lensing correlations).
In this case adding lensing improves the constraints on $A_s$ and $b_0$ by only a few percents, and $A_{\rm L}$ is less well determined than in a spectroscopic survey, where the degeneracy between $b_0$ and $A_s$ is already broken by RSD.

\section{Combining angular power spectra and correlation function}
\label{s:main_result}

One drawback of the correlation function is that it does not account for correlations between different redshift bins. In this section, we propose an improved analysis that uses the multipoles of the correlation function in each redshift bin, and the angular power spectrum for the cross-correlations between different redshift bins. These two estimators can be considered as nearly independent, since the correlation function probes the auto-correlation of density and RSD within a bin, and neglects correlation of these quantities between different redshift bins. On the other hand, the angular power spectrum probes the density and lensing cross-correlations between different bins, but neglects their correlation within the same redshift bin. 
Assuming independence, the Fisher matrix of the combined data is simply the sum.
While it is not entirely correct, this assumption is often used in the literature for forecasts with combined probes such as spectroscopic galaxy clustering and photometric clustering analysis, see e.g. Ref.~\cite{Blanchard:2019oqi}.
In Appendix~\ref{s:cross} we estimate the error which we make by neglecting the cross-correlation of the photometric survey used for the $C_\ell$'s with the spectroscopic one used for the correlation functions inside individual bins. We find that the size of the covariance is affected by three distinct effects. First, since the spectroscopic survey has much better redshift resolution than the photometric one, its variance contains more small scale power and is significantly larger than both the covariance and the variance of the photometric survey. This effect strongly suppresses the covariance with respect to the variance of the spectroscopic survey, but not with respect to the variance of the photometric survey. The second relevant effect is shot noise, that affects both the variance of the spectroscopic survey and the variance of the photometric survey (in a lesser extend), but not the covariance between the two surveys since it contains correlations between different populations of galaxies. This effect therefore suppresses the covariance with respect to both variances. In Appendix~\ref{s:cross} we estimate the importance of these two effects combined (resolution and shot noise), and we find that they are not enough to make the covariance negligible with respect to the variance of the photometric survey, in the case where the auto-correlations of the $C_\ell$'s are included in the photometric sample. Therefore neglecting the covariance between a spectroscopic and a photometric survey may not be justified in this case, contrary to what is usually assumed in the literature~\cite{Blanchard:2019oqi}.

For this reason, in our analysis we remove the auto-correlations of the $C_\ell$'s and only include cross-correlations between different redshift bins. In this case, the covariance is always suppressed with respect to the variance of each of the surveys, due to the fact that cross-correlations between different redshift bins are smaller than auto-correlations. Neglecting the covariance between the two surveys and summing the Fisher matrices is therefore well justified in our analysis.

Finally, let us note that even though our approximation may slightly overestimate the gain from lensing, our approach is still conservative in many ways. First, we use a conservative cut in $\ell$, which tends to underestimate the lensing contribution, that is increasing with $\ell$, and second, as explained above, we remove the auto-correlations of the $C_\ell$'s from the signal in the photometric sample, which also contain information.

With our approximation, the Fisher matrix for the combined sample is simply the sum of the Fisher matrix for the correlation function in the spectroscopic survey, and the Fisher matrix for the cross-correlation in the photometric survey
\be
F^\text{comb}_{\alpha\beta} =
F^\text{auto}_{\alpha\beta}
+ F^\text{cross}_{\alpha\beta}.
\label{combined-fisher}
\ee
In Eq. \eqref{combined-fisher} $F^\text{auto}_{\alpha\beta}$ denotes the Fisher matrix for the 
auto-correlations of the bin, which uses the multipoles of the correlation function as the estimator, and is computed from Eq. \eqref{fisher-xi}, while $F^\text{cross}_{\alpha\beta}$ encodes the information of the cross-correlations between different bins, estimated through the angular power spectrum. 
$F^\text{cross}_{\alpha\beta}$ is computed 
as the Fisher matrix for the angular power spectrum analysis which is well-known in the literature (see for example \cite{Villa_2018})
\begin{equation} \label{Fishdef}
F^\text{cross}_{\alpha\beta}=  \sum^{\ell _{\rm max}}_{\ell={\ell _{\rm min}}} \sum_{(ij)(pq)} \frac{\partial C^{ij}_{\ell,{\rm th}}}{\partial\theta_\alpha}  \frac{\partial C^{pq}_{\ell, {\rm th}}}{\partial\theta_\beta} {\rm cov}^{-1}_{C_{\ell}\,[(ij),(pq)]}\,,
\end{equation}
where ${\rm cov}^{-1}_{C_{\ell}\,[(ij),(pq)]}$ is the inverse of the covariance for the $C_\ell$, $\theta_\alpha$ denotes the $\theta$-th parameter, the second sum runs over the indices $(ij)$ and $(pq)$ with $i \leq j$ and  with $p \leq q$ which range from $1$ to the total number of redshift bins $N_{\rm bin}$.
All auto- and cross-correlations between the redshift bins are included in the sum, however, we set the derivative of the auto-correlation with respect to all cosmological parameters to zero, i.e. $\dd C_{\ell,\text{th}}^{ii} / \dd \theta_\alpha = 0$.
This way, we include only the cross-bin information in $F^\text{cross}_{\alpha\beta}$.

We consider the $\Lambda$CDM parameters of Table~\ref{table:planck_params}, plus the galaxy bias of SKA2.

We perform an analysis where we include lensing in the angular power spectrum but neglect it in the multipoles of the correlation function. We first study the shift inferred on cosmological parameters in this analysis. The motivation for this analysis is that it is much easier to include the lensing contribution in the $C_\ell(z,z')$ than in the correlation function. We want therefore to understand if this is enough to remove the shift of the whole analysis. We find that this is not the case, and that the shift is the same with and without the cross-correlations (see Fig.~\ref{fig:ska2_with_cross} in Appendix~\ref{s:cross}).

We then compare the constraints on $\Lambda$CDM cosmological parameters of this analysis with the ones obtained when only the correlation function is used.
We find that adding the $C_\ell$'s improves the constraints by less than 6 \%.

\begin{figure}
\centering
\includegraphics[width=\linewidth]{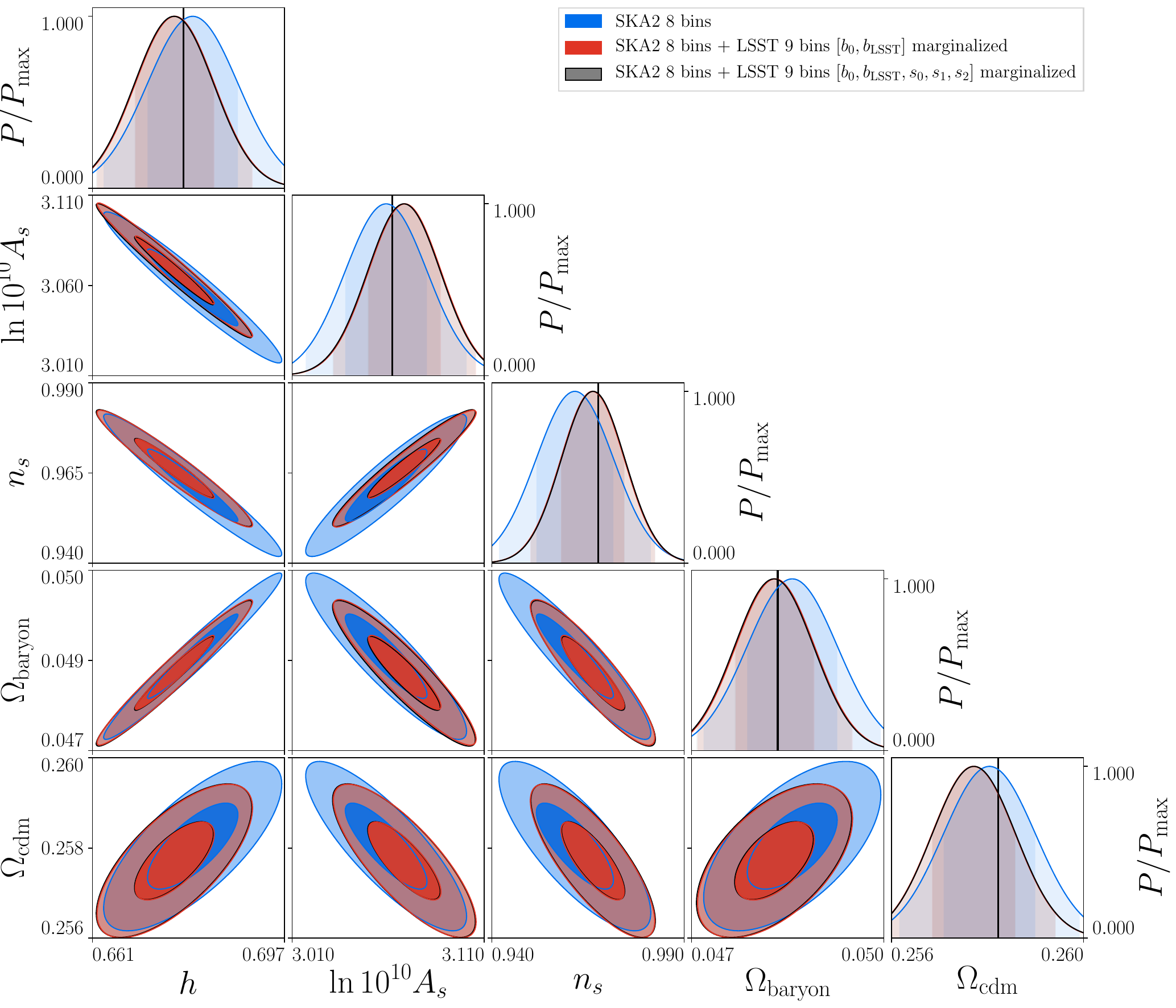}
\caption{
The constraints on cosmological parameters from SKA2 (blue) and SKA2 + LSST only cross-correlation (red) using a configuration with 8 redshift bins as described in~\ref{a:SKAII}, plus the additional redshift bin for LSST with $\bar z = 2.25$ and $\Delta z = 0.5$.
The shifts of the contours are computed neglecting lensing from the correlation function (SKA2), and consistently including it in the angular power spectrum (LSST).
The black lines on the diagonal plots denote the fiducial values.
}
\label{fig:ska2_with_lsst}
\end{figure}

We then extend this combined analysis to the case where we have a spectroscopic and a photometric survey. For this we use equal redshift bins for both types of surveys. Inside one redshift bin we then consider only the spectroscopic survey with the correlation function, while for the cross-correlations of different bins we consider only the photometric survey with the angular cross-power spectrum, $C_\ell(z,z')$. To this we add the auto and cross-correlations of the $C_\ell(z,z')$ for the high redshift bins, $z\in[2,2.5]$, that are observed with LSST but not with SKA2~\footnote{We also consider the cross-correlations of these high bins with the ones common to LSST and SKA2.}. As before we can neglect the covariance between the $C_\ell$'s and the correlation function. In principle, we could also add to this combined analysis the auto-correlation of the angular power spectrum for $z\leq 2$, which brings additional information since the sample of galaxies in the spectroscopic and photometric surveys are different. However, to do this in a consistent way we would have to account for the covariance between the $C_\ell$'s and the correlation function within the same redshift bin, since the different populations of galaxies trace the same underlying density field. This is beyond the scope of our paper.

We use the 8 bin configuration from SKA2, since the redshift resolution of LSST does not allow us to use the 11 bin configuration, and we add an extra bin for LSST with $\bar z = 2.25$ and $\Delta z = 0.5$.
We have tested that the results for SKA2 are very similar for the 8 and 11 bins configurations.
We consider the $\Lambda$CDM parameters of Table~\ref{table:planck_params}, plus the bias of SKA2 and the bias of LSST.
Since the bias of SKA2 and the bias of LSST are independent, we cannot constrain the bias of LSST using the auto-correlation of SKA2.
To solve this problem, we include in the entry of the Fisher matrix related to the bias of LSST the auto-correlation of the $C_\ell$'s.
This can be consistently done since the two biases are independent and there is therefore no covariance between the $C_\ell$'s and the correlation function for this entry.

As before, we compute the shift generated by an analysis where we include lensing in the angular power spectrum but not in the correlation function. In Fig.~\ref{fig:ska2_with_lsst} and Table~\ref{table:comparison_surveys_constraints} we compare this analysis with the one of Section~\ref{s:shiftLCDM}, where we consider only SKA2.
We have marginalized over the clustering biases of the two surveys.
{For the parameters, $n_s$ and $\Omega_{\rm baryon}$, adding the angular power spectrum reduces the shift generated by neglecting lensing.
However, it does not completely remove it.
Moreover, for $\Omega_{\rm cdm}$, $A_s$ and $h$, the shift even though still quite small, actually increases. Interestingly for $h$ and $A_s$ the shift  changes sign. This is possible since the parameters are not independent. For example decreasing $\Omega_{\rm cdm}$ requires a larger amplitude $A_s$ and in the present case, the best fit actually request an $A_s$ which is about $0.35\si$ too large, while when only considering SKA2 one obtains a value which is $0.15\si$ too small.} {In Table~\ref{table:comparison_surveys_constraints} we also compare the shifts when the magnification bias of LSST is fixed, with the shifts when the magnification bias is parameterized with three parameters. We see that the shifts are very similar in the two cases.}
These results show that including lensing only in the angular power spectrum is not a satisfactory solution. One needs to include it both in the angular power spectrum and in the multipoles of the correlation function.

\begin{table}
\caption{
Constraints (in percent) and shifts in unit of $1\sigma$ on $\Lambda$CDM parameters for SKA2 (8 bins), SKA2 (8 bins) + LSST-cross (9 bins) with magnification bias fixed, SKA2 (8 bins) + LSST-cross (9 bins) with three parameters for the LSST magnification bias marginalized over.
}
\begin{center}
\begin{tabular}{|c|c|ccccc|}
\hline
\backslashbox{survey}{parameter} & & $h$ & $\ln 10^{10} A_s$ & $n_s$ & $\Omega_\mathrm{baryon}$ & $\Omega_\mathrm{cdm}$
\\
\hline
SKA2  & $\sigma\,[\%]$ &1.255&0.697&1.041&1.312&0.372\\
& $\Delta / \sigma$ & 0.198&-0.145&-0.606&0.320&-0.194
\\
\hline
SKA2 + LSST-cross  &$\sigma\,[\%]$ &1.096&0.619&0.853&1.148&0.337 \\
 $s(z)$ fixed &$\Delta / \sigma$ & -0.225&0.347&-0.183&-0.069&-0.594\\
\hline
SKA2 + LSST-cross
&$\sigma\,[\%]$ &1.096&0.619&0.854&1.149&0.337 \\
$s(z)$ marginalized   & $\Delta / \sigma$ & -0.243&0.327&-0.165&-0.094&-0.585 \\
\hline
\end{tabular}
\end{center}
\label{table:comparison_surveys_constraints}
\end{table}

We then compare the constraints in the best-fit parameters obtained from the SKA2 analysis, with the ones in the combined analysis. The results are shown in Table~\ref{table:comparison_surveys_constraints} for the two cases, i.e.\ when the magnification bias of LSST is fixed and when it is parameterized with three parameters.~\footnote{Note that
we have verified that, for SKA2, neglecting lensing or parametrizing the magnification bias $s_\mathrm{SKA2}(z)$ and then marginalizing over it has very little impact on the constraints.} In both cases, we find that including the cross-correlations from LSST, the constraints on the parameters improve by 10--20\%.
{We find that, marginalizing over the magnification bias parameters does not have a significant impact on the combined analysis. This is due to the fact that the improvement in the constraints in the combined analysis is mainly driven by the density correlations between adjacent redshift bins and not by lensing.}

Note that here we have only studied the $\Lambda$CDM case. We defer an analysis, where we consider the growth of structure in each redshift bin as a free parameter, to a future paper. This will require to rewrite the angular power spectrum in terms of this growth rate and to determine how well the constraints on this quantity are improved by adding the $C_\ell$'s to the correlation function.

\section{Conclusion}\label{s:con}

In this paper we have studied the impact of gravitational lensing on the number counts of galaxies. We have considered both spectroscopic and photometric surveys. We have used a Fisher matrix analysis to determine the precision with which parameters can be constrained, and to compute the shift of these parameters when we neglect lensing. We have only considered quasi-linear scales. Including smaller scales with good theoretical control over the non-linear corrections will certainly improve the capacity of future surveys.

For a photometric survey like LSST, we have confirmed the results previously derived in the literature~\cite{Montanari:2015rga, Cardona:2016qxn, DiDio:2016ykq, Lorenz:2017iez, Villa_2018}, showing that lensing cannot be neglected in the measurement of standard $\Lambda$CDM parameters since this would induce a shift of these parameters as high as 1.3$\sigma$.

Our analysis for spectroscopic surveys is completely new, since lensing has never been included in the modelling of the multipoles of the correlation function so far. We have found that the importance of lensing on parameters estimation depends strongly on the cosmological model. For a $\Lambda$CDM analysis, neglecting lensing in a survey like SKA2 generates a shift of at most $0.6\sigma$.
However, we argue that this is non-negligible, because such a shift could hide (or enhance) deviations from General Relativity. If instead we perform a model-independent analysis, where the parameters to measure are the growth rate of structure and the bias in each redshift bin, then neglecting lensing in SKA2 generates a shift of the growth rate as large as $2.3\sigma$ in the highest bins of the survey. Since the growth rate is directly used to test the theory of gravity, such a large shift would be wrongly interpreted as the breakdown of General Relativity. It is therefore of crucial importance to develop fast and efficient codes that include the lensing contribution, and that can be used in the analysis of future spectroscopic surveys.

Contrary to SKA2, we have found that lensing has almost no impact on a survey like DESI. This is mainly due to the fact that in DESI the value of the prefactor, $5s(z)-2$, is 6 times smaller than for SKA2 at $z>1$, where the lensing contribution could become relevant.
However, a more detailed modelling of $s(z)$ is needed in order to confirm this result.

We have also compared the constraining power of DESI, SKA2 and LSST. A somewhat surprising result is that LSST promises the best constraints on {three of the standard cosmological parameters: $\Omega_{\rm cdm}, h$ and $n_s$} In particular, we have found that LSST can achieve about twice smaller error bars than the most futuristic spectroscopic survey presently planned, SKA2, {for $h$ and $n_s$}. The main reason for this is the number of galaxies which is about 10 times higher in LSST ($N\sim 10^{10}$) than in SKA2 ($N\sim 10^9$) yielding 3 times smaller shot noise errors. This more than compensates for the reduced redshift accuracy which is not so crucial for {$\Omega_{\rm cdm}, h$ and $n_s$. On the contrary, SKA2 constrains better $\Omega_{\rm baryon}$ (which relies on a good resolution of the baryon acoustic oscillations), $A_s$ and the bias. These last two parameters are degenerate in the density contribution, and therefore RSD, which are very prominent in a spectroscopic survey like SKA2, are important to break this degeneracy. We have seen that lensing does help breaking this degeneracy in LSST, and consequently improves the constraints on $A_s$ and the bias by a factor of 3. However, this is not sufficient to be competitive with SKA2. Interestingly, we have also found that the amplitude of the lensing potential, $A_{\rm L}$, is better determined with SKA2 than with LSST.}

{This comparison shows that spectroscopic and photometric surveys are highly complementary to probe $\Lambda$CDM, since they are affected differently by degeneracies between parameters.} Using photometric surveys not only for shear (weak lensing) measurements but also for galaxy number counts is therefore a very promising direction. We however stress that the best constraints for LSST are achieved if we use 12 redshift bins, hence good photometric redshifts are important. The errors on parameters using 8 redshift bins for LSST are substantially larger than those of SKA2 and comparable or larger than those of DESI.

The real advantage of spectroscopic surveys over photometric surveys is their capability to measure the growth rate of structure in a model-independent way, by combining measurements of the monopole, quadrupole and hexadecapole. For DESI we have found that for the redshift bins centered around $z=0.4$ and higher, DESI can measure the growth rate with an accuracy of 2--3\% and the bias to better than 1\%. SKA2 will improve on these constraints, measuring the growth rate in each bins with an accuracy of 0.5--3\% for the growth rate, and of 0.2--1\% for the bias.

We have not studied the constraint on the growth rate for photometric surveys since for the bin widths of $\De z\geq 0.2$ considered in LSST, RSD are washed out and the sensitivity to $f(z)$ is lost. However, as shown in~\cite{Jalilvand:2019brk} (see their Fig. 9), already for bin widths of $\De z\leq 0.1$ the signal-to-noise ratio of RSD for a photometric survey with negligible shot noise can reach 10 for $z=0.5$ and larger for higher redshifts (up to 100 for $z=2$).
Therefore, with sufficiently good photometric redshifts and sufficiently high numbers of galaxies, photometric surveys can also be used to determine the growth rate and provide a stringent test of gravity.

Finally, we have proposed a way of combining spectroscopic and photometric surveys, by using the multipoles of the correlation function within each redshift bin, and the angular power spectrum to cross-correlate different redshift bins.
We have found that for SKA2 and LSST, neglecting lensing in the multipoles of the correlation function but including it in the $C_\ell$'s is not enough to completely remove the shift, which remains of the order of 0.6$\sigma$.

Therefore, lensing has to be included in the modelling of both the correlation function and the angular power spectrum.
Moreover, for standard $\Lambda$CDM parameters, we have found that adding cross-correlations from LSST to SKA2 improves the constraints by 10--20\%. This shows that some information is present in the cross-correlations between different redshift bins, that should not be neglected.
{We have found that, for $\Lambda$CDM parameters, this information mainly comes from density correlations between neighbouring bins, rather than from lensing correlations.}

At the moment, such a combined analysis does not allow us to measure the growth rate in a model-independent way, since it is not clear how this growth rate can be modelled in the angular power spectrum. This would however be an optimal analysis, that would combine model-independent measurements of the growth rate and of the lensing amplitude, which can be measured with good accuracy from the cross-correlations of the $C_\ell$'s. Such a measurement will significantly increase our capability to test gravity, by probing the relation between density, velocity and gravitational potentials.

\section*{Acknowledgments}
We thank Benjamin Bose for fruitful discussions.
The confidence ellipses of Figs.~\ref{fig:ska2_param_errors_std_bins11}, \ref{fig:lsst-fisher-12bins}, \ref{fig:comparison}, \ref{fig:ska2_with_lsst}, \ref{fig:ska2_with_cross} were plotted using a customized version of the CosmicFish package for cosmological forecasts~\cite{raveri2016cosmicfish}.
This work is supported by the Swiss National Science Foundation.

\section*{Disclaimer}
This is the Accepted Manuscript version of an article accepted for publication in Journal of Cosmology and Astroparticle Physics.
Neither SISSA Medialab Srl nor IOP Publishing Ltd is responsible for any errors or omissions in this version of the manuscript or any version derived from it.
The Version of Record is available online at \url{https://doi.org/10.1088/1475-7516/2021/04/055}.

\vspace*{1cm}
\noindent{\LARGE \bf Appendix}
\appendix

\section{Survey specifications and biases}\label{ap:surveyspec}
In this appendix we specify the redshift bins, $z_i$, as well as the galaxy bias, $b(z_i)$, the magnification bias, $s(z_i)$, and the number of galaxies per bin, $N(z_i)$, for the three surveys considered in this paper. For LSST and SKA2 we can refer to the literature for estimations of the magnification bias $s(z)$.
For DESI we derive an effective $s(z)$ by studying the three different galaxy populations which the survey will observe.

\subsection{Magnification bias: generalities}

In order to compute the magnification bias for a galaxy population, we need to estimate the luminosity function, i.e.\ the comoving number density of sources in a certain luminosity range, for the type of galaxies under consideration.
The luminosity function is
modelled analytically with a
Schechter function. In terms of the absolute magnitude $M$ it can be expressed as~\cite{Montanari:2015rga}
\begin{equation}
\Phi(M, z) dM = 0.4 \ln{(10)} \phi^* \left(10^{ 0.4 (M^* - M)}\right)^{\alpha + 1} \exp{\left[- 10^{0.4 (M^* - M)} \right]} dM\,.
\end{equation}
The parameters $\phi^*$, $M^*$ and $\alpha$ are redshift dependent and they can be estimated from data for different types of galaxies.

The magnification bias, at a given
redshift, is computed as in~\cite{Montanari:2015rga}
\begin{equation}
s(z, M_\text{lim}) = \frac{1}{\ln{10}}
\frac{\Phi(M_\text{lim}, z)}{\bar{n}(M<M_\text{lim})}\, ,
\label{eq:s-bias}
\end{equation}
where $\bar{n}(M < M_\text{lim})$ is the cumulative luminosity function
\begin{equation}
\bar{n}(M < M_\text{lim}) = \int^{M_\text{lim}}_{-\infty}
\Phi(M, z) \integrand M\, .
\end{equation}

The limiting magnitude, or flux, is fixed in
the observer frame-band. Therefore, for each
survey and galaxy sample we need to convert
the apparent limiting magnitude $m_\text{lim}$
into the absolute magnitude at the redshift
of the sources $M_\text{lim}$.
These two quantities are related
as follows~\cite{Alonso:2015uua}
\begin{equation}
M = m - 25 - 5 \log_{10}{\left[\frac{D_\text{L}(z)}{10 \,\,\text{Mpc}\,\, h^{-1}}\right]} + \log_{10}{h}
- K(z)\,.
\end{equation}
Here $D_\text{L}$ is the luminosity distance
and $K$ is the $K$-correction
\cite{Peacock:879495}
\begin{equation}
10^{0.4 K(z)} = \frac{\int T(\lambda) f_{\log}(\lambda) \integrand\ln{\lambda}}{\int T(\lambda) f_{\log}(\lambda/(1+z)) \integrand\ln{\lambda}}\, ,
\label{k-corr}
\end{equation}
where $f_{\log} = \lambda f_\lambda$ is the logarithmic flux density and $T$ is the effective filter transmission function in a given band.
The galaxy spectra are observed in a fixed waveband, while absolute magnitudes are affected by a shift of the spectrum in frequency.
The $K$-correction accounts for this effect: it is an estimate of the difference between the observed spectrum at a given redshift and what would be observed if we could measure true bolometric magnitudes.

To sum up, in order to estimate the magnification bias, we need the luminosity function of the galaxy population, the limiting magnitude for a galaxy survey, and an estimate of the $K$-correction for the survey's observed wavebands.

\subsection{LSST}
\label{a:LSST}

The Large Synoptic Survey Telescope (LSST)~\cite{LSST} is a wide-angle deep photometric survey expected to be operational from 2022.
In our forecast, we adopt the specifics for LSST described in
Ref.~\cite{Alonso:2015uua}, i.e.\
we consider the so-called LSST 'gold' sample: we assume a redshift range $z \in [0, 2.5]$ and a sky
fraction $f_\text{sky} = 0.5$.
The galaxy sample here considered
includes approximately 3 billions
galaxies.

The luminosity function of the sample is modelled
as a Schechter function, with constant slope and redshift-dependent $M^*$ and $\phi^*$, while the $K$-correction is assumed to be proportional to the redshift of the sources (see Ref.~\cite{Alonso:2015uua} for details).

The computation of the redshift distribution of the sources, the galaxy bias and the magnification bias has been implemented by the authors of Ref.
\cite{Alonso:2015uua} in a public routine\footnote{\url{http://intensitymapping.physics.ox.ac.uk/Codes/ULS/photometric/}}. The specifics for
LSST here described are computed
using this code.

\begin{figure}
\centering
\includegraphics[width=0.7\textwidth]{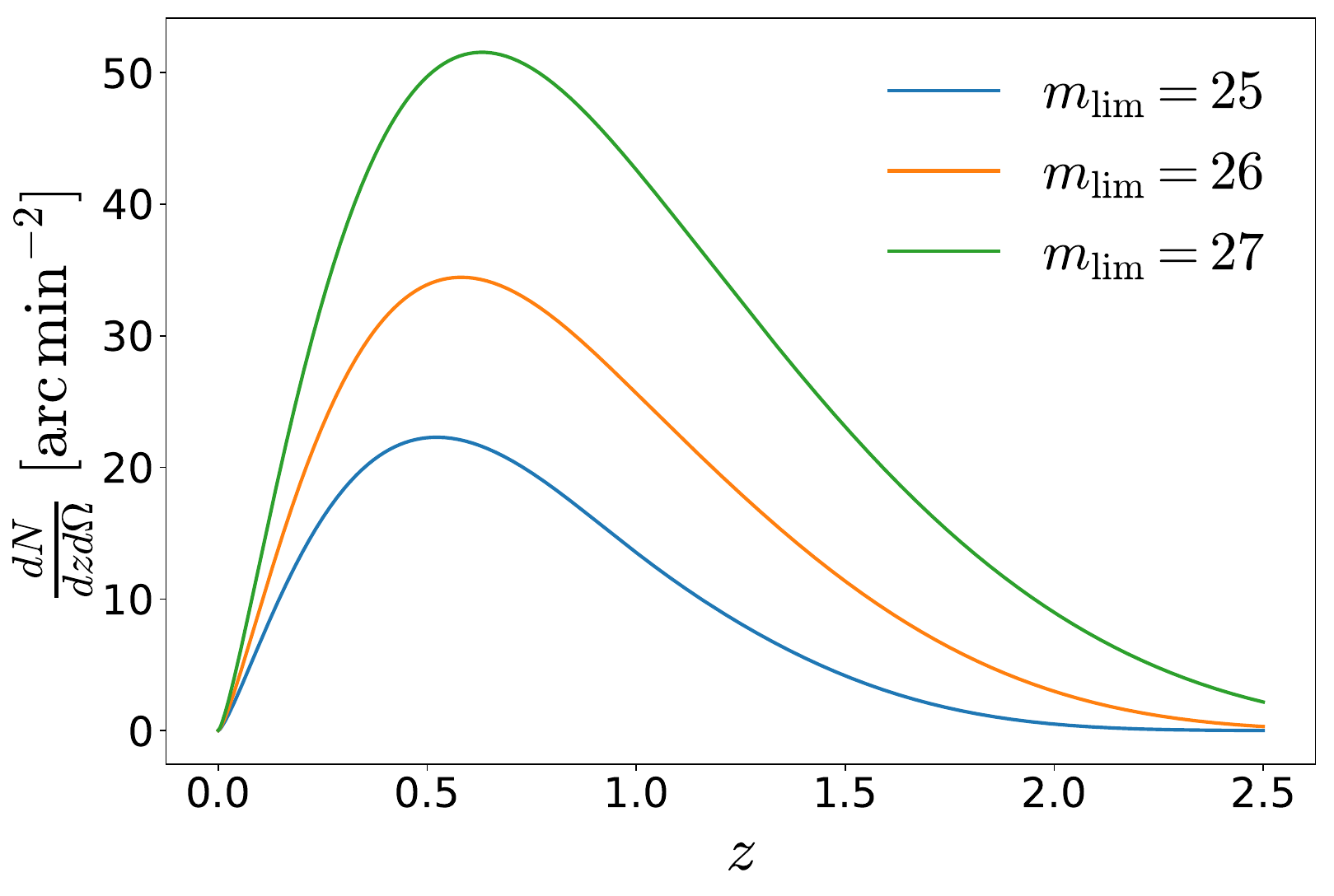}
\caption{Redshift distribution for the LSST galaxy sample, for three values of the magnitude cut
in the $r$-band $m_\text{lim}$.
}
\label{fig:lsst-dNdz}
\end{figure}

In Fig.~\ref{fig:lsst-dNdz} we show the redshift distribution for the LSST galaxy sample, for different values of the magnitude cut $m_\text{lim}$.
The 'gold' sample, which we use in our analysis, will adopt a magnitude-cut $m_\text{lim} = 26$.

\begin{figure}
\centering
\begin{subfigure}[b]{0.49\textwidth}
\includegraphics[width=\textwidth]{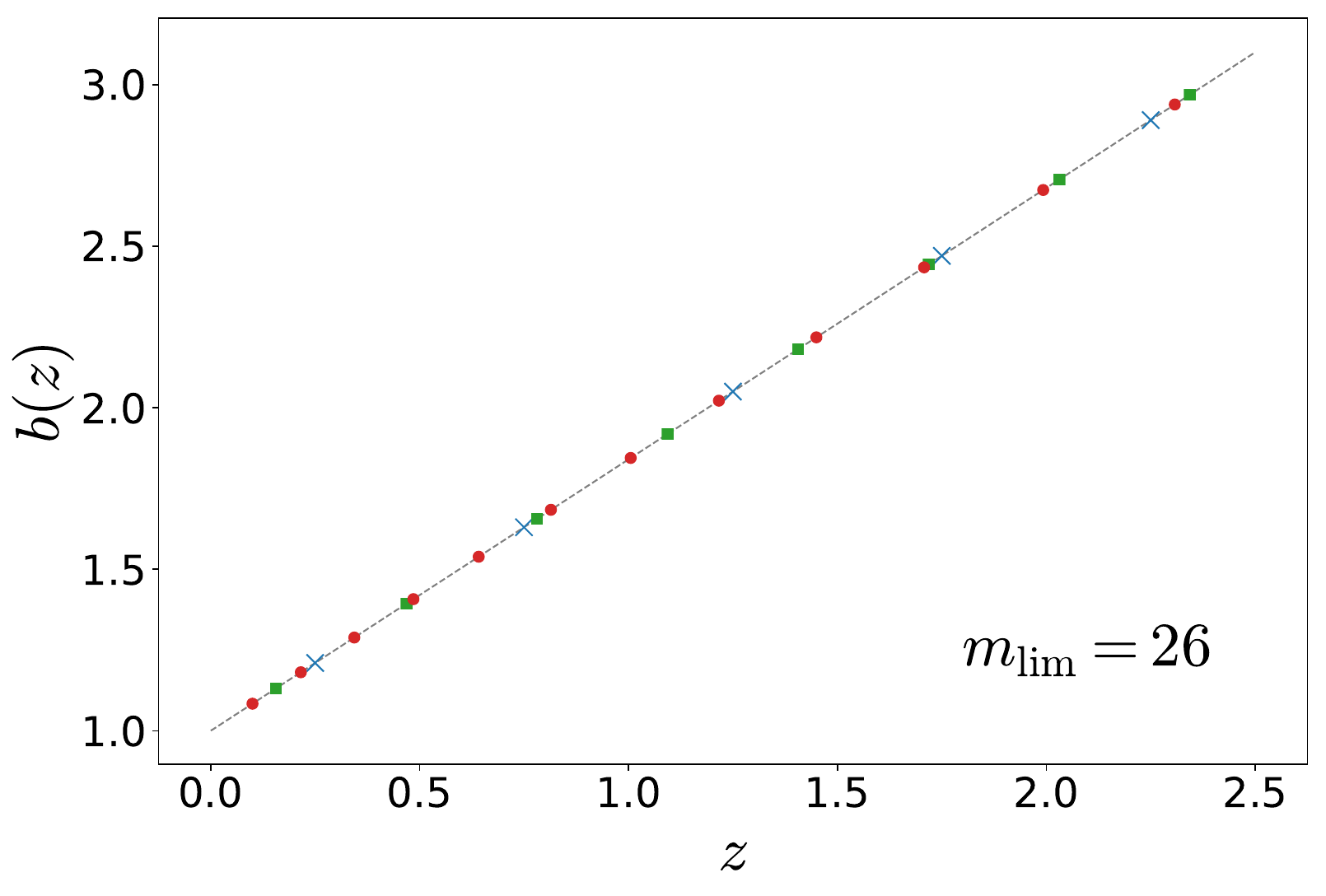}
\caption{Galaxy bias.}
\label{fig:b-lsst}
\end{subfigure}
\begin{subfigure}[b]{0.49\textwidth}
\includegraphics[width=\textwidth]{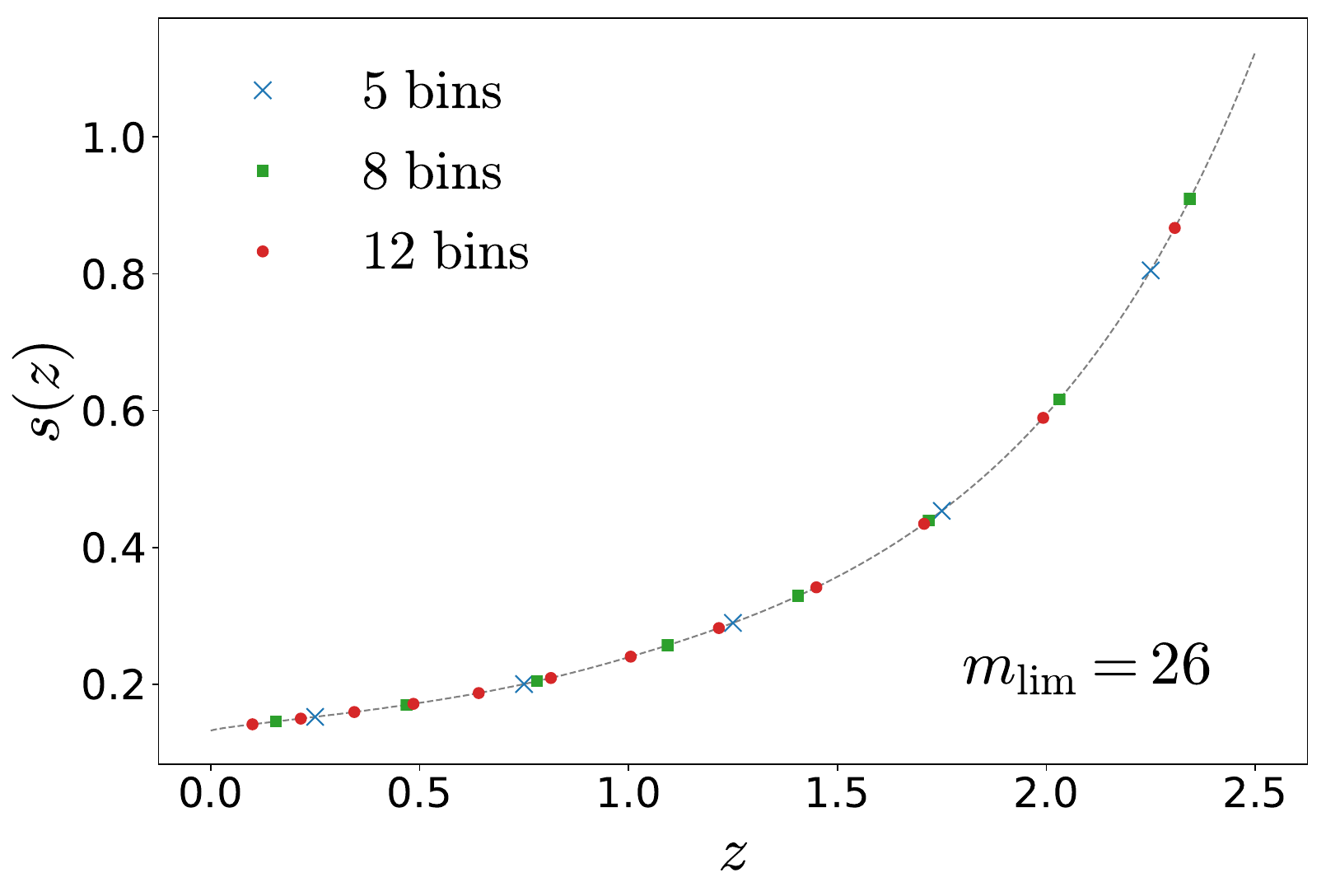}
\caption{Magnification bias.}
\label{fig:s-lsst}
\end{subfigure}
\caption{Clustering and magnification bias for LSST.}
\label{fig:lsst-biases}
\end{figure}

In Fig.~\ref{fig:lsst-biases} we show the redshift dependence of
galaxy bias (panel~\ref{fig:b-lsst})
and magnification bias (panel~\ref{fig:s-lsst}).
The galaxy bias is modelled as
$b(z) = 1 + 0.84\,\,z$. In our Fisher forecasts, we multiply this bias by a parameter $b_0$ with fiducial value $b_0=1$, that we let vary.
The different markers denote the values of the galaxy bias $b(z)$ at the mean redshifts for the three configurations studied in Appendix~\ref{ap:photo-alone}.

In our Fisher analyses we consider two cases for the magnification bias: one where we fix it to its fiducial value, and one where we model it with some parameters, that we let vary. For this purpose, we fit $s(z)$ from Fig.~\ref{fig:s-lsst} with an exponential
\be
s(z)=s_0 e^{s_1 z + s_2 z^2}\,,
\label{e:fit_s_LSST}
\ee
where $s_0, s_1$ and $s_2$ are three free parameters with fiducial values $s_0 = 0.1405663$, $s_1 = 0.30373535$ and $s_2 = 0.2102986$.

\subsection{DESI}\label{a:DESI}
\begin{figure}
\centering
\includegraphics[width=0.7\textwidth]{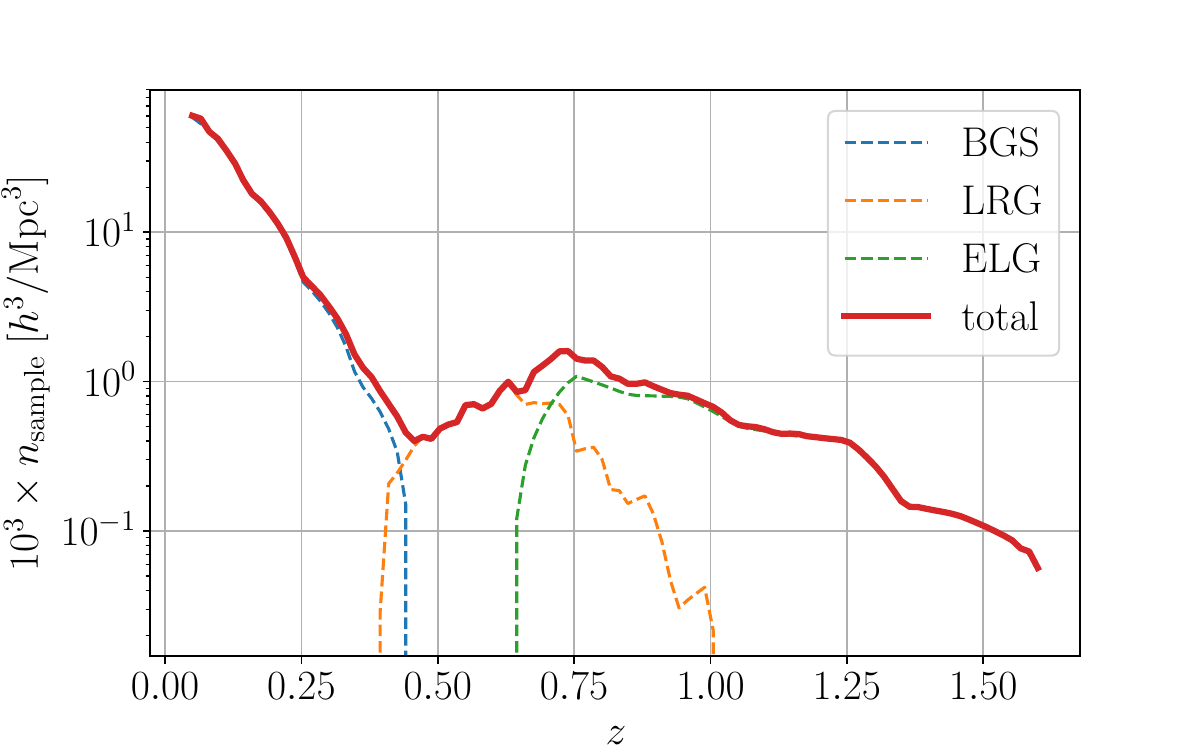}
\caption{
Number density of the three types of galaxies for the DESI survey.
}

\label{fig:desi_number_density}
\end{figure}

The DESI (Dark Energy Spectroscopic Instrument) survey will be split into three types of galaxies: emission line galaxies (ELG), luminous red galaxies (LRG), and the bright galaxy sample (BGS), the distribution of which is shown in Fig.~\ref{fig:desi_number_density}, with a total number of galaxies $N_\mathrm{tot} \sim 3.4 \times 10^7$.

Their galaxy biases are given by~\cite{Aghamousa:2016zmz}
\bea
b_{\rm ELG}(z) &=& 0.84 \times D_1(z=0)/D_1(z)\\
b_{\rm LRG}(z) &=& 1.7 \times D_1(z=0)/D_1(z)\\
b_{\rm BGS}(z) &=& 1.34 \times D_1(z=0)/D_1(z)\,,
\eea
where $D_1(z)$ denotes the growth factor.

We estimate the magnification bias for the three samples independently.
For the emission line galaxies, we assume that DESI will mainly target OII galaxies.
The OII luminosity function has been measured in Ref.~\cite{Comparat:2014xza}, and we use the best-fit parameters for the Schechter function reported in their Table 6.
We estimate the ELG $K$-correction from Eq. \eqref{k-corr}, where $f_\text{log}(\lambda)$ is the
typical rest-frame spectrum of an ELG galaxy and $T(\lambda)$ is the filter profile in a given band.
The typical ELG spectrum and the filters profile have been extracted
from Figure 3.9 in Ref.
\cite{Aghamousa:2016zmz}, for the
$grz$ optical filters.
For the three filters, we find that the ELG $K$-correction is proportional to the redshift $K(z) \sim -0.1\,z$.
We estimate the magnification bias for the three optical filters from Eq.~\eqref{eq:s-bias}, with magnitude
limits $m_\text{lim} = 24, \,23.4,\, 22.5$ for the $g$, $r$, $z$ bands, respectively~\cite{Aghamousa:2016zmz}.
The resulting magnification bias
does not depend on the filter.

The luminosity function of the LRG sample has been measured from data in Ref.~\cite{Cool:2008nv} at $z = 0.15, 0.25, 0.35, 0.8$.
We approximate the Schechter parameters by fitting the estimated luminosity function after passive evolution correction (Figure 9 and Table 1 in Ref.~\cite{Cool:2008nv}).
Note that the $K$-correction has already been applied.
\begin{table}
\caption{
Best fit parameters to the luminosity function for luminous red galaxies used to determine the magnification bias for DESI.
}
\begin{center}
\begin{tabular}{|c c c c| }
 \hline
 $z$ & $\alpha$ & $M^*$ & $\phi^* [h^3\,\mathrm{Mpc}^{-3}]$\\
 \hline
 0.15   & -0.170 & -22.39 & $2.4 \times 10^{-3}$ \\
 0.25   & -1.497 & -22.75  & $3.9 \times 10^{-3}$ \\
 0.35   & -1.593 & -22.81 & $3.1 \times 10^{-3}$    \\
 0.8   &  -3.186 &  -23.49& $1.1 \times 10^{-3}$  \\
 \hline
\end{tabular}
\end{center}
\label{table:fit-lrg}
\end{table}

In Table~\ref{table:fit-lrg} we show
the values of the best-fit parameters. Since DESI will
detect luminous red galaxies up to $z = 1$,
we assume that the luminosity function for LRG
does not evolve significantly between $z = 0.8$ and $z = 1$.
DESI will detect LRG in the $r$, $z$ and $W1$ bands with different luminosity cuts~\cite{Aghamousa:2016zmz}.
Therefore, the magnification bias for the LRG catalogue will depend on the observed band.
In our forecast, we use the $r$-band magnitude cut $m_\text{lim} = 23$.

The luminosity function for the bright galaxies sample is modelled following
\cite{Blanton:2000ns}, i.e. a Schechter function with constant parameters $\alpha = -1.20$, $\phi^* = 1.46 \times 10^{-2} h^{3}\, \mathrm{Mpc}^{-3}$ and $M^* = -20.83$.
The $K$-correction has been estimated from~\cite{Blanton:2000ns}, by fitting the measured $K$-correction in their Figure 4 to a linear relation.
Assuming a typical galaxy color $(g^*-r^*) = 0.6$, we find $K(z) \sim 0.87 \,z$.
The magnitude limit for this sample is $m_\text{lim} = 19.5$~\cite{Aghamousa:2016zmz}.

\begin{figure}
\centering
\includegraphics[width=0.7\textwidth]{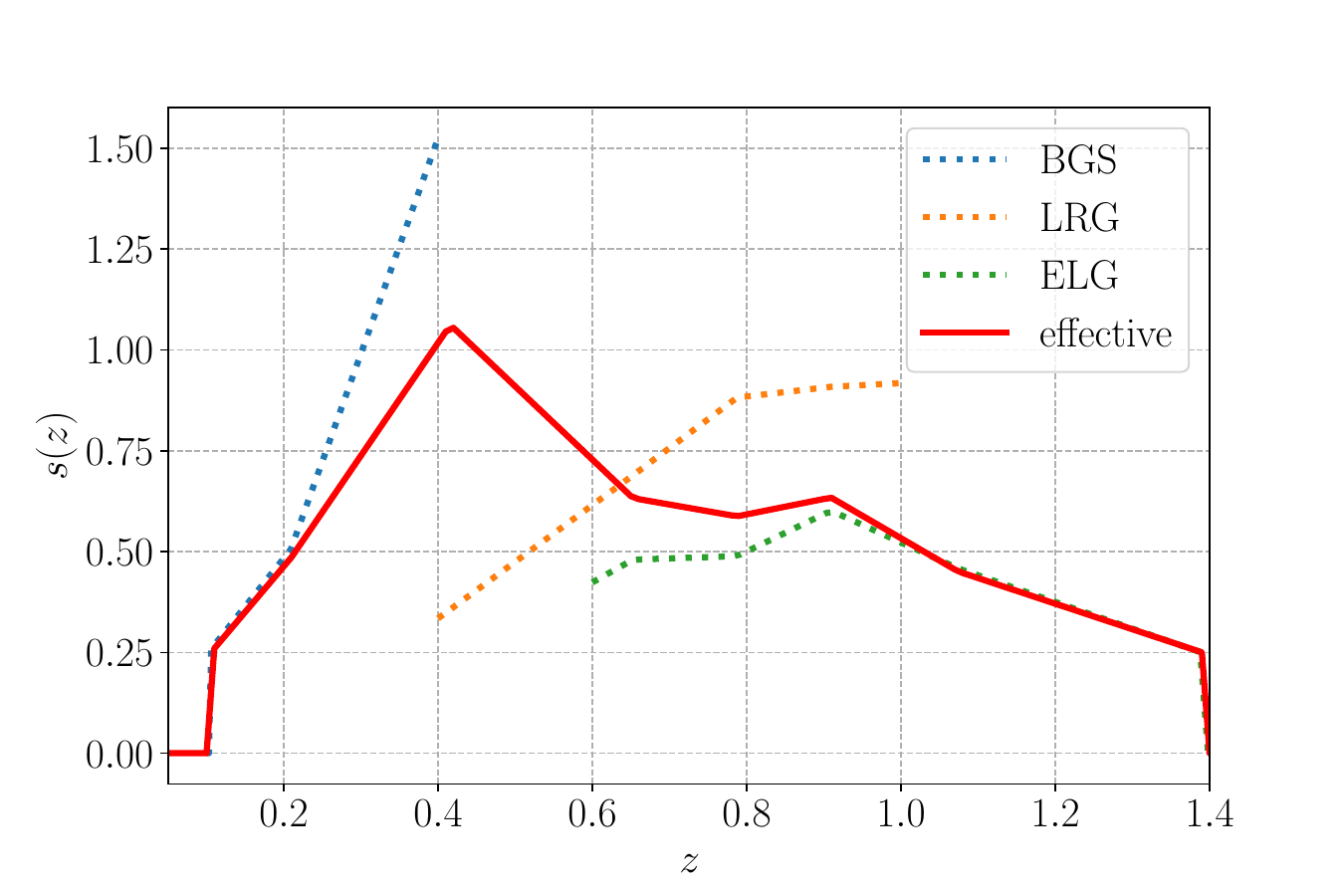}
\caption{
Magnification bias for the three galaxy populations that will be detected with DESI: emission line galaxies (ELG), luminous red galaxies (LRG), bright galaxy sample (BGS), as well as the weighted (effective) magnification bias.
}
\label{fig:sz-galaxysamples}
\end{figure}

In Fig.~\ref{fig:sz-galaxysamples} we show the magnification bias for the three galaxy populations targeted by DESI. While for the LRG and BGS the magnification bias is a monotonic function of redshift,
the ELG sample peaks at $z \simeq 0.9$ and decreases at larger redshift.

As there is significant overlap in the redshift range of the different types of galaxies in the survey, their Fisher matrices cannot simply be added together because they are not independent measurements.
One way of combining the data would be to compute the cross-correlation between all the different galaxy types in an overlapping bin, including both even and odd multipoles for consistency.
The total Fisher matrix would then include the covariance between the different populations.
Instead, we opt for a simpler strategy, by computing an effective galaxy and magnification bias for the entire sample in the survey.
More precisely, the effective galaxy and magnification bias in a given redshift bin centered at $\bar z$ and with width $\Delta z$ are computed as a weighted sum of the different types of galaxies:
\bea
s_\mathrm{eff}(\bar z) &= \frac{\sum\limits_{i = \mathrm{BGS, LRG, ELG}} N_i(\bar z, \Delta z) s_i(\bar z)}{\sum\limits_{i = \mathrm{BGS, LRG, ELG}} N_i(\bar z, \Delta z)}\, ,
\\
b_\mathrm{eff}(\bar z) &= \frac{\sum\limits_{i = \mathrm{BGS, LRG, ELG}} N_i(\bar z, \Delta z) b_i(\bar z)}{\sum\limits_{i = \mathrm{BGS, LRG, ELG}} N_i(\bar z, \Delta z)}\, , \label{e:DESIbias}
\eea
where $N_i(\bar z, \Delta z)$ denotes the number of galaxies of type $i$ in that redshift bin detectable by the survey.
It is clear that within a bin where only one type of galaxy has a nonzero number density, the effective bias reduces to the true bias of that particular type of galaxy. As for LSST, in our forecasts we multiply Eq.~\eqref{e:DESIbias} by a parameter $b_0$ with fiducial value $b_0=1$ that we let vary.

We explore two redshift binning configurations, with 5 and 8 bins respectively, which are detailed in Tables~\ref{table:desi_5bin} and~\ref{table:desi_8bin}. These bins have been chosen such that they contain a similar number of galaxies.

\begin{table}[H]
\caption{DESI, 5 bin configuration}
\begin{center}
\begin{tabular}{|clllll|}
\hline
$\bar{z}_i$  & 0.13 & 0.42 & 0.72 & 0.93 & 1.32 \\
\hline
$\Delta z_i$ & 0.16 & 0.4  & 0.2  & 0.22 & 0.56 \\
\hline
\end{tabular}
\end{center}
\label{table:desi_5bin}
\end{table}

\begin{table}[H]
\caption{DESI, 8 bin configuration}
\begin{center}
\begin{tabular}{|cllllllll|}
\hline
$\bar{z}_i$  & 0.1  & 0.21 & 0.42 & 0.65 & 0.79 & 0.91 & 1.07 & 1.39 \\
\hline
$\Delta z_i$ & 0.1 & 0.1 & 0.32 & 0.16 & 0.1 & 0.14 & 0.2  & 0.42 \\
\hline
\end{tabular}
\end{center}
\label{table:desi_8bin}
\end{table}

\subsection{SKA2}\label{a:SKAII}

For the SKA2 survey, we consider two redshift binning strategies, with 8 and 11 redshift bins respectively, which are summarized in Tables~\ref{table:ska2_8bin} and~\ref{table:ska2_11bin}, where $\Delta z_i$ again denotes the width of a redshift bin $i$ centered at a mean redshift $\bar z_i$.
The first $N - 4$ redshift bins were constructed in such a way that the total number of galaxies per bin is equal in all of them, while the last 4 bins are equal sized in redshift space to avoid having bins which are too wide, for which we would need to use a more general estimator, for instance the $\Xi_\ell$ estimator, described in~\cite{Tansella:2018sld}.

\begin{table}[H]
\caption{SKA2, 8 bin configuration}
\begin{center}
\begin{tabular}{|cllllllll|}
\hline
$\bar{z}_i$  & 0.22 & 0.35 & 0.48 & 0.66 & 0.92 & 1.23 & 1.54 & 1.85 \\ \hline
$\Delta z_i$ & 0.14 & 0.12 & 0.14 & 0.22 & 0.3 & 0.3 & 0.3 & 0.3 \\ \hline
\end{tabular}
\end{center}
\label{table:ska2_8bin}
\end{table}

\begin{table}[H]
\caption{SKA2, 11 bin configuration}
\begin{center}
\begin{tabular}{|clllllllllll|}
\hline
$\bar{z}_i$  & 0.2  & 0.28 & 0.36 & 0.44 & 0.53 & 0.64 & 0.79 & 1.03 & 1.31 & 1.58 & 1.86 \\ \hline
$\Delta z_i$ & 0.1 & 0.08 & 0.08 & 0.08 & 0.1 & 0.12 & 0.2  & 0.28 & 0.28 & 0.28 & 0.28 \\ \hline
\end{tabular}
\end{center}
\label{table:ska2_11bin}
\end{table}

\begin{figure}
\centering
\begin{minipage}[b]{0.495\linewidth}
\centering
\includegraphics[width=\linewidth]{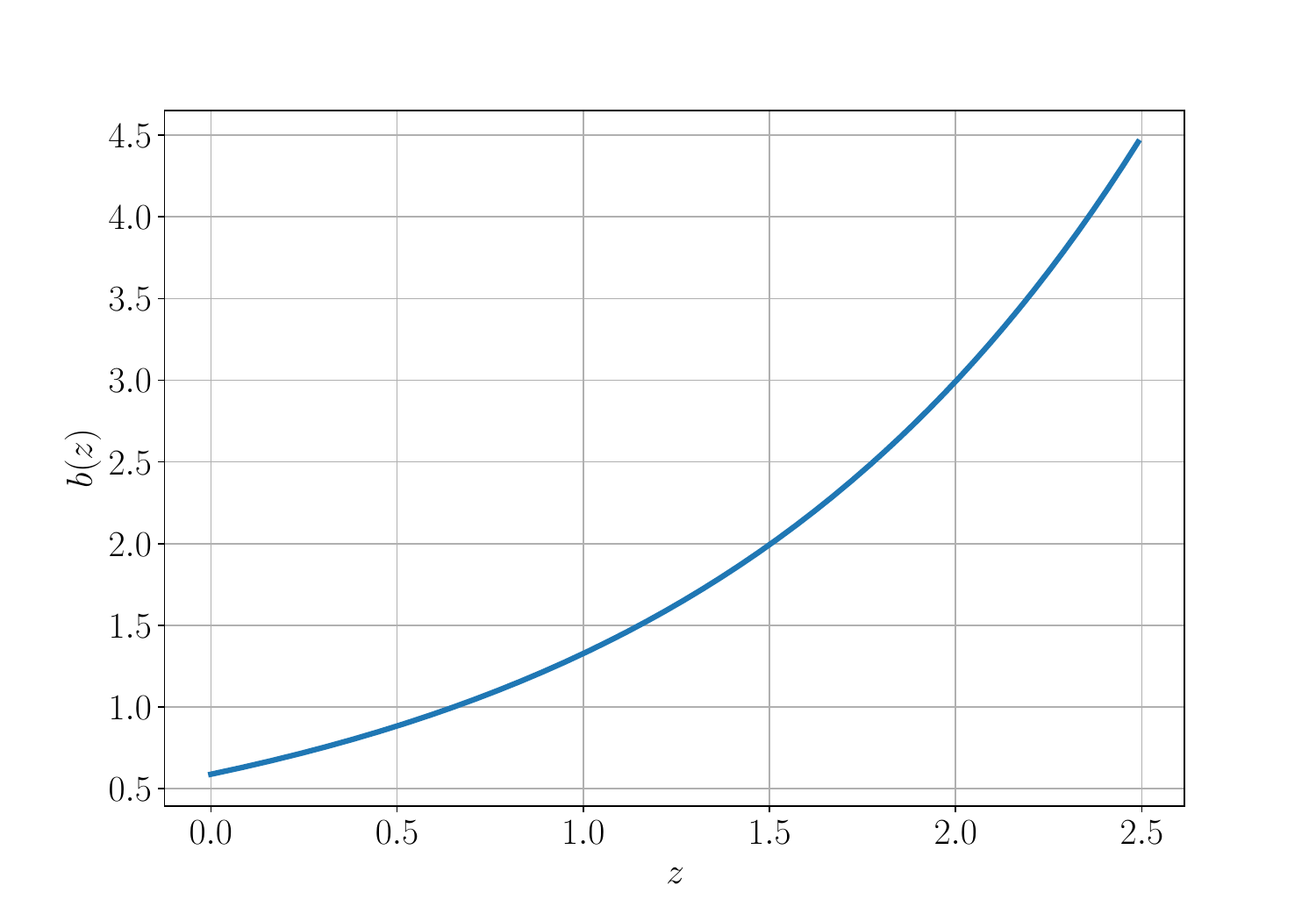}
\end{minipage}
\begin{minipage}[b]{0.495\linewidth}
\centering
\includegraphics[width=\linewidth]{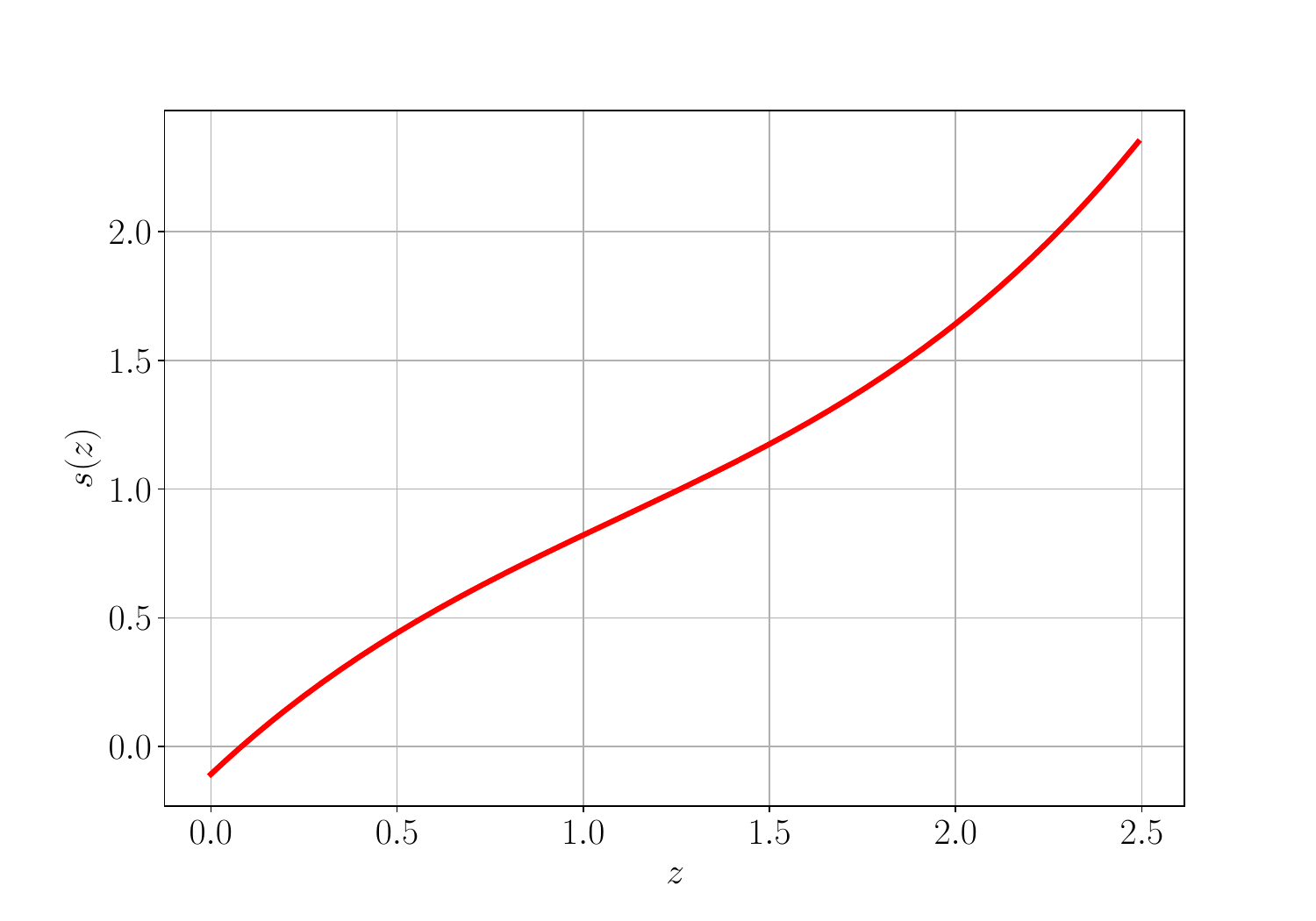}
\end{minipage}
\caption{
The galaxy bias (left panel) and magnification bias (right panel) for SKA2 used in the analysis.
}
\label{fig:ska2_biases}
\end{figure}

\begin{figure}
\centering
\includegraphics[width=0.65\linewidth]{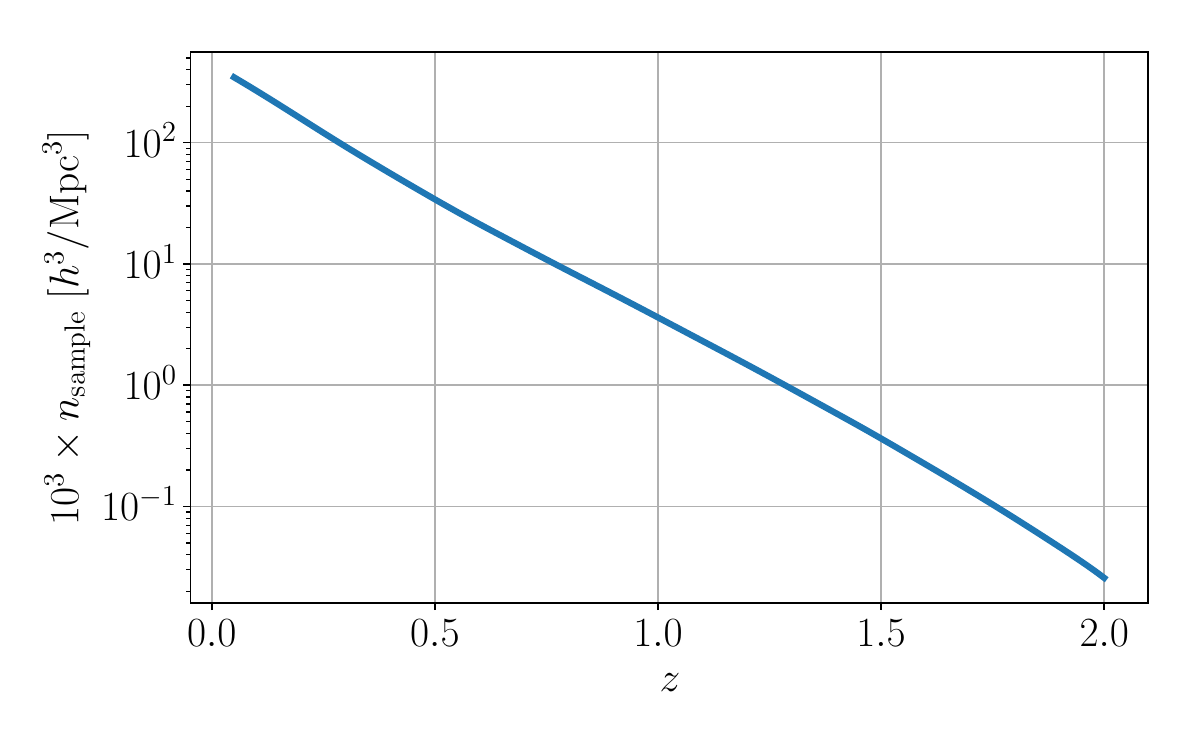}
\caption{
The number density of galaxies for SKA2.
}
\label{fig:ska2_number_density}
\end{figure}

The other specifications for the survey follow~\cite{Villa_2018} and~\cite{Bull_2016}, which we repeat here for completeness:

\begin{align}
f_\mathrm{sky} &= 0.73\, ,
\\
b(z) &= C_1 \exp{(C_2 z)}\, ,\label{e:biasSKA}
\\
s(z) &= s_0 + s_1 z + s_2 z^2 + s_3 z^3\, ,\label{e:magSKA2}
\end{align}
with $s_0 = -0.106875$, $s_1 = 1.35999$, $s_2 = -0.620008$, and $s_3 = 0.188594$, as well as $C_1 = 0.5887$ and $C_2 = 0.8130$.
The galaxy and magnification bias for SKA2 are shown in Fig.~\ref{fig:ska2_biases}, while the number density is shown in Fig.~\ref{fig:ska2_number_density}.
The total number of galaxies observed is predicted to be $N_\mathrm{tot} \sim 9 \times 10^8$.

\section{Fisher matrix analysis for a photometric survey}\label{ap:photo-alone}

In this Appendix we present a Fisher matrix analysis for
the LSST galaxy survey.
We want to address the following questions:
\begin{enumerate}
\item
How accurately will LSST estimate cosmological parameters?
\item
What is the impact of neglecting lensing magnification on the estimation of cosmological parameters and their errors?
\item
Will the survey be able to measure the lensing potential with high significance?
\end{enumerate}

The specifics assumed for the LSST galaxy survey have been described in Appendix~\ref{a:LSST}.
We consider 3 binning configurations in the redshift range $z \in [0, 2.5]$:
\begin{itemize}
    \item 5 redshift bins, equally spaced with width $\Delta z = 0.5$.
    \item 8 redshift bins, equally spaced with width $\Delta z =0.3125$.
    \item 12 redshift bins, with width
    $\Delta z = 0.1 (1+z)$. This configuration is the optimal binning for LSST, i.e.\ the half-width of the bins is the expected photometric redshift uncertainty for this galaxy sample.
\end{itemize}

\begin{figure}
\centering
\includegraphics[width=0.6\linewidth]{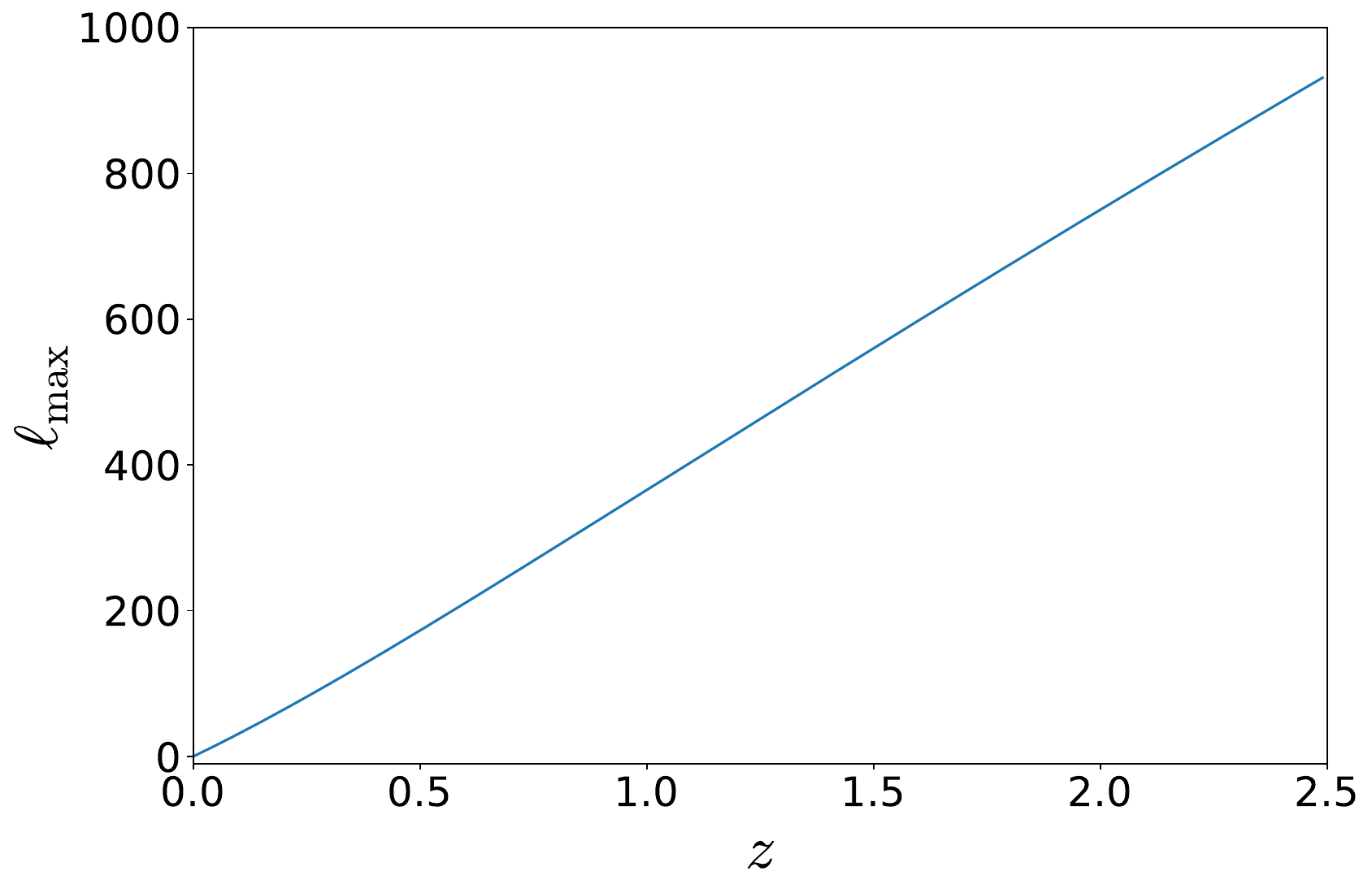}
\caption{The maximal multipole, $\ell_{\max}$ for which linear perturbation theory is applicable.
}
\label{fig:lsst-lmax}
\end{figure}

The angular power spectra for the estimation of the derivatives and the covariance have been computed using the Cosmic Linear Anisotropy Solving System ({\sc class}) code~\cite{class1, class2, CLASSgal}.
Since we are modelling a photometric survey, we used a Gaussian window function to model the redshift binning.
The non-linearity scale $d_\mathrm{NL}(z)$ of Eq.~\eqref{e:rNL} translates into a redshift-dependent maximal multipole, $\ell_{\max}(z)$ given by
\be
\ell_{\max}(z) =\frac{\pi}{\theta_\mathrm{NL}(z)}=\frac{\pi r(z)}{d_\mathrm{NL}(z)} \,,
\ee
where $r(z)$ is the comoving distance to redshift $z$. In Fig.~\ref{fig:lsst-lmax} we plot $\ell_{\max}(z)$ as a function of redshift, which represents the maximum angular power spectrum that we are using in our analysis.

{We adopted the Fisher matrix approach described, 
for example, in Ref. \cite{Villa_2018}, where the lensing contribution is always included in the covariance.
}

\subsection{Impact of lensing on parameter constraints and best-fit estimation.}

\label{s:LSST_constraints}

We consider a $\Lambda$CDM cosmology + a bias parameter $b_0$ with fiducial value $b_0 = 1$,
i.e.\ the galaxy bias in each redshift bin is $b_\text{gal}(z) = b_0 \times b_\text{LSST}(z)$.

\begin{table}
\caption{$1\sigma$ relative error (in percent) estimated including lensing in our theoretical model (odd rows) and ratio between $1\sigma$ error estimated including and neglecting lensing in our model (even rows) for three different binning configurations. We include in the analysis the standard $\Lambda$CDM parameters and  galaxy bias parameter.}
\begin{center}
\begin{tabular}{| c |  c c c c c c |}
 \hline
 $N_\text{bins}$ & $h$ & $\Omega_\text{baryon}$ & $\Omega_{\rm cdm}$ & $\ln{10^{10}\, A_s}$ & $n_s$ & $b_0$ \\
 \hline
5 bins - $1\sigma_\text{lens}\,[\%]$   &  $6.2\, \%$  & $8.5\, \%$ & $1.3\, \%$ & $3.6\, \%$  & $3.0\, \%$ &  $2.4\, \%$ \\
5 bins - $\sigma_\text{lens}/\sigma_\text{no-lens} $  & 0.996   & 0.994 & 1.004 & 0.275 & 0.991 & 0.119 \\
\hline
8 bins - $1\sigma_\text{lens}\,[\%]$   & $3.0\, \%$  & $3.4\, \%$ & $0.5\, \%$ & $2.3\, \%$  & $1.7\, \%$ & $2.4\, \%$ \\
8 bins - $\sigma_\text{lens}/\sigma_\text{no-lens} $   & 0.993   & 0.998 & 1.011 & 0.351 & 0.999 & 0.240 \\
\hline
12 bins - $1\sigma_\text{lens}\,[\%]$
 & $0.7\, \%$  & $2.4\, \%$ & $0.2\, \%$ & $1.6\, \%$  & $0.5\, \%$ & $2.4\, \%$ \\
12 bins - $\sigma_\text{lens}/\sigma_\text{no-lens} $  & 1.000   & 1.000 & 0.998 & 0.325 & 0.998 & 0.321 \\
 \hline
\end{tabular}
\end{center}
\label{table:lsst-error-st}
\end{table}

We estimate the Fisher matrix and the constraints on cosmological parameters for two theoretical models:
\begin{enumerate}[label=\alph*)]
\item
a model that consistently includes lensing in the galaxy number counts
\item
a model that neglects lensing in the galaxy number counts
\end{enumerate}

In Table~\ref{table:lsst-error-st} we
show the $1\sigma$ uncertainty on cosmological parameters when lensing is included in the theoretical model (rows 1, 3 and 5 for the 5, 8 and 12 bin configurations respectively) and the ratio between the constraints estimated when
including or neglecting lensing (rows 2, 4 and 6).

The errors on the best-fit parameters decrease significantly as the number of bins increase. In fact,
for large redshift bins, both the density and the RSD contribution to the angular power spectra are highly suppressed. Therefore, we expect an optimal analysis to adopt the largest number of bins allowed by the redshift resolution. In the optimal 12 bin configuration, we expect LSST to measure standard cosmological parameters at the percent/sub-percent level.
In {all configurations, including lensing significantly improves the constraints on
$A_s$ and on the galaxy bias $b_0$.  The errors are largest for the 5 bin configurations and significantly smaller for the 12 bin one.}
This is due to the fact that for the 5 and 8 bins configuration the RSD contribution to the angular power spectra is highly suppressed and, therefore, the amplitude of the primordial power spectrum is strongly degenerate with the galaxy bias.
Indeed, the density contribution is only sensitive to $b^2_0A_s$, but not to each of the parameters separately.
Including lensing in the analysis helps breaking this degeneracy, since the lensing-lensing correlation is sensitive to $A_s$, whereas the density-lensing correlation is sensitive to $b_0A_s$. This is clearly visible from Table~\ref{table:lsst-error-st}, where we see that the errors on $b_0$ increase by a factor of 8 (5 bins), 4 (8 bins) and 3 (12 bins) when neglecting lensing in the analysis. The fact that the improvement decreases when the number of bins increases is due to the fact that RSD contributes more for thin redshift bins and that it helps breaking the degeneracy between $A_s$ and $b_0$. The impact of lensing magnification on the parameter constraints is therefore smaller in this case.
See also~\cite{Cardona:2016qxn} for a discussion of these points.

\begin{figure}
\centering
\includegraphics[width=\linewidth]{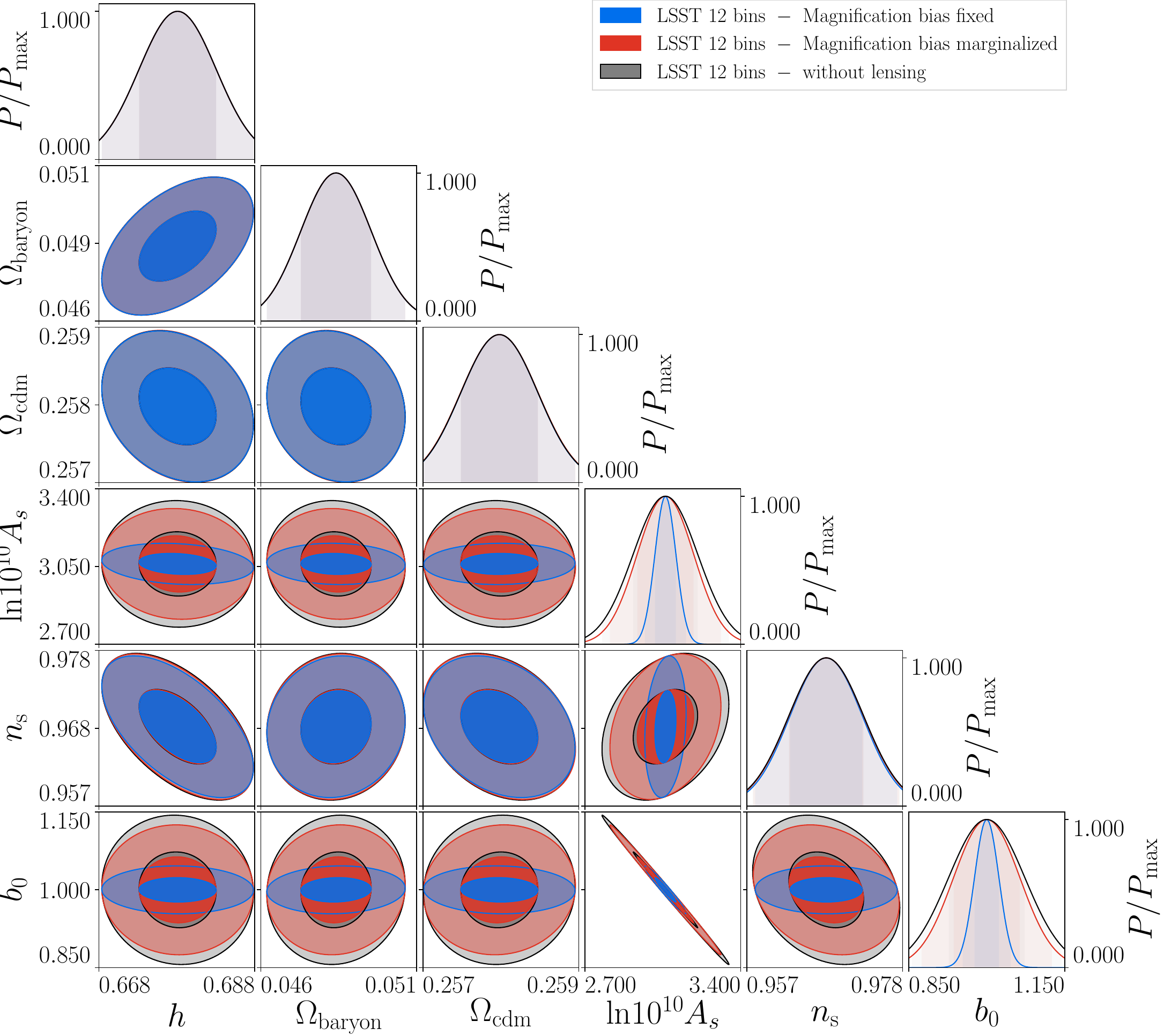}
\caption{Constraints on parameters for LSST with 12 redshift bins where: both the signal and the model have no lensing (black contours), lensing is included in the signal and in the model and the magnification bias parameter $s(z)$ is fixed (blue contours); and lensing is included in the signal and in the model and the magnification bias parameter $s(z)$ is modelled by four free parameters that are marginalized over (red contours).
}
\label{fig:lsst-fisher-sbias-margin}
\end{figure}

We study how this improvement changes if instead of fixing $s(z)$, we model it with three parameters, as proposed in Eq.~\eqref{e:fit_s_LSST}, and we let these parameters vary in our Fisher analysis. In Fig.~\ref{fig:lsst-fisher-sbias-margin}, we compare the constraints on parameters in this case. To ease the comparison of the error bars, we have removed the shift in the case "without lensing", i.e.\ we assume there that there is no lensing in the signal and no lensing in the modelling.
\begin{table}
\caption{
Shift in the best-fit parameters that we expect if lensing is neglected in the theoretical model for LSST, in units of $1\sigma$.
In the analysis we include the standard $\Lambda$CDM parameters and a galaxy bias parameter $b_0$.
}
\begin{center}
\begin{tabular}{| c |  c c c c c c |}
 \hline
 $N_\text{bins}$ & $h$ & $\Omega_\text{baryon}$ & $\Omega_{\rm cdm}$ & $\ln{10^{10}\, A_s}$ & $n_s$ & $b_0$ \\
 \hline
5 bins   &  $-0.18$  & $-0.16$ & $1.68$ & $-0.17$  & $0.17$ &  $0.13$ \\
8 bins   &  $0.01$  & $-0.22$ & $1.56$ & $0.66$  & $0.79$ &  $-0.69$ \\
12 bins   & $0.49$  & $0.24$ & $0.077$ & $1.19$  & $1.31$ &  $-1.28$ \\
 \hline
\end{tabular}
\end{center}
\label{table:lsst-shift}
\end{table}

In Table~\ref{table:lsst-shift}
we report the shifts of the best-fit parameters (in units of $1\sigma$), due to neglecting lensing in the theoretical model.
The parameters most affected and the size of the shift depend strongly on the number of redshift bins.
For 5 bins, $\Omega_{\rm cdm}$ experiences the largest shift, whereas for 12 bins it is $n_s$ which is more shifted.
For all configurations, we see that a positive shift in $b_0$ requires a negative shift in $A_s$ and vice versa.
This is due to the fact that the amplitude of the density term is given by $b^2_0A_s$.

For the optimal 12-bins analysis, $A_s$, $n_s$ and $b_0$ are all significantly shifted.
Note, however, that shifts of order $1\si$ and larger cannot be trusted since our analysis gives the first term of a series expansion in $\De\theta_i/\si_i$ which is only reliable if $|\De\theta_i/\si_i|\ll 1$.
Nevertheless, our findings show that neglecting lensing in LSST is not a valid option if we want to reach reliable \% or sub-\% accurate cosmological parameters.

\subsection{Detection of the lensing potential with LSST}

\label{s:LSST_AL}

\begin{figure}
\centering
\includegraphics[width=\linewidth]{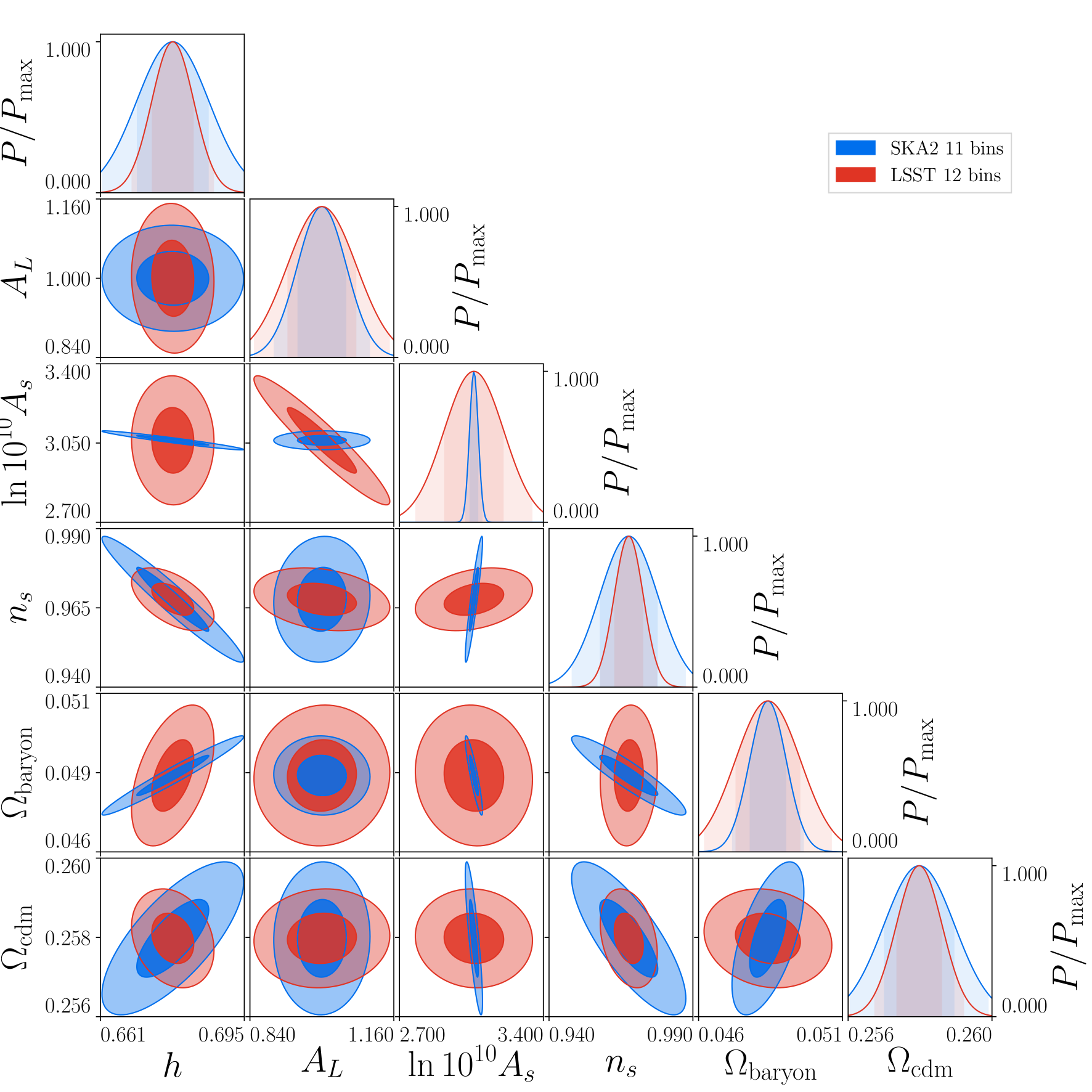}
\caption{Contour plot for the Fisher analysis (SKA vs.\ LSST) which includes an extra parameter for the amplitude of the lensing potential.}
\label{fig:lsst-fisher-12bins-AL}
\end{figure}

We extend the standard $\Lambda$CDM model adopted in the previous section, adding an extra parameter: the amplitude of the lensing potential $A_\text{L}$, which multiplies the lensing term and whose fiducial value in General Relativity is $A_\text{L} = 1$.
The results are shown in Table~\ref{table:lsst_errors} and Fig.~\ref{fig:lsst-fisher-12bins-AL}.
We see that increasing the number of redshift bins strongly decreases the error on $A_{\rm L}$.
For the optimal, 12 bins configuration, $A_{\rm L}$ can be measured with an accuracy of $7.7$\%.
Note that this result assumes that we know $s(z)$ perfectly well.

\begin{table}[h!]
\caption{
$1\sigma$ relative error (in percent) for standard $\Lambda$CDM parameters + galaxy bias + amplitude of the lensing potential from LSST.
We compare different binning configurations.
}
\begin{center}
\begin{tabular}{| c | c c c c c c c|}
 \hline
 $N_\text{bins}$ & $h$ & $\Omega_\text{baryon}$ & $\Omega_\text{cdm}$ & $\ln{10^{10}\, A_s}$ & $n_s$ & $b_0$ & $A_\mathrm{L}$ \\
 \hline
5 bins   &  $6.2\, \%$  & $8.5\, \%$ & $1.3\, \%$ & $13.1\, \%$  & $3.0\, \%$ &  $20.1\, \%$ & $20.2\, \%$ \\
8 bins   & $3.0\, \%$  & $3.4\, \%$ & $0.5\, \%$ & $6.5\, \%$  & $1.7\, \%$ & $9.9\, \%$  & $10.2\, \%$ \\
12 bins   & $0.7\, \%$  & $2.4\, \%$ & $0.2\, \%$ & $4.7\, \%$  & $0.5\, \%$ & $7.3\, \%$  & $7.7\, \%$\\
 \hline
\end{tabular}
\end{center}
\label{table:lsst_errors}
\end{table}

Comparing Table~\ref{table:lsst_errors} with Table~\ref{table:lsst-error-st}, we see that adding $A_{\rm L}$ as a free parameters degrades significantly the constraints on $A_s$ and $b_0$ for all configurations. For the 12-bins configuration, the degradation is however less severe. This can be understood by the fact that for a small number of bins adding $A_{\rm L}$ worsen the degeneracy between $A_s$ and $b_0$, since in the lensing contribution $A_{\rm L}$ is degenerated with $A_s$. For 12 bins, RSD partially break the degeneracy between $A_s$ and $b_0$. This in turns helps breaking the degeneracy between $A_{\rm L}$ and $A_s$ in the lensing contribution.

\begin{table}[h!]
\caption{
Constraints on $\Lambda$CDM parameters + lensing amplitude (in percent), with marginalization over the galaxy bias (see Fig.~\ref{fig:lsst-fisher-12bins-AL})
}
\begin{center}
\begin{tabular}{| c | c c c c c c|}
\hline
parameter &$h$&$A_\mathrm{L}$&$\ln 10^{10} A_s$&$n_s$&$\Omega_\mathrm{baryon}$&$\Omega_\mathrm{cdm}$\\
\hline
SKA2 11 bins &1.259&5.462&0.699&1.046&1.317&0.372\\
LSST 12 bins &0.733&7.708&4.748&0.522&2.341&0.241\\
\hline
\end{tabular}
\end{center}
\label{table:ska2_lsst_errors}
\end{table}

{As discussed in the main text, the multipoles of the correlation function can also be used to measure the amplitude of the lensing potential, $A_{\rm L}$. From Table~\ref{table:ska2_lsst_errors}, we see that $A_{\rm L}$ is better measured with SKA2 than with LSST, even though the lensing contribution is more important in LSST. This is due to the fact that RSD in SKA2 help breaking the degeneracy between $A_{\rm L}$, $A_s$ and $b_0$.}

\section{Combining angular power spectra and correlation function}
\label{s:cross}

{As described in the main text, we also perform an analysis where we combine the multipoles of the correlation function from SKA2 in each redshift bin, with the cross-correlation of the $C_\ell$'s between {\it different} redshift bins. We include lensing in the $C_\ell$'s but not in the correlation function. The main results are presented and discussed in detail in Section~\ref{s:main_result}. 

\begin{figure}
\centering
\includegraphics[width=\linewidth]{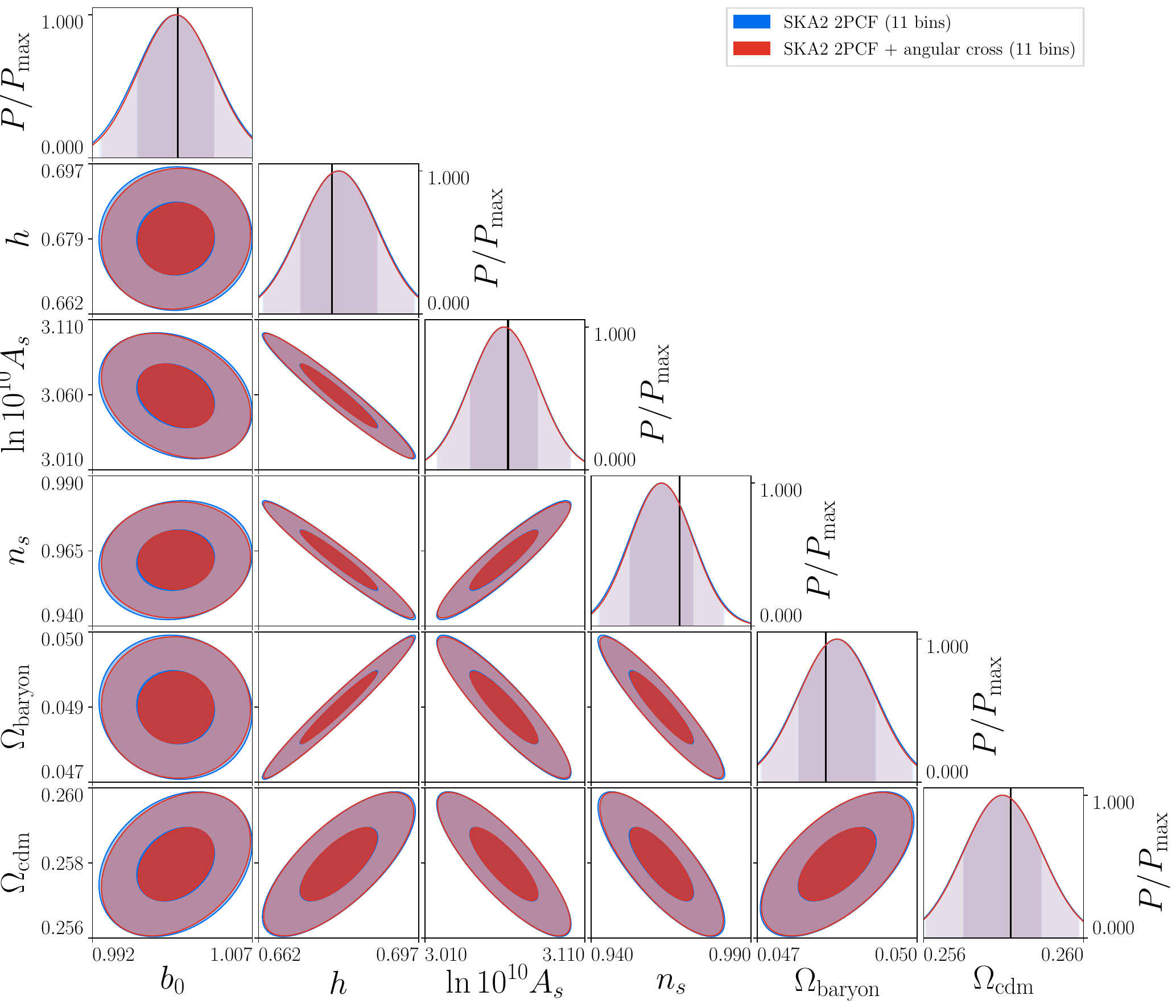}
\caption{
The constraints on cosmological parameters from SKA2 using the correlation function (blue) and SKA2 from the correlation function + $C_\ell$'s for the  cross-correlations (red) using a configuration with 11 redshift bins as described in~\ref{a:SKAII}.
The shifts of the contours are computed neglecting lensing from the correlation function, and consistently including it in the angular power spectrum.
The black lines on the diagonal plots denote the fiducial values.
}
\label{fig:ska2_with_cross}
\end{figure}
In Fig.~\ref{fig:ska2_with_cross} we compare the constraints and shifts when we consider the correlation function of SKA2 alone, or when we combine the correlation function of SKA2 with the cross-correlation $C_\ell$'s for different redshift bins. We see that adding the $C_\ell$'s has very little impact on the constraints or the shifts.}

\begin{figure}
\centering
\includegraphics[width=\linewidth]{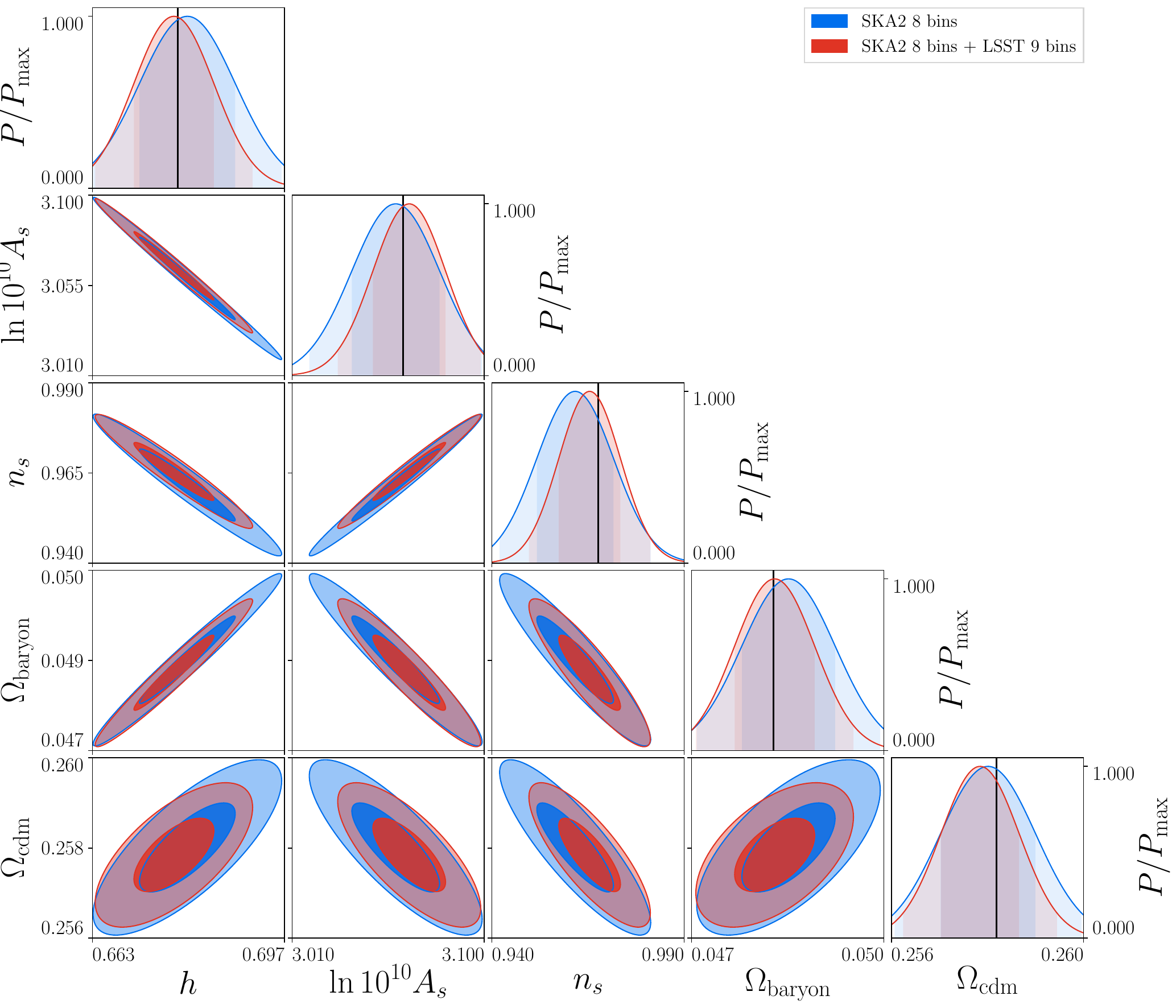}
\caption{
Contour plot for the Fisher analysis (SKA2 8 bins vs. SKA2 8 bins + LSST 9 bins) for $\Lambda$CDM parameters.
The black solid lines denote fiducial values.
{The difference of this w.r.t. Fig~\ref{fig:ska2_with_lsst} is that here the galaxy bias has been fixed and not marginalized over. As one can compare, the constraints obtained are virtually identical.}
}
\label{fig:ska-lsst-fisher-89}
\end{figure}

{In Fig.~\ref{fig:ska-lsst-fisher-89} and Table~\ref{fig:ska-lsst-fisher-89}, we compare the constraints from SKA2 alone, with the combined constraints from SKA2 and the $C_\ell$'s in LSST. Contrary to the results presented in the main text, we fix here the value of the bias $b_0=1$. Comparing with Table~\ref{table:comparison_surveys_constraints}, we see that fixing the bias has almost no impact on the constraints or on the shifts. This is due to the fact that in SKA2, RSD are strong enough to break the degeneracy between $A_s$ and $b_0$, and both are very well measured in these surveys. }

\begin{table}[h!]
\caption{
Constraints (in percent) and shift, for SKA2 and SKA2 + LSST, for the standard $\Lambda$CDM parameters, {when the biases are fixed to their fiducial value $b_0=1$} (see also Fig.~\ref{fig:ska-lsst-fisher-89})
}
\begin{center}
\begin{tabular}{| c | c c c c c|}
\hline
parameter &$h$&$\ln 10^{10} A_s$&$n_s$&$\Omega_\mathrm{baryon}$&$\Omega_\mathrm{cdm}$\\
\hline
SKA2 8 bins &1.255&0.675&1.036&1.309&0.356\\
$\Delta / \sigma$ & 0.199&-0.169&-0.602&0.316&-0.181\\
\hline
SKA2 8 bins + LSST 9 bins &1.048&0.558&0.833&1.124&0.293\\
$\Delta / \sigma$ & -0.096&0.165&-0.282&0.023&-0.423\\
\hline
\end{tabular}
\end{center}
\label{table:ska2_lsst_errors_bias}
\end{table}

In this analysis we add the two Fisher matrices which is equivalent to neglecting correlations between the spectroscopic number counts inside a redshift bin done with SKA and the photometric cross correlations of different redshift bins done with LSST. While the two surveys see mostly different galaxies such that the shot noise can be neglected in the covariance, they trace the same underlying density field which does lead to correlations. To assess their importance, we compare them with the variance of the spectroscopic and photometric surveys. The estimator for the correlation function, measured in a spectroscopic bin with mean redshift $\bar z$ is
\begin{align}
\xi_n(d,\bar z)=\alpha_n\sum_{ij\in\bar z}\Delta_{\rS}(\bx_i)\Delta_{\rS}(\bx_j)P_n(\cos\sigma_{ij})\delta_K(d_{ij}-d)\, , \end{align}
where the sum is over all pixel positions inside the spectroscopic redshift bin $\bar z$, the angle $\sigma_{ij}$ is the angle between the orientation of the pair of pixels $ij$ and the direction $\bn$, $\alpha_n$ is a normalisation factor, and the subscript S refers to galaxies in the spectroscopic survey. The estimator for the angular power spectrum between two photometric redshift bins $\bar z'$ and $\bar z''$ is given by
\begin{align}
C_\ell(\bar z',\bar z'')=\frac{\beta}{2\ell+1}\sum_{m=-\ell}^\ell \sum_{k\in \bar z'}
\sum_{q\in \bar z''}Y^\star_{\ell m}(\hat \bx_k)Y_{\ell m}(\hat \bx_q)\Delta_{\rP}(\bx_k)\Delta_{\rP}(\bx_q)\, ,
\end{align}
where $\beta$ is a normalisation factor and the subscript P refers to galaxies in the photometric survey.
The covariance between the spectroscopic and photometric survey can be written as
\begin{align}
{\rm cov}\Big[\xi_n(d,\bar z),C_\ell(\bar z',\bar z'') \Big]=&\frac{2\alpha_n\beta}{2\ell+1}\sum_{m=-\ell}^\ell\sum_{ij\in\bar z} \sum_{k\in \bar z'}
\sum_{q\in \bar z''}Y^\star_{\ell m}(\hat \bx_k)Y_{\ell m}(\hat \bx_q)P_n(\cos\sigma_{ij})\delta_K(d_{ij}-d)\nonumber\\
&\times \langle\Delta_{\rS}(\bx_i)\Delta_{\rP}(\bx_k)\rangle\langle\Delta_{\rS}(\bx_j)\Delta_{\rP}(\bx_q)\rangle\, .
\end{align}
We want to compare this with the variance of the spectroscopic survey, which is given by
\begin{align}
{\rm var}\big[\xi_n(d,\bar z) \big]=& 2\alpha_n^2 \sum_{ijkq\in\bar z}P_n(\cos\sigma_{ij})P_n(\cos\sigma_{kq})\delta_K(d_{ij}-d)\delta_K(d_{kq}-d)\\
&\times\langle\Delta_{\rS}(\bx_i)\Delta_{\rS}(\bx_k)\rangle\langle\Delta_{\rS}(\bx_j)\Delta_{\rS}(\bx_q)\rangle\, .\nonumber
\end{align}
We see that the covariance is related to the variance of the spectroscopic survey by the ratio
\begin{align}
{\rm ratio_S}=\frac{\langle\Delta_{\rS}(\bx_i)\Delta_{\rP}(\bx_k)\rangle\langle\Delta_{\rS}(\bx_j)\Delta_{\rP}(\bx_q)\rangle}{\langle\Delta_{\rS}(\bx_i)\Delta_{\rS}(\bx_k)\rangle\langle\Delta_{\rS}(\bx_j)\Delta_{\rS}(\bx_q)\rangle}\propto
\frac{C^{\rS\rP}_\ell(\bar z,\bar z')C^{\rS\rP}_\ell(\bar z,\bar z'')}{\left(C_\ell^{\rS}(\bar z)\right)^2}\, .
\label{eq:ratioS}
\end{align}
For the last proportionality sign we use the fact that the $C_\ell$'s contain the same information as the correlation function. We therefore can discuss these ratios in terms of the $C_\ell$'s.
Similarly, the covariance can be compared to the variance of the photometric survey, which reads
\begin{align}
{\rm var}\Big[C_\ell(\bar z',\bar z'') \Big]=&\frac{2\beta^2}{(2\ell+1)^2}\sum_{m=-\ell}^\ell\sum_{m'=-\ell}^\ell\sum_{ik\in\bar z'} \sum_{jq\in \bar z''}
Y^\star_{\ell m}(\hat \bx_i)Y_{\ell m}(\hat \bx_j)Y^\star_{\ell m'}(\hat \bx_k)Y_{\ell m'}(\hat \bx_q)\nonumber\\
&\times \langle\Delta_{\rP}(\bx_i)\Delta_{\rP}(\bx_k)\rangle\langle\Delta_{\rP}(\bx_j)\Delta_{\rP}(\bx_q)\rangle\, . \end{align}
The ratio between the covariance and the variance of the photometric survey is now given by
\begin{align}
{\rm ratio_P}=\frac{\langle\Delta_{\rS}(\bx_i)\Delta_{\rP}(\bx_k)\rangle\langle\Delta_{\rS}(\bx_j)\Delta_{\rP}(\bx_q)\rangle}{\langle\Delta_{\rP}(\bx_i)\Delta_{\rP}(\bx_k)\rangle\langle\Delta_{\rP}(\bx_j)\Delta_{\rP}(\bx_q)\rangle}\propto
\frac{C^{\rS\rP}_\ell(\bar z,\bar z')C^{\rS\rP}_\ell(\bar z,\bar z'')}{C_\ell^{\rP}(\bar z')C_\ell^{\rP}(\bar z'')}\, .
\label{eq:ratioP}
\end{align}

The ratios in Eqs.~\eqref{eq:ratioS} and~\eqref{eq:ratioP} are largest when $\bar z=\bar z'=\bar z''$. In this case they are mainly determined by two effects: first a geometric effect, due to the fact that spectroscopic and photometric surveys have different redshift resolutions, i.e. in a photometric survey  small-scales fluctuation are smeared over a wider radial bin, due to the poor redshift information.
\begin{figure}[ht]
\centering
\begin{subfigure}[b]{0.49\textwidth}
\includegraphics[width=\linewidth]{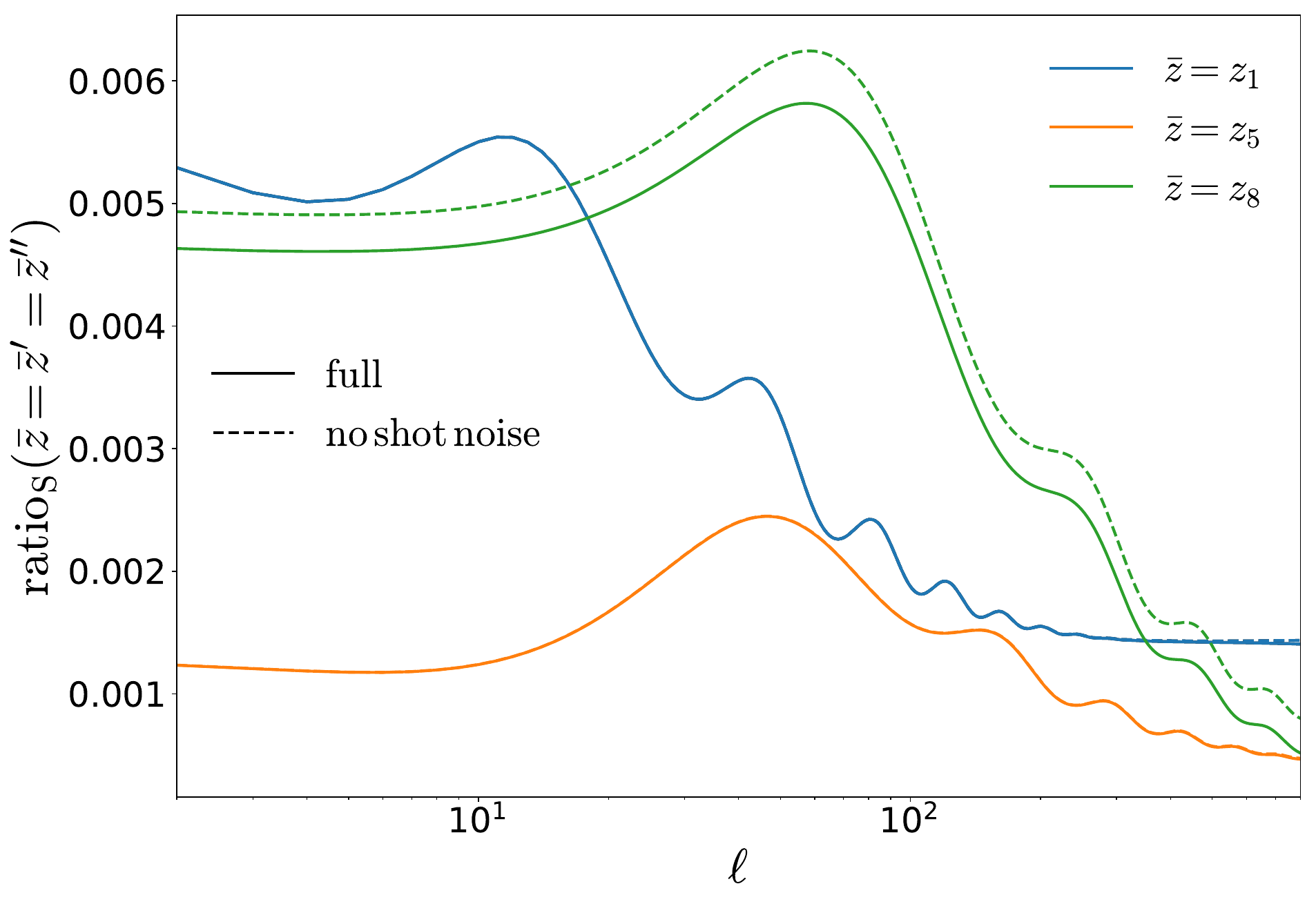}
\end{subfigure}
\begin{subfigure}[b]{0.49\textwidth}
\includegraphics[width=\textwidth]{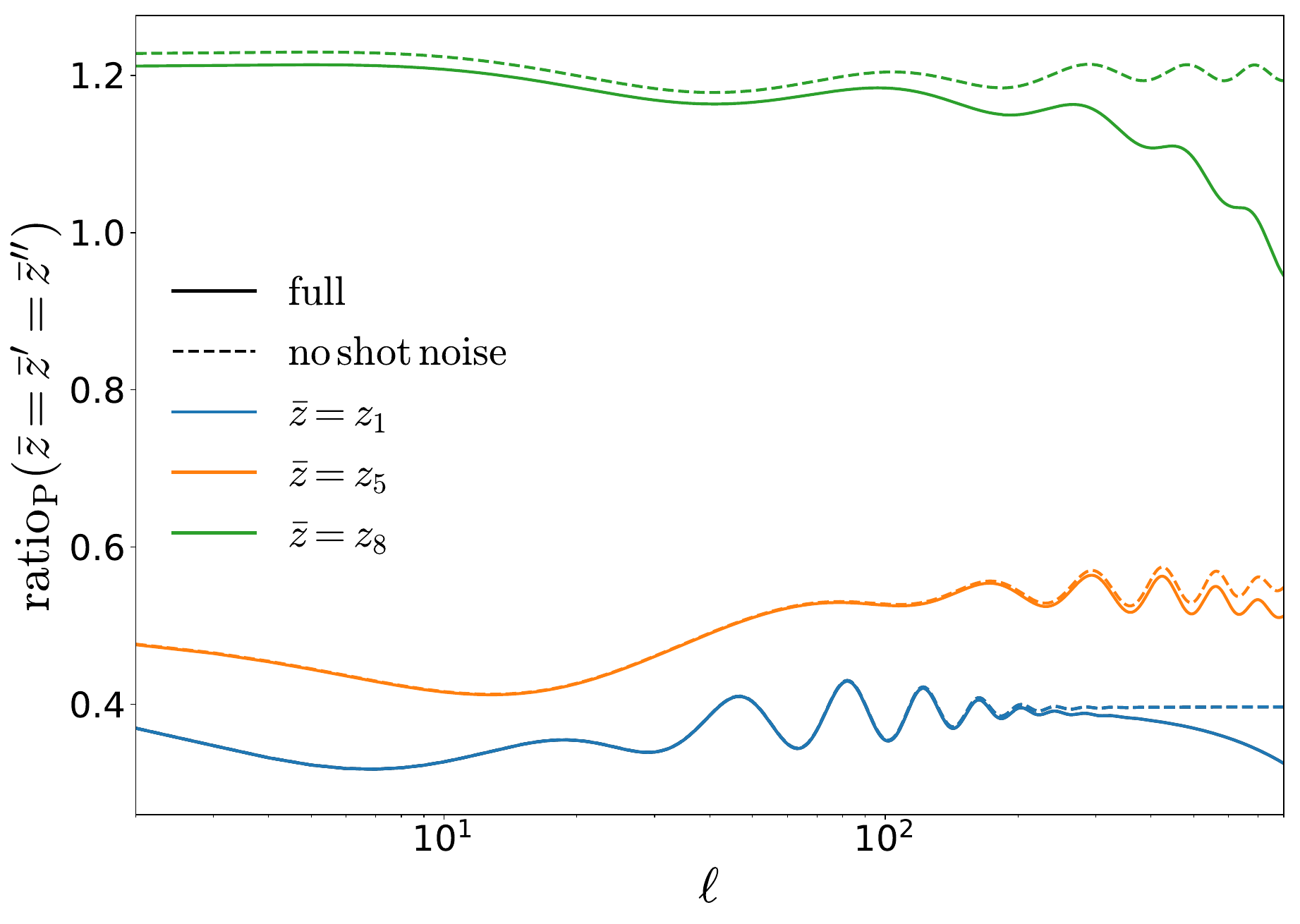}
\end{subfigure}
\caption{We show the ratios of the covariance with respect to the variance of the spectroscopic survey (left panel) and of the photometric survey (right panel), for the case $\bar z=\bar z'=\bar z''$.}
\label{fig:ratio}
\end{figure}
This geometric effect is plotted in Fig.~\ref{fig:ratio} for the cross-correlation of SKA with LSST (in dashed lines). We see that it always strongly suppresses ${\rm ratio_S}$ but it can enhance ${\rm ratio_P}$ for some values of $\bar z$. On top of this effect, we have a shot-noise suppression: $C^{\rS}_\ell$ and $C^{\rP}_\ell$ contain a shot noise contribution, whereas $C^{\rS\rP}_\ell$ is not affected by shot noise since it correlates galaxies from different surveys. This additional suppression is plotted in Fig.~\ref{fig:ratio} (continuous lines). Even though the geometric effect and the shot noise suppression strongly reduce ${\rm ratio_S}$, we cannot conclude that the covariance can be neglected, because ${\rm ratio_P}$ even becomes larger than 1 in some cases, see  Fig.~\ref{fig:ratio}, right panel. It is therefore in general not a good assumption to neglect the covariance between spectroscopic and photometric surveys, if all correlations are included, as is done in ~\cite{Blanchard:2019oqi}. In our case however, we take a much more conservative approach, and we remove from the photometric signal the angular power spectra with $\bar z'=\bar z''$, i.e. we only consider cross-correlations between different redshift bins for the photometric survey.
Therefore, we have an additional suppression due to the fact that the correlation between different redshift bins $\bar z\neq \bar z''$ (that enters in the covariance) is significantly smaller than the auto-correlation at redshift $\bar z$ (that contributes to the variance). This suppression is given by
\begin{align}
{\rm ratio_{S, cross}}=\frac{C^{\rS\rP}_\ell(\bar z)}{C_\ell^{\rS}(\bar z)}\times\frac{C^{\rS\rP}_\ell(\bar z,\bar z'')}{C_\ell^{\rS}(\bar z)}\, .\label{eq:ratioScross}
\end{align}
Here we consider the largest contributions to the covariance, coming from $\bar z=\bar z'$. The suppression becomes even stronger when $\bar z\neq \bar z'$.
\begin{figure}[ht]
\centering
\includegraphics[width=0.7\textwidth]{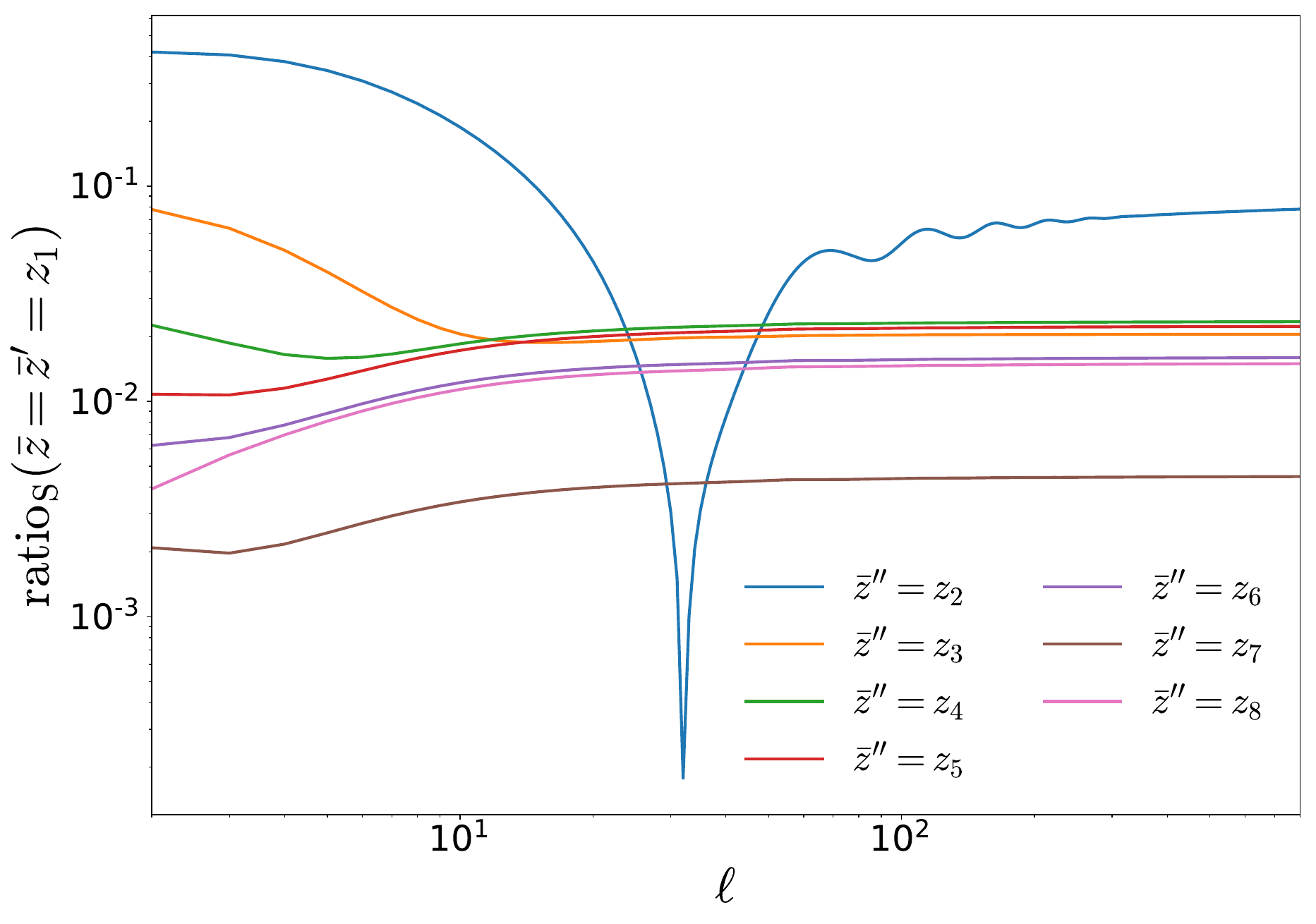}
\caption{We show the ratio of the covariance with respect to the variance of the spectroscopic survey, for the case $\bar z=\bar z'=z_1\neq\bar z''$.}
\label{ratioScross}
\end{figure}
In Fig.~\ref{ratioScross} we show the suppression due this effect. In order to disentangle this effect from the two effects considered in Fig.~\ref{fig:ratio}, the spectra have been computed using the same redshift dependent bin-size for the photometric and the spectroscopic number count. 
We find that these terms that we neglect in the covariance are suppressed by a factor 100 or more for most configurations. For $\ell<20$ in neighbouring bins the suppression can become less relevant, only a factor $2.5$ to $10$ (see the blue line in Fig.~\ref{fig:ratio}). Due to cosmic variance, these low $\ell$'s do not contribute significantly to parameter constraints. Most importantly, this cross-redshift suppression affects both ${\rm ratio_S}$ and ${\rm ratio_P}$ in the same way. Therefore, thanks to this suppression the covariance is always subdominant with respect to both the variance of the spectroscopic survey and the variance of the photometric survey. It is therefore a reasonable approximation to consider the two surveys as independent, provided that we remove the auto-correlation of the $C_\ell$ from the photometric sample.

\section{The lensing contribution to the correlation function}
\label{s:lensing_2pcf}
{
For completeness, here we explicitly write down the relevant lensing contributions to the 2 point correlation function, namely those given by eq.~\eqref{eq:lensing_contrib}.
The details of the calculation can be found in the literature, and here we follow the notation of~\cite{Tansella:2018sld}.
The lensing-lensing, density-lensing, and rsd-lensing contributions are respectively given by:

\begingroup
\allowdisplaybreaks
\begin{align*}
Z \big|_\text{len} = \qquad & \frac{9 \Omega_m^2}{4}\HH_0^4\frac{(2-5s_1)(2-5s_2)}{\chi_1\chi_2} \int\limits_0^{\chi_1} \!\integrand\la \int\limits_0^{\chi_2} \!\integrand\la' \frac{(\chi_1-\la)(\chi_2-\la')}{\la \la'} \frac{D_1(\la)D_1(\la')}{a(\la)a(\la')} \\
&\bigg\{ \frac{2}{5} (\cos^2\theta-1) \la^2 \la'^2 I^0_0(r) 
 +\frac{4 r^2 \cos\theta \la \la'}{3} I^2_0(r) +\frac{4 \cos\theta \la \la' (r^2 +6 \cos\theta \la \la')}{15} I^1_1(r)\\
& \qquad +\frac{2(\cos^2\theta -1)\la^2\la'^2(2r^4 +3 \cos\theta r^2 \la \la')}{7 r^4} I^0_2(r) \\
& \quad +\frac{2 \cos\theta \la \la' \left(2 r^4 +12\cos\theta r^2 \la\la' +15 (\cos^2\theta-1)\la^2\la'^2 \right)}{15 r^2} I^1_3(r)\\
&\quad +\frac{(\cos^2\theta-1)\la^2\la'^2 \left(6 r^4 +30\cos\theta r^2\la\la' +35 (\cos^2\theta -1)\la^2\la'^2 \right)}{35r^4} I^0_4(r)\bigg\} \,,\\
Z \big|_\text{den-len} =& - \frac{3\Omega_m}{2} b_1 \HH_0^2 \frac{2-5s_2}{\chi_2}D_1(z_1) \int\limits_0^{\chi_2} \integrand\la \frac{\chi_2-\la}{\la}\frac{D_1(\la)}{ a(\la)} \bigg\{ 2\chi_1\la\cos\theta I^1_1(r) -\frac{\chi_1^2\la^2(1-\cos^2\theta)}{r^2} I^0_2(r)\bigg\} \,,\\
Z \big|_\text{rsd-len} =& \frac{3\Omega_m}{2} f_1 \HH_0^2 \frac{2-5s_2}{\chi_2}D_1(z_1) \int\limits_0^{\chi_2} \integrand\la \frac{\chi_2-\la}{\la}\frac{D_1(\la)}{ a(\la)}  \bigg\{ \frac{ \la}{15} (\la-6 \chi_1 \cos\theta+3 \la \cos2 \theta)I^0_0(r)\\ \quad &-\la\frac{6 \chi_1^3 \cos\theta-\chi_1^2 \la \left(9 \cos ^2\theta+11\right)+\chi_1 \la^2 \cos\theta (3 \cos2 \theta+19)-2 \la^3 (3 \cos2 \theta+1)}{21 r^2}I^0_2(r)\\ \quad&-\frac{\la}{35r^4} \bigg[-4 \chi_1^5 \cos\theta-\chi_1^3 \la^2 \cos\theta (\cos2 \theta+7)+\chi_1^2 \la^3 \left(\cos ^4\theta+12 \cos ^2\theta-21\right)\\ \quad&-3 \chi_1 \la^4 \cos\theta (\cos2 \theta-5)-\la^5 (3 \cos2 \theta+1)+12 \chi_1^4 \la\bigg]I^0_4(r) \bigg\} \,,\\
\end{align*}
\endgroup
where $D_1$ represents the scale-independent matter growth function normalized so that $D_1(z = 0) = 1$, $r(\lambda, \lambda') = \sqrt{\lambda^2 + \lambda'^2 - 2 \lambda \lambda' \cos\theta}$ is the comoving distance between the two lines of sight along which we perform the integration, $\theta$ is the angle between the two galaxies as seen by the observer, $\chi_1$ and $\chi_2$ are the comoving distances of the two galaxies, located at redshifts $z_1$ and $z_2$ respectively, and we have defined:
\[
I_\ell^n(r) \equiv \frac{1}{2 \pi^2} \int\limits_0^\infty \integrand k\; k^2\; P(k) \frac{j_\ell(k r)}{(k r)^n},
\]
where $P(k)$ is the linear matter power spectrum at redshift $z = 0$.
The contribution from the term $A B$ to the multipoles of the correlation function is simply given by:
\[
\xi_\ell^{AB}(d, \bar z) = \frac{2\ell + 1}{2} \int\limits_{-1}^1 \integrand\mu\, Z\big|_{AB}(d, \bar z, \mu) P_\ell(\mu),
\]
where $Z\big|_{AB}$ is one of the above.
}

\clearpage

\bibliographystyle{JHEP}
\bibliography{refs}

\end{document}